\documentclass[letterpaper]{article} %
\usepackage[margin = 1in]{geometry}
\usepackage{helvet}  %
\usepackage{courier}  %
\usepackage[hyphens]{url}  %
\usepackage{graphicx} %
\usepackage{natbib}  %
\usepackage{caption} %
\usepackage[linesnumbered,ruled,lined,noend,resetcount]{algorithm2e}
\usepackage{xcolor}
\usepackage{float}
\usepackage{booktabs}

\usepackage{newfloat}
\usepackage{listings}
\DeclareCaptionStyle{ruled}{labelfont=normalfont,labelsep=colon,strut=off} %
\floatstyle{ruled}
\newfloat{listing}{tb}{lst}{}
\floatname{listing}{Listing}
\usepackage{amsmath}
\usepackage{hyperref}
\usepackage{cleveref}

\usepackage{bm}
\let\oldbm\bm
\usepackage{chrkroer}
\usepackage{amssymb}
\usepackage{amsthm}
\usepackage{array}
\usepackage{svg}
\usepackage{pgf}
\usepackage{thm-restate}
\usepackage{multirow}
\usepackage{array}
\usepackage{longtable}
\usepackage{mathtools}
\usepackage{microtype}
\usepackage{colortbl}
\usepackage{inconsolata}

\newcommand\numberthis{\addtocounter{equation}{1}\tag{\theequation}}

\newcommand{\PreserveBackslash}[1]{\let\temp=\\#1\let\\=\temp}
\newcolumntype{C}[1]{>{\PreserveBackslash\centering}p{#1}}
\newcolumntype{R}[1]{>{\PreserveBackslash\raggedleft}p{#1}}
\newcolumntype{L}[1]{>{\PreserveBackslash\raggedright}p{#1}}
\crefname{assumption}{assumption}{assumptions}
\Crefname{assumption}{Assumption}{Assumptions}

\crefname{LT@tables}{table}{tables}
\Crefname{LT@tables}{Table}{Tables}

\crefname{ineq}{inequality}{inequalities}
\Crefname{ineq}{Inequality}{Inequalities}
\creflabelformat{ineq}{#2{\upshape(#1)}#3}

\input{algomacros}

\title{
Efficient Learning in Polyhedral Games via Best Response Oracles
}
\author{
Darshan Chakrabarti \\
IEOR Department \\
Columbia University \\
\texttt{dc3595@columbia.edu} \\
\and
Gabriele Farina \\
EECS Department \\
MIT \\
\texttt{gfarina@mit.edu} \\
\and
Christian Kroer \\
IEOR Department \\
Columbia University \\
\texttt{ck2945@columbia.edu} \\
}

\DeclarePairedDelimiter\parens()
\NewDocumentCommand \divi {mm} {\cD_{\vphi_i}\parens*{#1\,\|\,#2}}

\RenewDocumentCommand \div {mm} {\cD_\vphi\parens*{#1\,\|\,#2}}

\renewcommand\^[1]{^{(#1)}}

\NewDocumentCommand \fw    {} {{\normalfont\texttt{FW}}\xspace}
\NewDocumentCommand \afw    {} {{\normalfont\texttt{AFW}}\xspace}
\NewDocumentCommand \romd   {} {{\normalfont\texttt{ROMD}}\xspace}

\NewDocumentCommand \fwromd {} {{\normalfont\texttt{AFW-ROMD}}\xspace}
\NewDocumentCommand \rfwromd {} {{\normalfont\texttt{rAFW-ROMD}}\xspace}

\NewDocumentCommand \fwomd  {} {{\normalfont\texttt{AFW-OMD}}\xspace}
\NewDocumentCommand \rfwomd  {} {{\normalfont\texttt{rAFW-OMD}}\xspace}

\NewDocumentCommand \ftpl  {} {{\normalfont\texttt{FTPL}}\xspace}
\NewDocumentCommand \rftpl  {} {{\normalfont\texttt{rFTPL}}\xspace}
\NewDocumentCommand \oftpl  {} {{\normalfont\texttt{OFTPL}}\xspace}
\NewDocumentCommand \roftpl  {} {{\normalfont\texttt{rOFTPL}}\xspace}

\NewDocumentCommand \ftrl  {} {{\normalfont\texttt{FTRL}}\xspace}
\NewDocumentCommand \oftrl  {} {{\normalfont\texttt{OFTRL}}\xspace}

\NewDocumentCommand \fp  {} {{\normalfont\texttt{FP}}\xspace}
\NewDocumentCommand \rfp  {} {{\normalfont\texttt{rFP}}\xspace}
\NewDocumentCommand \ofp  {} {{\normalfont\texttt{OFP}}\xspace}
\NewDocumentCommand \rofp  {} {{\normalfont\texttt{rOFP}}\xspace}

\NewDocumentCommand \br  {} {{\normalfont\texttt{BR}}\xspace}
\NewDocumentCommand \rbr  {} {{\normalfont\texttt{rBR}}\xspace}
\NewDocumentCommand \obr  {} {{\normalfont\texttt{OBR}}\xspace}
\NewDocumentCommand \robr  {} {{\normalfont\texttt{rOBR}}\xspace}

\NewDocumentCommand \apo {} {\textsc{APO}\xspace}

\DeclarePairedDelimiter\norm\lVert\rVert
\DeclarePairedDelimiterX{\inp}[2]{\langle}{\rangle}{#1, #2}
\DeclarePairedDelimiter\brackets[]

\renewcommand{\vec}[1]{\oldbm{#1}}

\begin{document}

\onecolumn
\maketitle
\begin{abstract}
    We study online learning and equilibrium computation in games with polyhedral decision sets, a property shared by normal-form games (NFGs) and extensive-form games (EFGs), when the learning agent is restricted to utilizing a best-response oracle.
    We show how to achieve constant regret in zero-sum games and $O(T^{1/4})$ regret in general-sum games while using only $O(\log t)$ best-response queries at a given iteration $t$, thus improving over the best prior result, which required $O(T)$ queries per iteration.
    Moreover, our framework yields the first last-iterate convergence guarantees for self-play with best-response oracles in zero-sum games.
    This convergence occurs at a linear rate, though with a condition-number dependence. We go on to show a $O(1/\sqrt{T})$ best-iterate convergence rate without such a dependence.
    Our results build on linear-rate convergence results for variants of the Frank-Wolfe (\fw{}) algorithm for strongly convex and smooth minimization problems over polyhedral domains.
    These \fw{} results depend on a condition number of the polytope, known as facial distance. In order to enable application to settings such as EFGs, we show two broad new results:
    1) the facial distance for polytopes of the form $\{\vx \in \R^n_{\geq 0} \mid \bA\vx = \vb\}$ is at least $\gamma / \sqrt{k}$ where $\gamma$ is the minimum value of a nonzero coordinate of a vertex in the polytope and $k\leq n$ is the number of tight inequality constraints in the optimal face,
    and
    2) the facial distance for polytopes of the form $\bA\vx = \vb, \bC \vx \leq \vd, \vx \geq \mathbf{0}$ where $\vx \in \R^n$, $\bC \geq \mathbf{0}$ is a nonzero integral matrix, and $\vd \geq \vec{0}$, is at least $1/(\|\bC\|_\infty \sqrt{n})$.
    This yields the first such results for several problems, such as sequence-form polytopes, flow polytopes, and matching polytopes.

\end{abstract}

\vspace{-0.2em}
\section{Introduction}
Learning in games is a well-studied framework in which agents iteratively refine their strategies through repeated interactions with their environment. One natural way for agents to iteratively refine their strategies is by best-responding. This idea can be applied in many forms, the simplest and earliest instance of which was fictitious play (\fp{})~\citep{brown1951iterative}. This algorithm involves the agent observing the strategies played by the opponent and then playing a strategy that corresponds to the best response to the average of the observed strategies. This algorithm was shown to converge~\citep{robinson1951iterative}, but its convergence rate can, in the worst case, scale quite poorly with the number of actions available to each player~\citep{daskalakis2014counter}. It is then natural to ask what are the best convergence guarantees that can be obtained for the computation of Nash equilibria in two-player zero-sum games or coarse correlated equilibria in multiplayer games when agents are learning through a best-response oracle.

In the online learning community, methods based only on best-response oracles are special cases of methods based on a \emph{linear minimization oracle} (LMO), which can be queried for points that minimize a linear objective over the feasible set.
Such methods are known as \emph{projection-free} methods because they avoid potentially expensive projections onto the feasible set.
Projection-free online learning algorithms might perform multiple LMO calls per iteration, so our paper and related literature are concerned not only with the number of iterations $T$ of online learning but also the total number of LMO calls, which we will denote by $N$.
Because LMOs for polyhedral decision sets essentially correspond to a best-response oracle (BRO), we will use these two terms interchangeably.

Follow the Perturbed Leader (\ftpl{})~\citep{kalai2005efficient} was the first such algorithm. When used by both players in two-player zero-sum games, it yields a $O(1/\sqrt{T})$ convergence rate to Nash equilibrium, with a single LMO call in each iteration. More recently, \citet{suggala2020follow} propose an optimistic variant of \ftpl{} (\oftpl{}); \oftpl{} achieves a $O(1/T)$ rate of convergence to Nash equilibrium but requires $O(T)$ LMO calls per iteration, thus corresponding to a $O(1/\sqrt{N})$ rate as a function of the total number of LMO calls. Online Frank-Wolfe (\texttt{OFW})~\citep{hazan2012projection} is another projection-free algorithm, based on the well-studied Frank-Wolfe (\fw{}) algorithm~\citep{frank1956algorithm}. While it does not require multiple calls to an LMO, it can only achieve a $O(T^{-1/3})$ rate for two-player zero-sum games.
We aim to break the $O(1/\sqrt{N})$ barrier in terms of convergence towards two-player zero-sum equilibria and beyond.
We focus on the setting of polyhedral games, a class containing both normal-form and extensive-form games. Our primary contribution is an online learning method which enjoys $O(1/T)$ average individual regret in zero-sum games and $O(1/T^{3/4})$ average individual regret in general-sum games while only requiring $O(\log t)$ LMO calls in iteration $t$.
\Cref{table:algo_comparison} compares our algorithm with other algorithms with the best-known guarantees for the setting we consider. In the table, we also include a non-optimistic version of our algorithm; despite having worse theoretical guarantees than existing projection-free algorithms, it outperforms them in our numerical experiments.

Independent of work in projection-free online learning, there has also been substantial work developing BRO-based algorithms in the game-solving community. Most prominent is the Double Oracle (\texttt{DO}) algorithm~\citep{mcmahan2003planning} for computing Nash equilibria in two-player zero-sum games, which uses BROs and a meta-solver for computing Nash equilibria in a restricted game formed by the returned iterates. More recently, the Policy Space Response Oracle framework~\citep{lanctot2017unified}, a generalization of Double Oracle, has laid the foundation for the design of \texttt{DO}-variants. These algorithms only have theoretical guarantees for computing Nash equilibria in two-player zero-sum games, require a meta-solver for solving the restricted game composed of the strategies chosen by the players at each iteration, and thus are centralized, and do not have convergence rate guarantees to equilibria.

The optimization community has also done substantial work on developing projection-free methods, spurred by the work of~\citet{frank1956algorithm}. Guarantees for \fw{} typically assume the function $f$ being optimized is smooth (has Lipschitz gradient) and convex, the domain $\cX$ being optimized over is convex and compact, and that the algorithm has access to a first-order oracle for the function which returns gradients $\nabla f(\vx)$ at a queried point $\vx$ and an LMO which returns solutions to minimization problems of the form $\argmin_{\vx \in \cX} \langle \vc, \vx \rangle$ for any choice of $\vc \in \R^n$. Given an initial iterate $\vx\^0$, it produces new iterates given by the following update rule:
\begin{align*}
    \vx\^{t+1} & = \frac{t}{t+2} \vx\^{t} + \frac{2}{t+2} \argmin_{\vx' \in \cX} \langle \nabla f(\vx\^t), \vx' \rangle.
\end{align*}

In recent years, there has been work on developing \fw{}-based approaches to saddle-point computation (e.g., \citet{gidel2017frank, lan2016conditional}). However, ~\citet{gidel2017frank} only has fast convergence guarantees for strongly convex-concave objectives
and \citet{lan2016conditional} are only able to provide $O(1/\sqrt{N})$ convergence to saddle-points. On the other hand, our method is able to leverage a \fw{} variant, away-step Frank-Wolfe (\afw{}), to achieve faster convergence rates.

\vspace{-0.2em}
\subsection{Contributions}
We present a projection-free online learning method, Approximate Reflected Online Mirror Descent, using away-step Frank-Wolfe (\fwromd{}), for learning over compact and convex polyhedral decision sets.
\fwromd{} uses reflected online mirror descent (\romd{}), an optimistic online learning algorithm which requires a prediction of the next loss, instantiated with the Euclidean regularizer. The proximal problem for this \romd{} setup is thus strongly convex and smooth. Using the linear convergence of \afw{} for polyhedral domains, we implement approximate steps of \romd{} using only a logarithmic number of \afw{} iterations. We show that even with approximate steps, \romd{} still yields an approximate \emph{regret bounded by variation in utilities} (RVU) bound~\citep{syrgkanis2015fast}, a form of regret guarantee that depends on how much the observed utilities vary, and which has enabled proving constant regret of \emph{optimistic} learning dynamics in two-player zero-sum games and $O(T^{1/4})$ regret in general games~\citep{syrgkanis2015fast}. While using \fw{}-based methods to approximate proximal steps has been previously studied, pioneered by work of ~\citet{lan2016conditional}, it is a surprising blind spot in the literature that the connection to regret guarantees for games has not previously been made.

We then apply our \fwromd{} algorithm to learning in games using only BROs (as opposed to typical learning algorithms, which require access to Euclidean projection or other proximal oracles). We show that when every player employs \fwromd{}, it is possible to converge to a Nash equilibrium in a two-player zero-sum game at a rate of $O(\log N / N)$, where $N$ is the number of BRO calls.
In contrast, the best prior results converged at a rate of $O(1/\sqrt{N})$~\citep{suggala2020follow}.
More generally, we show that \fwromd{} requires only $O(\log t)$ best-response queries at each self-play iteration $t$ while guaranteeing constant social regret, as well as $O(T^{1/4})$ regret for each player after $T$ total iterations of self-play.
We go on to study the \emph{last-iterate convergence} properties of \fwromd{} in self-play in zero-sum settings. We show that, indeed it is possible to retain last-iterate convergence under \fwromd{}, and in fact, it converges at a linear rate (up to error induced by the approximate proximal computation). To the best of our knowledge, these are both the first last-iterate convergence and the first linear-rate convergence results for self-play dynamics that purely rely on best-response oracles.
As with existing linear-rate results in the literature~\citep{tseng1995linear,gilpin2012first,wei2021last}, we have a dependence on a condition number inherent to the game, which is generally hard to evaluate. We show that if one wishes to avoid this condition-number dependence, then self-play with \fwromd{} still achieves a $O(1/\sqrt{T})$ \emph{best-iterate} convergence. Our convergence results to Nash equilibria and coarse correlated equilibria are summarized below using the following two informal theorems:

\begin{theorem}[Informal; Full version in \Cref{thm:ne_regret_bound,thm:asymptotic_last_iterate_approx_sc_omd,thm:linear_rate_approx_sc_omd}] Using \fwromd{} a $\eps'$-Nash equilibrium in a two-player zero-sum game can be computed in $O(1/\eps')$ iterations and $O(\frac{1}{\eps'}\log{\frac{1}{\eps'}})$ LMO calls.
Furthermore, in two-player zero-sum games, \fwromd{} will produce an iterate which is a $\eps'$-Nash equilibrium in $O(\frac{1}{\eps'^2}\log{\frac{1}{\eps'}})$ iterations without a dependence on a problem-dependent constant.
\fwromd{} will produce an iterate which is a $\eps'$-Nash equilibrium in $O(\log{\frac{1}{\eps'}})$ iterations, with this convergence rate having a dependence on a game-dependent constant. Exact dependence on problem parameters can be found in the formal theorems.
\end{theorem}

\begin{theorem}[Informal; Full version in \Cref{thm:cce_regret_bound}] In multiplayer games, \fwromd{} can be used to compute a $\eps'$-coarse correlated equilibrium in $O(\frac{1}{\eps'^{\frac{4}{3}}} \log \frac{1}{\eps'})$ iterations. Exact dependence on problem parameters can be found in the formal theorem.
\end{theorem}

The linear convergence of \afw{} on polyhedral domains is crucial to our result, but the particular rate depends on the facial distance constant of the polytope in question (in addition to the strong convexity and smoothness constants).
To that end, we show two novel lower bounds on the facial distance of a polytope.
Our first result concerns polytopes that can be described in the form $\bA\vx = \vb, \vx \geq \mathbf{0}$ where $\vx \in \bbR^n$. Let $\gamma$ be the minimum value of a nonzero coordinate of a vertex in the polytope. Then, we show that the facial distance is at least $\gamma/\sqrt{n}$.

Moreover, if the optimal solution lies in a face such that $k\leq n$ inequality constraints $x_i\geq 0$ are tight, then the rate dependence in \afw{} can be tightened to $\frac{\gamma}{\sqrt{k}}$.
This theorem immediately implies several useful results, including a bound $1/\sqrt{n}$ on the facial distance of the \emph{sequence-form polytope}, which is the polytope describing the set of feasible strategies of a player in an EFG, where $n$ is the number of sequences for the player. It also implies similar results for flow polytopes and matching polytopes. The fact that the facial distance is only square-root power small in the dimension of the problem ensures that the convergence rate of linearly convergent \fw{} variants over these polytopes does not scale poorly as the ambient dimension of the problem increases.
Our second result concerns an \emph{integral} polytope $\cX$ given by $\bA\vx = \vb, \bC \vx \leq \vd, \vx \geq \mathbf{0}$ where $\vx \in \R^n$, where $\bC \geq \vec{0}$ is a nonzero integral matrix, and $\vd \geq \vec 0$. In particular, for integral polytopes, we are able to handle inequality constraints. In that case, we show that the facial distance is at least $1/(\|\bC\|_\infty \sqrt{n})$.

Finally, we conduct experiments to demonstrate that our algorithm performs well in practice relative to other projection-free algorithms when computing Nash and coarse correlated equilibria in polyhedral games.
\begin{table*}
    \centering
    \setlength\tabcolsep{2mm}
    \scalebox{0.97}{\begin{tabular}{lcccc}
                                                      & \bf $\nabla$ computations & \bf LMO calls        & \bf Social regret  & \bf Avg. social regret         \\
            \bf Algorithm                             & \bf at iteration $t$      & \bf at iteration $t$ & $\sum_i \Reg_i\^T$ & $\sum_i \frac{1}{T} \Reg_i\^T$ \\[1mm]
            \toprule
            \parbox[m][][c]{4.5cm}{\ftpl{}                                                                                                                        \\\citep{kalai2005efficient}}                               & $O(1)$                    & $O(1)$               & $O(\sqrt{T})$     & $O(1/\sqrt{N})$                \\
            \midrule
            \parbox[m][][c]{4.5cm}{Optimistic \ftpl{} (\oftpl)                                                                                                             \\\citep{suggala2020follow}} & $O(1)$                    & $O(T)$               & $O(1)$            & $O(1/\sqrt{N})$                \\
            \midrule
            \rowcolor{gray!20} \fwomd{} [this paper]  & $O(1)$                    & $O(\log t)$          & $O(\sqrt T)$       & $O(\sqrt{\log N/N})$           \\
            \rowcolor{gray!20} \fwromd{} [this paper] & $O(1) $                   & $O(\log t)$          & $O(1)$             & $O(\log N/N)$                  \\
            \bottomrule
        \end{tabular}}
    \caption{A comparison of the number of gradient ($\nabla$) computations, number of LMO calls, cumulative regret (as a function of the total number of iterations $T$), and average regret (as a function of total LMO calls $N$) of various projection-free algorithms. In two-player zero-sum games, average social regret upper bounds the duality gap to Nash equilibrium for the averaged iterates.} 
    \vspace{-3mm}
    \label{table:algo_comparison}
\end{table*}

\subsection{Related Work}\label{sec:additional_related_work}

\paragraph{Frank-Wolfe algorithm.}
\citet{frank1956algorithm} presented the original Frank-Wolfe (\fw{}) algorithm (also known as conditional gradient descent), a projection-free first-order method for solving smooth constrained convex minimization problems. Vanilla \fw{} provides a $O(\frac{1}{t})$ rate of convergence for smooth objectives, but has stronger guarantees for special classes of functions; in particular, for strongly convex functions,\mbox{~\citet{garber2015faster}} showed a $O(\frac{1}{t^2})$ convergence rate. Many \fw{} variants have been developed since Frank and Wolfe's original presentation of the algorithm, including the away-step variant~\citep{wolfe1970convergence}. Importantly for us, \citet{lacoste-julien2015on} showed that \afw{} and several other variants achieve linear convergence for strongly convex and smooth objectives over polyhedral domains; see also \citet{pena2019polytope,beck2017linearly}. There are several excellent overviews of Frank-Wolfe algorithms, e.g., \citet{jaggi2013revisiting, bomze2021frank, braun2022conditional}.

\paragraph{\fw{} for saddle-point problems.}
There has been some work on extending \fw{} to saddle-point problems, prominently by ~\citet{gidel2017frank}. However, they only provide fast convergence guarantees for strongly convex-concave objectives (and moreover, require a significant degree of strong convexity, thus rendering their results incompatible with smoothing techniques). They do provide a convergence guarantee for the bilinear case when the feasible sets are polytopes, but it is extremely slow.
\citet{lan2016gradient} introduced the idea of ``gradient sliding,'' which involves computing approximate prox steps to save on the number of gradient computations required to optimize a composite function with a smooth and nonsmooth component. In the spirit of this work, \citet{lan2016conditional} introduced ``conditional gradient sliding,'' which involves using conditional gradient methods to compute these approximate prox steps. They combine this idea with Nesterov acceleration to achieve $O(1/\sqrt{\epsilon})$ gradient computations and $O(1/\epsilon)$ LMO calls for smooth functions. They present a smoothed version of their algorithm as well that can be applied to saddle-point problems, but the smoothing degrades the guarantee to $O(1/\epsilon)$ gradient computations and $O(1/\epsilon^2)$ LMO calls.

\paragraph{Projection-free online learning.}
The first projection-free online learning algorithm was \ftpl{}, introduced by~\citet{kalai2005efficient}. The algorithm involves randomly perturbing the sum of the observed losses (which serves as a form of regularization, see e.g., \citet{abernethy2016perturbation}) before computing the best response, and achieves $O(1/\sqrt{T})$ average regret for linear loss functions. ~\citet{suggala2020follow} introduced \oftpl{}, which achieves $O(1/T)$ average regret for players in zero-sum games but requires doing $O(T)$ LMO calls at every iteration. ~\citet{hazan2012projection} presented \texttt{OFW} which uses \fw{} to achieve $O(1/T^{1/4})$ average regret for Lipschitz convex losses.

\paragraph{Algorithms for game solving.}
There has been much work done to develop efficient algorithms for game-solving.
We only touch on the major trends related to discrete-time methods and sequential games.
A line of research has focused on constructing no-regret algorithms with ergodic convergence to equilibrium. Out of these, we highlight two categories: methods based on the \texttt{CFR} regret-decomposition framework~\citep{zinkevich2007regret,farina2021faster,tammelin2015solving,brown2019solving,lanctot2009monte}, and methods based on the \texttt{OMD} framework and more generally first-order methods~\citep{hoda2010smoothing,kroer2020faster,farina2021better,farina2022kernelized}. Some of these methods were key in solving large games, such as poker~\citep{bowling2015heads,brown2018superhuman,brown2019superhuman}.
A recent trend has focused on establishing learning algorithms with (poly)logarithmic per-player regret when used in self-play, including~\citet{anagnostides2023near,anagnostides2022on,daskalakis2015near,daskalakis2021near,wibisono2022alternating,anagnostides2022uncoupled,farina2022near}. These methods are able to compute equilibria in multiplayer games at the rate of $\tilde O(1/T)$. Finally, another recent trend in the literature has focused on establishing learning algorithms~\citep{wei2021last, lee2021last} and first-order methods~\citep{tseng1995linear,gilpin2012first} with guarantees for last-iterate convergence.

\vspace{-0.2em}
\section{Notation and Preliminaries} \label{sec:preliminaries}
We will use $\|\cdot \|_p$ to denote the $\bell_p$-norm and $\| \cdot \|$ without subscript to denote $\| \cdot \|_2$. Any norm-dependent quantity (e.g., diameter, facial distance, strong convexity, and smoothness) will be with respect to the Euclidean norm (which is self-dual) unless otherwise noted. Because we are principally concerned with using these algorithms for equilibrium computation in games, we will use subscripts to indicate a set or constant corresponding to a particular agent. We will use $[n]$ to denote the set $\{1, \dots, n \}$, and $L$-smoothness refers to Lipschitz continuity of the gradient, with modulus $L$.

\subsection{Online Linear Optimization}
In online learning, an agent $i$ repeatedly interacts with an environment, aiming to minimize its \emph{regret}. At each time $t$, the agent chooses a strategy $\vx_i\^{t}$ from a given feasible set $\cX_i \subseteq \R^{n_i}$ and then receives a loss vector $\bell\^{t}_i \in \cX_i \to \R$. The loss is allowed to depend adversarially on $\vx_i\^{t}$. The agent then pays a cost of  $\inp{\bell_i\^{t}}{\vx_i\^{t}}$. The (cumulative) regret $\Reg_i\^T$ after $T$ iterations is defined as $\max_{\vx' \in \cX_i} \sum_{t = 1}^{T} \inp{\bell_i\^{t}}{\vx_i\^{t}} - \inp{\bell_i\^{t}}{\vx'}$, and average regret is defined as regret divided by the number of iterations. We will assume that losses are bounded and normalized: $\| \bell\^{t}_i \| \leq 1$ for all $t \in [T]$.

In order to achieve desired regret guarantees, online learning algorithms typically require some form of regularization. While \ftpl{} achieves this regularization through randomization, the framework of algorithms utilizing approximate prox calls that we present will require access to a regularizer $\vphi_i : \cX_i \to \R$, which is 1-strongly convex and $L_i$ smooth on $\cX_i$. The Bregman divergence between $\vx, \vy\in \cX_i$ is denoted by $\divi{\vx}{\vy}$.
Furthermore, we define $\Omega_i \defeq \sup_{\vx, \vy \in \cX_i} \divi{\vx}{\vy}$ and $D_i \defeq \sup_{\vx, \vy \in \cX_i} \|\vx - \vy \|$. $\delta(\cX_i)$ will be used for the facial distance of $\cX_i$, defined in \Cref{subsec:frank_wolfe_facial_distance}. For a given set, $\cX$, and a point $\vx \in \cX$, we denote $\dist(\vx, \cX) \defeq \inf_{\vx' \in \cX} \|\vx - \vx' \|$ and in the case that $\cX$ is compact, define $\Pi_{\cX}(\vx) = \argmin_{\vx' \in \cX} \norm{\vx - \vx'}$.

Online Mirror Descent (\texttt{OMD}) is an algorithm which performs a single proximal computation at every iteration of the algorithm, generating iterates as follows:

\begin{equation*}
    \vx_i\^{t+1} = \argmin_{\vx_i \in \cX_i} \biggl\{ \langle \bell_i\^t, \vx_i \rangle + \frac{1}{\eta} \divi{\vx_i}{\vx_i\^t} \biggr\}.
\end{equation*}

It enjoys $O(1/\sqrt{T})$ average regret (e.g., ~\citet{hazan2016introduction,orabona2019modern}).

Reflected Online Mirror Descent (\romd{}) is an optimistic version of \texttt{OMD} which utilizes a prediction $\vm_i\^{t+1}$ of the next loss $\bell_i\^{t+1}$ to generate the iterate at time $t+1$.

\begin{align*}
    \vx_i\^{t+1} & = \argmin_{\vx_i \in \cX_i} \biggl\{ \langle \bell_i\^t + \vm_i\^{t+1} - \vm_i\^{t}, \vx_i \rangle + \frac{1}{\eta} \divi{\vx_i}{\vx_i\^{t}}\biggr\}.
\end{align*}

It is common to use the last observed loss as the prediction for the next loss: set $\vm_i\^{t+1}$ equal to $\bell_i\^{t}$. In this case, \romd{} achieves $O(1/T)$ average regret \citep{malitsky2015projected, joulani2017modular} in self-play. Since $\vm_i\^{t+1}$ is the prediction of a loss $\bell_i\^{t+1}$ which is assumed to have norm bounded by 1, we will assume that $\|\vm_i\^{t}\| \leq 1$ for all $t \in [T]$. 

\citet{syrgkanis2015fast} introduce the notion of Regret bounded by Variation in Utilities (RVU), recalled next, and demonstrate that algorithms with this property exhibit faster convergence to equilibria in games.
\begin{definition}[RVU~\citep{syrgkanis2015fast}]
    A learning algorithm for Player $i$ is said to satisfy the RVU property if for some $\alpha, \beta, \gamma > 0$, and all possible $\bell_i\^1, \dots, \bell_i\^T$,
    \begin{align*}
        \Reg_i\^T \leq \alpha + \beta \sum_{t=1}^{T} \|\bell_i\^t - \bell_i\^{t-1}\|^2 - \gamma \sum_{t=1}^{T} \| \vx_i\^t - \vx_i\^{t-1}\|^2.
    \end{align*}
\end{definition}

\romd{} satisfies this inequality with $\alpha = \Omega/\eta$, $\beta = \eta$, $\gamma = 1/4\eta$; we are not aware of a reference for this, but it can be shown very similarly to known results for optimistic \texttt{OMD}.
Later we will show in \Cref{lemma:approx_sc_omd_refined_rvu} that our approximate \romd{} in \Cref{algo:approx_sc_omd} still satisfies the RVU property.

\subsection{Game-Theoretic Notions}
\emph{Normal-form games} (NFGs) model single-shot simultaneous interactions among a set of agents denoted by $[N]$. The agents each have a set of possible actions $\cA_i$ and a normalized utility function $u_i: \prod_{i \in [N]} \cA_i \rightarrow [-1, 1]$, the latter specifying their payoff for a given choice of actions by each of the agents. The game is said to be \emph{zero-sum} if $\sum_{i \in [n]} u_i(\va)  = 0$ for all $\va \in  \prod_{i \in [N]} \cA_i$. A mixed-strategy $\vx_i$ for Player $i$, is a probability distribution over $\cA_i$; $\vx_i \in \Delta(\cA_i)$. We can extend the domain of $u_i$ to be over $\Delta(\cA_i)$ by taking the expectation of the utility function over the distribution over $\cA_i$ induced by $\vx_i \in \Delta(\cA_i)$.

Nash equilibrium is the de facto notion of equilibrium in NFGs, and the problem of computing a Nash equilibrium (NE) in two-player zero-sum games can be formulated as a bilinear saddle-point problem (BSPP):

\begin{equation*}\tag{BSPP}
    \min_{\vx \in \cX} \max_{\vy \in \cY} \langle \bA\vx, \vy \rangle.
    \label{eq:bspp}
\end{equation*}

In this case, $\cX$ and $\cY$ are the space of mixed strategies for Player 1 and Player 2, respectively, and $\bA$ encodes the utility of Player 2 for a given choice of strategies for both players. The duality gap $\xi$ of $\bar{\vx}$ and $\bar{\vy}$ for \eqref{eq:bspp} can be defined as $\max_{\vy \in \cY} \langle \bA \bar{\vx}, \vy \rangle - \min_{\vx \in \cX} \langle \bA\vx, \bar{\vy} \rangle$. This quantity is typically used to measure the quality of a solution; in the case of two-player zero-sum games, a duality gap of $\eps'$ corresponds to an $\eps'$-NE (and thus is also known as Nash gap).

\begin{definition}[$\eps'$-coarse correlated equilibrium]
    An $\eps'$-coarse correlated equilibrium ($\eps'$-CCE) is defined as $\vx \in \Delta(\prod_{i \in [N]} \cA_i)$ such that
    \begin{align*}
        \E_{\va \sim \vx} [u_i(\va)] \geq \E_{\va_{-i} \sim \vx} [u_i(\va'_i, \va_{-i})] - \eps'
    \end{align*}

    for all players $i \in [N]$, for all $\va'_i \in \cA_i$, for $\eps' \geq 0$; $\eps' = 0$ corresponds to an exact CCE.

\end{definition}
\emph{Extensive-form games} (EFGs) are a generalization of normal-form games, which also allow for modeling of sequential moves (and also private and/or imperfect information and stochasticity). Equilibrium computation for two-player zero-sum EFGs can also be formulated as \eqref{eq:bspp}, by letting $\cX$ and $\cY$ be convex polytopes known as \emph{sequence-form polytopes}~\citep{romanovskii1962reduction,koller1996efficient,stengel1996efficient}, and letting $\bA$ be a matrix representing the utility of Player 2 for a given choice of strategies for both players. An important property that sequence-form polytopes have is they can be expressed in the form $\bA \vx = \vb, \vx \geq 0$; this will allow us to characterize the facial distance of these polytopes. Additional background on EFGs can be found in \Cref{sec:additional_prelims}.

When we are analyzing multiplayer games involving $N$ agents, we will let $\cX_i \subset \R^{n_i}$ denote the convex and compact set of strategies for the $i^{th}$ player, where $i \in [N]$, and let $\vx_i \in \cX_i$ represent their strategy. For NFGs, $\cX_i$ is $\Delta(\cA_i)$, the set of mixed strategies for $i$, while for EFGs, $\cX_i$ is the sequence-form polytope for Player $i$. In the case of two-player zero-sum NFGs or EFGs, in which case Nash equilibrium computation corresponds to \eqref{eq:bspp}, we will let $\cX = \cX_1$, $\cY = \cX_2$, and $\cZ = \cX \times \cY$. In this case, we will use $\cZ^*$ to denote the set of solutions to the BSPP. We define the vector field $\vec{F}(\vz) = (\bA^T \vy, -\bA \vx)$ for $\vz \in \cZ$. Without loss of generality, we assume that $\vec{F}$ is smooth with constant 1 (the payoff matrix $\bA$ can be scaled to ensure this is the case).

\subsection{Saddle-Point Metric-Subregularity}
Problems of the form \eqref{eq:bspp} satisfy a condition known as \emph{Saddle-Point Metric Subregularity}~\citep{wei2021last} as long as $\cX$ and $\cY$ are convex polytopes (this is the case for NFGs and EFGs).

\begin{definition}[Saddle-Point Metric Subregularity]
    The SP-MS condition is satisfied if for any $\vz \in \cZ \setminus \cZ^*$ with $\vz^* = \Pi_{\cZ}(\vz)$
    for some $\beta \geq 0$ and $\nu > 0$,
    \begin{align}
        \tag{SP-MS}
        \sup_{\vz' \in \cZ} \frac{\langle \vec{F}(\vz) , \vz-\vz'\rangle}{\|\vz - \vz'\|} \geq \nu \|\vz - \vz^*\|^{\beta + 1}.
    \label[ineq]{ineq:spms}
    \end{align}
\end{definition}

For any given NFG or EFG, there exists $\nu \geq 0$ so that \Cref{ineq:spms} holds with $\beta = 0$; given a choice of a game, we will use $\nu$ to refer to this problem-dependent constant. \citet{wei2021last} use this condition to demonstrate linear last-iterate convergence of certain online learning algorithms. Earlier works~\citep{tseng1995linear, gilpin2012first} showed linear last-iterate convergence using error bounds, and \citet{wei2021last} note that there is a close correspondence between the \ref{ineq:spms} condition and error bound techniques for bilinear polyhedral settings.

\subsection{Frank-Wolfe and Facial Distance}
\label{subsec:frank_wolfe_facial_distance}
Frank-Wolfe is a projection-free algorithm that converges with rate $O(1/T)$ for smooth convex functions over convex compact sets. In certain situations, faster convergence rates can be obtained; for example, when the function is strongly convex and the optimal solution lies in the relative interior of the feasible set, the original \fw{} algorithm is linearly convergent~\citep{guelat1986some}. In the case where the optimal solution is not in the interior, away-step Frank Wolfe (\afw{}) is a variant of Frank-Wolfe which was shown to achieve linear convergence for strongly convex objectives over polyhedral sets~\citep{wolfe1970convergence, guelat1986some,lacoste-julien2015on}. Pseudocode for \afw{} is provided in~\Cref{sec:afw}. The linear rate of \afw{} and several other linearly convergent variants of \fw{} depends on a condition number of the polytope known as the facial distance.
It can be defined concisely using a theorem from~\citep{pena2019polytope}:
\begin{definition}[Facial distance,~\citet{pena2019polytope}]

    \begin{align*}
        \delta(\cP) = \min_{\substack{\cF \in \operatorname{faces}(\cP) \\ \emptyset \subseteq \cF \subseteq \cP}} \dist(\cF, \conv(\operatorname{Vert}(\cP) \setminus \cF) ).
    \end{align*}
\end{definition}

For linearly convergent methods for strongly convex functions, the ratio between the strong convexity modulus and smoothness constant appears as a term in the linear rate. For linearly convergent \fw{} variants, both the former ratio and the ratio between the facial distance and diameter over the polyhedral domain appear as terms in the linear rate.

\begin{theorem}[Convergence rate of \afw{} for strongly convex functions over polyhedral sets~\citep{lacoste-julien2015on}\footnote{\emph{Pyramidal width} was used in the original linear-rate result~\citep{lacoste-julien2015on}. It was shown to be equivalent to facial distance by~\citet{pena2019polytope}.}]
    In order to compute an $\eps$-optimal solution to a 1-strongly convex, $L$-smooth function over a convex polytope that has diameter $D$ and facial distance $\delta$, \afw{} requires $O(\frac{LD^2}{\delta^2}\log\frac{LD}{\eps})$ LMO calls.
\end{theorem}

This dependence on the facial distance to diameter ratio makes it desirable to have a lower bound on the facial distance, to ensure that the linear rate scales well with the size of the problem.

\vspace{-0.2em}
\section{New Results on Polyhedral Facial Distance}
In this section, we present two theorems which characterize lower bounds on the facial distance in special cases where the constraints of the polytope can be written in a certain form. The proofs are deferred to \Cref{sec:facial_distance}.

\begin{restatable}{theorem}{thmfacialdistanceone}
    Let $\cP$ be a polytope given by $\bA\vx = \vb, \vx \geq \mathbf{0}$ where $\vx \in \bbR^n$. Let $\gamma$ be the minimum value of a nonzero coordinate of a vertex. Then $\delta(\cP) \geq \frac{\gamma}{\sqrt{n}}$. Moreover, if the optimal solution lies in a face $\cF$ such that $k$ coordinates are zero (\textit{i.e.},
    $|\{i: x_i = 0\}| = k$), then $\delta(\cP) \geq \frac{\gamma}{\sqrt{k}}$.
    \label{thm:facial_distance_1}
\end{restatable}

\begin{corollary}
    For any $0/1$-polytope $\cX$, $\delta(\cX) \geq \frac{1}{\sqrt{n}}$.
\end{corollary}

The corollary follows by noting that \Cref{thm:facial_distance_1} holds with $\gamma = 1$ for $0/1$ polytopes. Note that simplices (the strategy spaces of NFGs), sequence-form polytopes (the strategy spaces of EFGs), flow polytopes, and matching polytopes are all $0/1$ polytopes. The fact that the facial distance decreases only at a square-root rate in the dimension of the problem ensures reasonable scaling of \afw{} as the ambient dimension of the problem increases.

In the case we are dealing with integral polytopes (polytopes with integral vertices), we can also handle inequality constraints beyond the non-negativity constraints.

\begin{restatable}{theorem}{thmfacialdistancetwo}
    Let $\cP$ be an integral polytope given by $\bA\vx = \vb, \bC \vx \leq \vd, \vx \geq \mathbf{0}$ where $\vx \in \R^n$, with $\bC \geq \vec{0}$ a nonzero integral matrix, and $\vd \geq \vec{0}$. Then $\delta(\cP) \geq \frac{1}{\|\bC\|_\infty \sqrt{n}}$.
    \label{thm:facial_distance_2}
\end{restatable}

We note that facial distance and essentially equivalent notions, have been considered non-trivial to evaluate~\citep{garber2016linear, bashiri2017decomposition, braun2022conditional} for most polytopes besides hypercubes, unit $\ell_1$ balls and simplices. Thus, our lower bounds contribute to a more complete characterization of convergence rates of linearly convergent \fw{} variants over a fairly broad class of polytopes.

\section{Approximate Reflected Online Mirror Descent using Away-step Frank-Wolfe}
In this section, we propose an algorithmic framework that uses approximate proximal updates instead of exact proximal updates.
First, we will show that such an approximate variant of \romd{} still retains many of the nice properties of \romd{}, up to the error in the approximation oracle.
Then, we propose the use of linearly convergent variants of \fw{} for implementing the approximate proximal step, specifically when the regularizer is smooth, and strongly convex (as is the case with the Euclidean regularizer) and the decision set is a convex polytope, which is the case in NFGs and EFGs.
We abstract away the concept of computing an approximate proximal update using what we call an approximate proximal oracle (APO).

\begin{definition}[Approximate proximal oracle]
    An \apo{}$_\cX$, given a choice of convex and compact set $\cX$, takes as input a function $f : \cX \to \R$, a $L$-smooth and 1-strongly convex regularizer $\vphi$, a prox center $\vx_c$, and a desired accuracy $\eps \geq 0$, and returns $\vx' \in \cX$ such that
    \begin{equation*}
        f(\vx') + \div{\vx'}{\vx_c} \leq \min_{\vx \in \cX} \Big\{ f(\vx) + \div{\vx}{\vx_c}\Big\} + \eps.
        \tag{APO}
        \label{eq:apo}
    \end{equation*}
\end{definition}

While our framework can be adapted to a variety of online learning algorithms, we illustrate the framework using \romd{} in \Cref{algo:approx_sc_omd}. $\vx_i\^t$ is the iterate returned by the algorithm at iteration $t$, $\bell_i\^t$ is the loss received at iteration $t$, and $\vm_i\^t$ is the prediction of the loss to be used at iteration $t$. Proofs for results in this section are deferred to \Cref{sec:approx_sc_omd_proofs}. The subscript $i$ is dropped in statements about a single regret-minimizing agent applying the algorithm; the only exceptions are this paragraph and the pseudocode in \Cref{algo:approx_sc_omd}.

\begin{figure}[H]
    \removelatexerror\begin{algorithm}[H]
        \caption{Reflected Gradient \texttt{OMD} with Approximate Proximal Computation (for a generic Player~$i$)}
        \label{algo:approx_sc_omd}
        \KwData{%
        $\cX_i \subseteq \R^n$: convex and compact set,
        $\vphi_i: \cX \to \R_{\geq 0}$: $L$-smooth, 1-strongly convex,
        $\eta_i > 0$: step-size parameter,
        $\eps_i\^{t}$: desired accuracy of prox call at each $t$,
        $\operatorname{\apo{}_\cX}$: an \apo{} for $\cX_i$,
        $\vx_i\^0 \in \cX_i$,
        $\bell_i\^0 = \vm_i\^0 = \mathbf{0}$ }
        \Fn{NextStrategy($\vm_i\^t$)}{
            \Return{$\operatorname{APO}_{\cX_i}\Big(-\eta_i \langle \bell_i\^{t-1} + \vm_i\^t - \vm_i\^{t-1}, \,\cdot\, \rangle,
                    \vphi_i, \vx_i\^{t-1}, \eps_i\^{t}\Big)$}
        }
    \end{algorithm}
\end{figure}

We present the following lemma, which characterizes the cumulative regret of \Cref{algo:approx_sc_omd} when using an \apo{}.
\begin{restatable}{lemma}{lemmaapproxscomdregret}
    Let $\vx^* = \argmax_{\vx' \in \cX} \sum_{t = 1}^{T} \inp{\bell\^{t}}{\vx\^{t}} - \inp{\bell\^{t}}{\vx'}$.
    \Cref{algo:approx_sc_omd} yields
    \vspace{-2mm}
    \begin{align*}
        \Reg\^T & \leq \sum_{t = 1}^{T} \|\bell\^{t}  - \vm\^{t}\| \cdot \|\vx\^{t+1} - \vx\^{t}\|     - \frac{1}{2\eta} \sum_{t=1}^{T} \|\vx\^{t+1} - \vx\^{t}\|^2                       \\
        &\eqskip + \frac{1}{\eta} \div{\vx^*}{\vx\^{0}}  + \inp[\big]{\vm\^1}{\vx\^2 - \vx^*} + \sum_{t=1}^{T} \frac{\eps\^{t}}{\eta}.
    \end{align*}
    \label{lemma:approx_sc_omd_regret}
    \vspace{-2mm}
\end{restatable}

In the rest of this section, for the sake of brevity, we state our results using $\eps\^t = \frac{1}{t^2}$, and the last observed loss as the prediction $\vm\^t = \bell\^{t-1}$.
In this case, we obtain the following refined result:
\begin{restatable}{lemma}{lemmarefinedrvu}
    \Cref{algo:approx_sc_omd} with $\eps\^{t} = \frac{1}{t^2}$-optimal prox computations at each time step and using $\vm\^t = \bell\^{t-1}$ yields
    \begin{align*}
        \Reg\^T & \leq \frac{\Omega + 2}{\eta} + {\eta} \sum_{t = 1}^{T}  \|\bell\^{t}  - \bell\^{t-1}\|^2  - \frac{1}{4\eta} \sum_{t=1}^{T} \|\vx\^{t+1} - \vx\^{t}\|^2.
    \end{align*}
    In particular, this satisfies the RVU property with $\alpha = \frac{\Omega+2}{\eta}, \beta = \eta, \gamma = \frac{1}{4\eta}$.
    \label{lemma:approx_sc_omd_refined_rvu}
\end{restatable}

We refer to \Cref{algo:approx_sc_omd} instantiated with \afw{} (\Cref{algo:afw}) as \textbf{\fwromd{}}. \Cref{lemma:approx_sc_omd_refined_rvu} immediately enables us to state several results on \Cref{algo:approx_sc_omd}'s ergodic convergence to equilibrium and characterize average regret in terms of LMO calls for \fwromd{}.

\begin{restatable}{theorem}{thmneregretbound}
    An $\eps'$-Nash equilibrium in any two-player zero-sum polyhedral game can be computed in $O(1/\eps')$ iterations of \Cref{algo:approx_sc_omd}. This corresponds to $O(\max_{i \in \{1,2\}} \frac{1}{\eps'} \frac{L_iD_i^2}{\delta_i^2} \log \left[ \frac{L_iD_i}{\eps'} \right])$ LMO calls when using \fwromd{}.
    \label{thm:ne_regret_bound}
\end{restatable}

\begin{restatable}{theorem}{thmcceregretbound}
    An $\eps'$-CCE in any $N$-player general-sum polyhedral game can be computed in $O(1/{\eps'}^\frac{4}{3})$ iterations of \Cref{algo:approx_sc_omd}. This corresponds to $O(\max_{i \in [N]} \frac{1}{{\eps'}^\frac{4}{3}} \frac{L_iD_i^2}{\delta_i^2} \log \left[{\frac{L_iD_i}{\eps'}}\right])$ LMO calls when using \fwromd{}.
    \label{thm:cce_regret_bound}
\end{restatable}

\subsection{Last Iterate Convergence}
Next, we obtain asymptotic last-iterate convergence to an approximate Nash equilibrium, when $\sum_{i=1}^{N} \Reg_i\^t \geq 0$ for any $t \in \N$, adapting a result from~\citet{anagnostides2022on}. A wide class of games, including two-player NFGs and EFGs, polymatrix zero-sum games, constant-sum polymatrix games, strategically zero-sum games, and polymatrix strategically zero-sum games satisfy this condition on social regret ~\citep{anagnostides2022on}; thus, our result holds for this class of games as well.

\begin{figure*}[t]
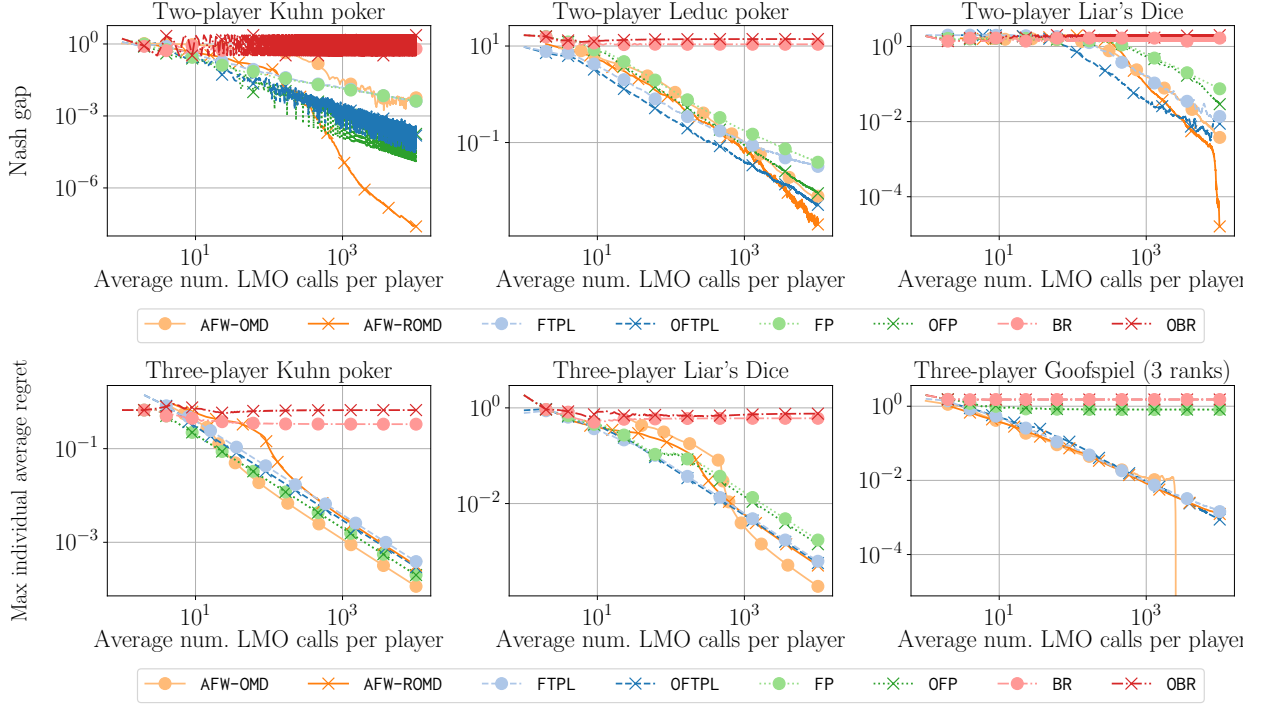

    \centering
    \resizebox{\textwidth}{!}{\input{new_figures/NE_no_restarts.pgf}}\\
    \resizebox{\textwidth}{!}{\input{new_figures/CCE.pgf}}
    \caption{Convergence to equilibrium (Nash eq. and CCE) as a function of average LMO calls per player for \fwomd{}, \fwromd{}, \ftpl{}, \oftpl{}, \fp{}, \ofp{}, \br{}, and \obr{}.}
    \vspace{-5mm}
    \label{fig:NE_CCE_alg_comparison}
\end{figure*}

\begin{restatable}{theorem}{thmasymptoticlastiterate}
    For any $N$-player general-sum polyhedral game, given $\eps \in (0,1)$, let Player $i$ employ the above framework with $\eps_i\^t =\eps^2$ and $\vm_i\^t = \bell_i\^{t-1}$. Let $\eta_{\mathrm{max}} \leq \frac{1}{2\sqrt{2}(N-1)}$ where $\eta_{\mathrm{max}} = \max_{i \in [N]} \eta_i$ and suppose $\sum_{i=1}^{N} \Reg_i\^t \geq 0$ for any $t \in \N$. 
    Define $\alpha_i = \parens*{\frac{1}{\eta_i} + \frac{2 \Omega_i}{\eta_i}(L_i + N - 1) + 1}$.
    Then, after $T > \left \lceil \frac{8 \eta_{\mathrm{max}}}{\eps^2} \sum_{i=1}^{N} \frac{(\Omega_i + 2)}{\eta_i} \right \rceil$ iterations, there exists $\vx\^{t}$ with $t \in [T]$ which is an
    $\eps \parens*{\max_{i \in [N] } \sqrt{2\eta_i}(\frac{2L_i D_i}{\eta_i} + 3) + \alpha_i}$-approximate Nash equilibrium.
    \fwromd{} will yield an iterate that is an $\eps'$-approximate Nash equilibrium in $O\parens*{ \max_{j \in [N]}\left\{\frac{\eta_{\mathrm{max}} \alpha_j^2}{\eps'^2} \sum_{i=1}^{N} (\frac{\Omega_i + 2}{\eta_i}) \frac{L_iD_i^2}{\delta_i^2} \log \left[\frac{L_iD_i\alpha_j}{\eps'}\right]\right\}}$ LMO calls when $\eps \leq \min_{i \in [N]} \frac{\eps'}{\alpha_i}$.
    \label{thm:asymptotic_last_iterate_approx_sc_omd}
\end{restatable}

In the two-player zero-sum case, we also obtain last-iterate linear-rate convergence to $\eps'$-equilibria when instantiated with the Euclidean regularizer, $\vphi_i(\vx_i) = \frac{1}{2}\|\vx_i\|_2^2$ for $i \in \{1, 2\}$, for any choice of $\eps'$.

\begin{restatable}{theorem}{thmlinearrate}
    In any two-player zero-sum polyhedral game, both players employing \Cref{algo:approx_sc_omd} with $\vm_i\^t = \bell_i\^{t-1}$, $\eps_i\^t = \eps $, $\vphi_i(\vx_i) = \frac{1}{2}\|\vx_i\|_2^2$, and $\eta_i = \eta \leq \frac{1}{4}$ yields linear last-iterate convergence to a $\frac{(16 + C_1)\eps + 32\max_{i \in \{1, 2\}} \sqrt{2\eta\eps} (2L_iD_i + 3 \eta)}{C_2}$-approximate Nash equilibrium, where $\nu$ is a game-dependent constant associated with the \ref{ineq:spms} condition, $C_1 = 2(1 +  \frac{4\eta^2 \nu^2}{25})$, and $C_2 = \min(\frac{1}{2}, \frac{\eta^2 \nu^2}{25})$:
    \[\operatorname{dist}(\vz\^{t}, \cZ^*)^2 \leq 2 \parens*{1 + \frac{C_2}{4}}^{-t} \dist(\vz\^1 , \cZ^*)^2 + \frac{(16 + C_1)\eps + 32\max_{i \in \{1, 2\}} \sqrt{2\eta\eps} (2L_iD_i + 3 \eta)}{C_2}. \]
    In the same setting ($\vm_i\^t = \bell_i\^{t-1}$, $\eps_i\^t = \eps $, and $\eta_i  = \eta \leq \frac{1}{4}$), if it is assumed that both players are applying \fwromd{}, then they can achieve linear last-iterate convergence to a $\frac{48+C_1}{C_2}\eps$-approximate Nash equilibrium, with the same definitions for $\nu, C_1, C_2$.
    \[\operatorname{dist}(\vz\^{t}, \cZ^*)^2 \leq 2 \parens*{1 + \frac{C_2}{4}}^{-t} \dist(\vz\^1 , \cZ^*)^2 + \frac{48 + C_1}{C_2} \eps. \]
    \fwromd{} requires
    \[O\left(\max_{i \in \{1,2\}} \frac{\log \frac{2C_2+ 48 + C_1}{C_2\eps'}}{\log \frac{4+C_2}{4}} \frac{L_iD_i^2}{\delta_i^2} \log \left[\frac{(2C_2+ 48 + C_1)L_iD_i}{C_2\eps'} \right] \right).
    \]
    LMO calls to compute an $\eps'$-NE. Furthermore, the approximate solution it returns will have support of size $O(\max_{i \in \{1,2\}}\frac{L_iD_i^2}{\delta_i^2}\log \left[\frac{L_iD_i(2C_2+ 48 + C_1)}{C_2\eps'}\right])$.
    \label{thm:linear_rate_approx_sc_omd}
\end{restatable}
\begin{figure*}[t]
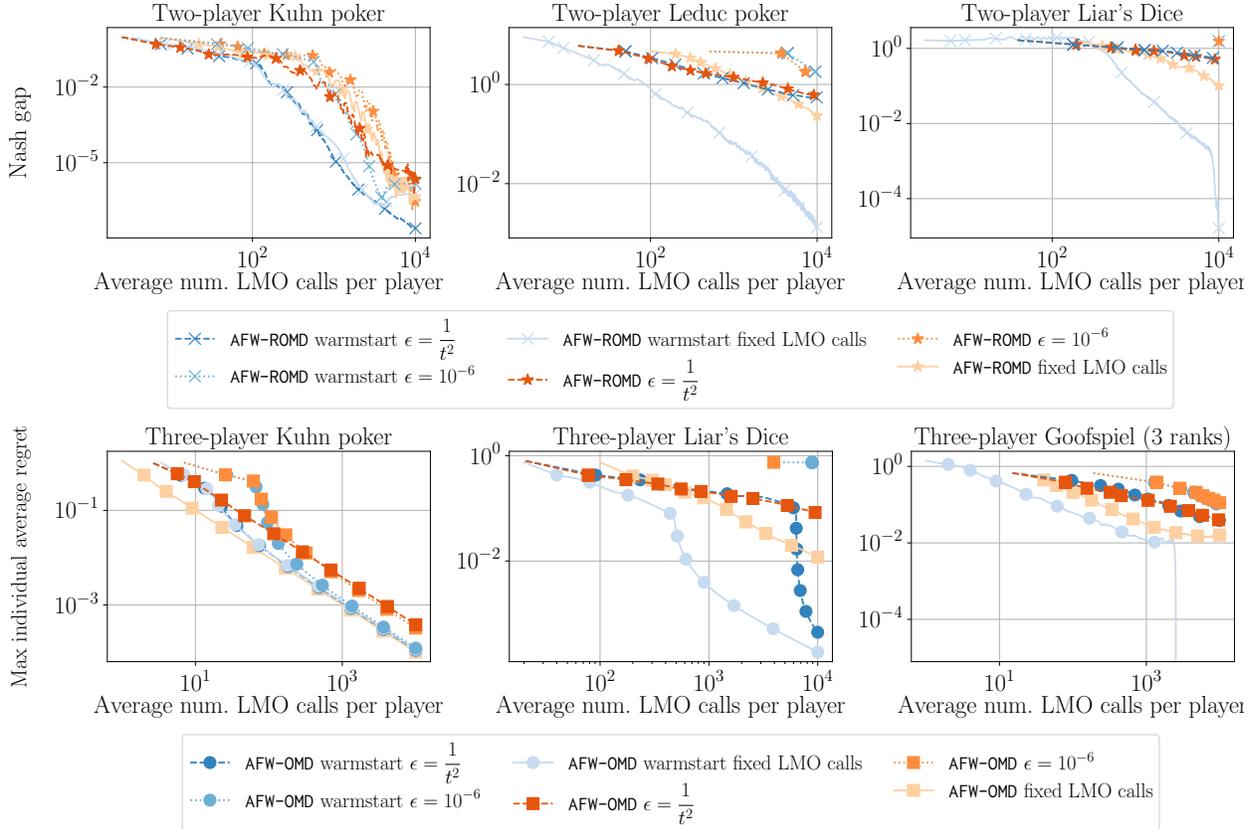

    \centering
    \resizebox{\textwidth}{!}{\input{new_figures/NE_no_restarts_ablation.pgf}}\\
    \resizebox{\textwidth}{!}{\input{new_figures/CCE_ablation.pgf}}
    \caption{Convergence to equilibrium (Nash and CCE) as a function of average per-player LMO calls for \fwromd{} (top row, Nash) and \fwomd{} (bottom row, CCE) for different \apo{} termination criteria, and with and without warmstarting. The best averaging scheme is chosen for each warmstart and termination criterion setting.}
    \vspace{-3mm}
    \label{fig:NE_CCE_ablation}
\end{figure*}

\section{Experimental Results and Discussion}
\label{sec:experiments}
We conduct experiments on standard EFG benchmarks to demonstrate the numerical performance of our algorithm relative to known algorithms from the literature. Details of games are provided in \Cref{sec:game_descriptions}.
In addition to evaluating \fwromd{}, we also consider its non-optimistic variant (taken by letting $\vm\^t = \mathbf{0}$ for all $t \in [T]$). We call this algorithm \fwomd{} since it corresponds to using \afw{} as an \apo{} in vanilla \texttt{OMD}. We use the Euclidean regularizer, $\vphi_i(\vx_i) = \frac{1}{2}\|\vx_i\|_2^2$ for $i \in [N]$.

We compare against \ftpl{} and \oftpl{}, fictitious play (\fp{})~\citep{brown1951iterative} and best-response dynamics (\br{}), the latter two being unregularized variants of \ftrl{}/\ftpl{} and \texttt{OMD}, respectively. Finally, we also compare to \emph{optimistic} versions of fictitious play (\ofp{}) and best-response dynamics (\obr{}). We provide pseudocode for all of these algorithms in \Cref{sec:experimental_algos_pseudocode}; the pseudocode for these algorithms explicitly demonstrates that only one LMO call is required per iteration of these algorithms.
Because \fp{} and \br{} represent unregularized variants of \ftrl{}/\ftpl{} and \texttt{OMD}, they can be thought of as letting the stepsize be arbitrarily large for \ftrl{} and \texttt{OMD} respectively (or letting the noise be zero for \ftpl{} in the case of the former); we choose to depict their performance to evaluate the limiting behavior of \ftrl{}/\ftpl{} and \texttt{OMD}.

We note that, unlike \ftpl{}, \oftpl{}, and our algorithms, \fp{} may converge at a \emph{very} slow rate even in two-player zero-sum games~\citep{daskalakis2014counter}, and \br{} may not converge at all.

For \fwomd{}, \fwromd{}, \ftpl{}, and \oftpl{}, we try $\eta \in 0.01 \cdot 2^{[14]}$, where $\eta$ is the stepsize for our algorithms, while $\eta$ is the noise used for \ftpl{} and \oftpl{}. Additionally, we try uniform, linear, and quadratic iterate averaging for all algorithms, as well as last-iterate. Non-uniform averaging schemes are known to often outperform uniform averaging when solving BSPPs~\citep{tammelin2015solving,gao2021increasing}. Note that we demonstrate theoretical guarantees for last-iterate convergence of \fwromd{}, whereas the other algorithms are not known to have such guarantees. Moreover, in the case of averaging, we examine the effects of applying \emph{adaptive restarting} in \Cref{sec:further_experimental_details}. Adaptive restarts are known to lead to linear convergence for polyhedral BSPPs for some algorithms, as they satisfy a \emph{sharpness} property~\citep{applegate2023faster,fercoq2023quadratic,tseng1995linear,gilpin2012first}.

For our algorithms and \normalfont{(\texttt{O})\texttt{FTPL}}, we restrict the number of LMO calls per iteration to be in $\{1, 2, 3, 4, 5, 10, \allowbreak 20, 100, 200\}$. Note that the original presentations of \oftpl{} and \ftpl{} set the termination criterion for an iteration of the algorithm based on the number of LMO calls ($4T$ and $1$, respectively, for the convergence guarantees provided for each algorithm). On the other hand, the more natural termination criterion for our algorithm is the accuracy to which the approximate proximal call is to be computed. Nevertheless, we find that using the number of LMO calls as a termination criterion generally works best for our algorithms as well. Furthermore, we use warmstarting for our algorithm, which involves initializing the active set of \afw{} in the current iteration of our algorithms with the active set of \afw{} in the previous iteration. We provide complete pseudocode for adaptive restarting and various iterate averaging schemes in \Cref{sec:experimental_algos_pseudocode}.
In the \emph{Warmstart and Termination Criteria Ablation} subsection, we demonstrate that restricting the number of LMO calls per outer iteration and warmstarting generally leads to better performance for our algorithms. We conduct further ablation on the averaging scheme in \Cref{sec:further_experimental_details}.

For each of the six algorithms, we use the choice of step size, number of LMO calls, and averaging, which generally leads to the best performance for each game. We provide additional graphs in \Cref{sec:further_experimental_details} demonstrating that the performance of our algorithms relative to the others generally holds irrespective of the choice of the averaging. All of our experiments are run until the average number of LMO calls for each player is $10^4$.

We show the results of running our algorithms on two-player Kuhn poker, two-player Leduc poker, two-player Liar's Dice, three-player Kuhn poker, three-player Liar's Dice, and three-player Goofspiel (3 ranks) in \Cref{fig:NE_CCE_alg_comparison}, seeking to compute Nash equilibria in the former three games and CCE in the latter three games. In the case of NE computation, \fwromd{} outperforms existing algorithms in all three games. In Kuhn and Liar's Dice, we observe that (\texttt{O})\ftpl{} erratically reach small Nash gaps before returning to an iterate which has a high duality gap. This erratic behavior is because the last-iterate averaging is used for \ftpl{} in Kuhn and for both \oftpl{} and \ftpl{} in Liar's Dice. Despite \normalfont{(\texttt{O})\texttt{FTPL}} often achieving low duality gap in those two games, \fwromd{} performs well relative to them (and the other algorithm). As noted above, we evaluate all of these algorithms for fixed choices of averaging schemes in \Cref{sec:further_experimental_details}. Optimism is clearly helpful for all of the algorithms besides \br{}. In the case of CCE computation, we measure the maximum individual player's average regret since a bound of $\eps'$ on all players' average regrets corresponds to an $\eps'$-CCE. Again, in all of the games, our algorithms are competitive with existing algorithms. However, this time it is our non-optimistic algorithm, \fwomd{}, that performs best. Interestingly, for the other sets of algorithms, the effect of optimism is minimal but does not hurt, whereas it hurts for our algorithm.

\subsection{Warmstart and Termination Criteria Ablation}
We also test the performance of our algorithm when using different choices of termination criteria for the approximate prox call and the choice of whether to warmstart. We only show the ablation for \fwromd{} for NE computation and for \fwomd{} for CCE computation since these were our respective best-performing algorithms for each of these sets of experiments. The ablation of \fwromd{} for CCE computation and \fwomd{} for NE computation is deferred to \Cref{sec:further_experimental_details}. It can be seen in \Cref{fig:NE_CCE_ablation}, that in two-player Leduc poker and two-player Liar's Dice, using warmstarting and a fixed number of LMO calls per iteration leads to the best performance. In the multiplayer setting, it can be observed again that using warmstarting and a fixed number of LMO calls leads to best performance of our algorithm in three-player Liar's Dice and three-player Goofspiel (3 ranks). For both two-player and three-player Kuhn, the choice of using warmstarting and a fixed number of LMO calls is competitive.

\section{Conclusions}

We investigated projection-free (linear-optimization-based) algorithms for learning in games with polyhedral strategy sets, including normal-form and imperfect-information extensive-form games. For those settings, we introduced the first projection-free algorithm that attains a $O(\log N/N)$ convergence to Nash equilibrium in two-player zero-sum games, where $N$ is the number of best-response oracle queries, thus breaking the long-standing $O(1/\sqrt{N})$ bound known for this setting. Moreover, our algorithm achieves $O(T^{1/4})$ per-player regret in multiplayer general-sum settings while only requiring $O(\log t)$ best responses per iteration.

\section*{Acknowledgements}
Darshan Chakrabarti was supported by the National Science Foundation Graduate Research Fellowship Program under award number DGE-2036197.
Christian Kroer was supported by the Office of Naval Research awards N00014-22-1-2530 and N00014-23-1-2374, and the National Science Foundation awards IIS-2147361 and IIS-2238960.

\bibliography{refs.bib}
\bibliographystyle{plainnat}
\appendix
\onecolumn

\section{Away-step Frank-Wolfe Pseudocode}\label{sec:afw}

Below, we present pseudocode for away-step Frank-Wolfe, based on the presentation by~\citet{guelat1986some}. We assume that the polytope is expressed as the convex hull of a set of \emph{atoms} and that the LMO for the polytope always returns an atom (since there always exists an atom that minimizes a given linear objective over the polytope). This is equivalent to assuming that we have an LMO over the set of atoms themselves.

\begin{figure}[H]
    \removelatexerror\begin{algorithm}[H]
        \DontPrintSemicolon
        \caption{Away-step Frank-Wolfe (\afw{})~\citep{guelat1986some}}
        \label{algo:afw}
        \KwData{%
            $\cP \subseteq \R^n$: convex polytope\newline
            $\cA \subseteq \R^n$: set of atoms, such that $\conv(\cA) = \cP$\newline
            LMO$_\cA$: linear minimization oracle over $\cA$ \newline
            $f: \cP \to \R$: $L$-smooth, convex function to be optimized \newline
            $\vx\^0\in \cP$ \newline
            $\eps$: desired accuracy
            $T$: maximum number of LMO calls
        }
        $\cS\^0 = \{\vx\^0\}$\;
        $\alpha_{\vx\^0}\^0 = 1$\;
        \For{$t = 0 \dots T-1$}{
        $\vs\^t = \operatorname{LMO}_{\cA}(\nabla f(\vx\^t)),\, \vd\^t_{\mathrm{FW}} = \vs\^t - \vx\^t$ \;
        $\vv_{\mathrm{A}}\^t \in \argmax_{\vv \in \cS\^t} \inp[\big]{ \nabla f(\vx\^t)}{\,  \vv }, \vd\^t_{\mathrm{A}} = \vx\^t - \vv_{\mathrm{A}}\^t$ \;
        \uIf{$\vg\^t_{\mathrm{FW}} = \inp[\big]{ -\nabla f(\vx\^t)}{ \vd\^t_{\mathrm{FW}} } \leq \eps$}{\Return $\vx\^t$\;}
        \uIf{$\inp[\big]{ - \nabla f(\vx\^t)}{ \vd\^t_{\mathrm{FW}} } \geq \inp[\big]{ - \nabla f(\vx\^t)}{ \vd\^t_{\mathrm{A}} }$}{
        $\vd\^t = \vd\^t_{\mathrm{FW}}, \,\gamma\^t_\mathrm{max} = 1$ \;
        }\Else{
        $\vd\^t = \vd\^t_{\mathrm{A}},\, \gamma\^t_\mathrm{max} = \frac{\alpha_{\vv_{\mathrm{A}}\^t}}{1 - \alpha_{\vv_{\mathrm{A}}\^t}}$ \;
        }
        $\gamma\^t = \min \biggl({\frac{\inp[\big]{ -\nabla f(\vx\^t)}{ \vd\^t }}{L \norm[\big]{ \vd\^t}^2}, \gamma\^t_\mathrm{max}} \biggr)$ \;
        $\vx\^{t+1} = \vx\^t + \gamma\^t \vd\^t$ \;

        \uIf{$\inp[\big]{ - \nabla f(\vx\^t)}{ \vd\^t_{\mathrm{FW}} } \geq \inp[\big]{ - \nabla f(\vx\^t)}{ \vd\^t_{\mathrm{A}} }$}{
        $\alpha\^{t+1}_{\vv} = (1- \gamma\^t) \alpha\^{t}_{\vv}$ for all $\vv \in \cS\^t \setminus \{\vs\^t\}$\;
        $\alpha\^{t+1}_{\vs\^t} = \begin{cases}
                \gamma\^t + (1-\gamma\^t) \vs\^t & \mathrm{if } \vs\^t \in \cS\^t \\
                \gamma\^t                        & \mathrm{otherwise}
            \end{cases}$\;
        $\cS\^{t+1} = \begin{cases}
                \cS\^t \cup \{\vs\^t\} & \mathrm{if } \gamma\^t < 1 \\
                \{\vs\^t \}            & \mathrm{if } \gamma\^t = 1
            \end{cases}$\;
        }\Else{
        $\alpha\^{t+1}_{\vv} = (1 + \gamma) \alpha\^{t}_{\vv}$ for all $\vv \in \cS\^t \setminus \{\vv\^t_{\mathrm{A}}\}$\;
        $\alpha\^{t+1}_{\vs\^t} = \begin{cases}
                \gamma\^t + (1-\gamma\^t) \vs\^t & \mathrm{if } \vv\^t_{\mathrm{A}} \in \cS\^t \\
                \gamma\^t                        & \mathrm{otherwise}
            \end{cases}$\;
        $\cS\^{t+1} =
            \begin{cases}
                \cS\^t                                & \mathrm{if } \gamma\^t < \gamma\^t_{\mathrm{max}} \\
                \cS\^t \setminus\{\vv_{\mathrm{A}}\^t\} & \mathrm{if } \gamma\^t = \gamma\^t_{\mathrm{max}}
            \end{cases}$\;

        }
        }
    \end{algorithm}
\end{figure}

\section{Additional Preliminaries about Extensive-form Games}\label{sec:additional_prelims}
Extensive-form games can be represented using a game tree. Each of the internal nodes of the tree corresponds to points at which one of the players or nature (corresponding to stochastic outcomes independent of the players' choices) takes an action. The leaves of the game tree correspond to termination of the game and are associated with utilities for each player. Information sets correspond to partitions of the nodes, such that all the nodes in an information set correspond to a single player, and the actions available at each node of the information set are the same; players cannot distinguish between different nodes in an information set, so the actions must be the same. The set of information sets for a Player $i$ is denoted $\cI_i$.

We are concerned with the perfect recall setting, in which, when Player $i$ is asked to make a decision at a given information set $I \in \cI_i$, he or she remembers the entire history of information sets visited and actions taken at those information sets; each information set has a sequence of previously visited information sets and previously taken actions associated with it.

A strategy for a player is defined as a distribution over the choices of actions at each information set in the tree belonging to the player. Similar to NFGs, a utility function can be defined over the space of strategies for the players by taking an expectation with respect to the joint distribution over the leaves induced by the players' strategies.

The sequence-form representation allows for a compact representation of the strategies of a player in an EFG and allows for formulation of the utility function as a linear function of the players' strategies. A sequence for a player is defined as a choice of action and information set for that player. For a given strategy, the value associated with a given sequence in the corresponding sequence-form strategy is the probability that the player reaches that information set and plays that action given that nature, and the other plays play in such a way that allows the player to reach that sequence. The set of all sequences for a player is denoted $\Sigma_i$.

The space of all sequence-form strategies is a convex polytope: $Q_i = \{\vy \in \mathbb{R}^{|\Sigma_i|}
    : \bF_i \vy = \vec f_i
    , \vy \geq \vec 0\}$, where $\bF_i$
is a sparse $|\cI_i| \times |\Sigma_i|$ matrix with entries in
$\{0, 1, -1\}$, and $\vec f_i$ is a vector with entries in $\{0, 1\}$. Each row of the matrix $\bF_i$ corresponds to a probability flow constraint, which ensures that the sum of the probability of all the sequences associated with an information set sum up to the probability associated with the parent sequence. Note that the vertices of this polytope correspond to making deterministic choices at each information set, which means that the vertices have binary coordinates, and thus, the minimum non-zero entry in a vertex is 1.

\section{Facial Distance of Polytopes Proofs} \label{sec:facial_distance}

\thmfacialdistanceone*
\begin{proof}
    Consider a face $\cF$ of the polytope $\cP$. The face is generated by making a subset of the inequalities that define $\cP$ tight, that is, setting $x_i = 0$ for a subset $S \subseteq [n]$ of indices. Now consider the complement polytope
    \[
        \cP' \defeq \conv(\operatorname{Vert}(\cP) \setminus \cF).
    \]
    Necessarily, the sum of the coordinates corresponding to the indices $S$ of any vertex $\vx' \in \operatorname{Vert}(\cP) \setminus \cF$, that is,
    $
        \sum_{i \in S} \vx'_i,
    $
    must be at least $\gamma$ since at least one of the $k$ coordinates of a vertex in the complement polytope must be nonzero (otherwise, it would be in $\cF$).
    Hence, any convex combination of points in $\operatorname{Vert}(\cP) \setminus \cF$ must also put total mass at least $\gamma$ on coordinates $S$, implying that $\sum_{i \in S} \vx'_i \geq \gamma$ for any $\vx' \in \cP'$.

    On the other hand, by construction, any point on the chosen face $\cF$ satisfies $x_i = 0$ for all $i \in S$. Lower bounding distances by focusing only on the coordinates in $S$ means that the distance between any point on the chosen face $\cF$ and any point in the complement polytope $\cP'$ is at least
    \[
        \min_{\substack{\vx' \in \mathbb{R}^n \\ \sum_{i\in S}\vx'_i \ge \gamma}} \sqrt{\sum_{i\in S}(\vx'_i - 0)^2} = \sqrt{|S| \cdot \left(\frac{\gamma}{|S|}\right)^2} = \frac{\gamma}{\sqrt {|S|}},
    \]
    where the minimum of the objective was obtained by setting all $|S| = k$ coordinates to be equal.
\end{proof}

\thmfacialdistancetwo*
\begin{proof}
    Consider a face $\cF$ of the polytope $\cP$. The face is generated by tightening some subset of the inequalities, that is, setting $x_i = 0$ for a subset $S \subseteq [n]$ of indices, and setting $\vc^\top_j \vx = \vd_j$ for a subset $T \subseteq [n]$ of indices. We can call the submatrices obtained from $\bC$ and $\vd$ by collecting the rows whose indices are in $T$ as $\bC'$ and $\vd'$, respectively.

    Consider the complement polytope
    \[
        \cP' \defeq \conv(\operatorname{Vert}(\cP) \setminus \cF)
    \]
    and let $\vv$ be a vertex of $\cP'$. Now note that since $\vv$ does not lie on $\cF$, it must be the case that there exists an index $i \in S \cup T$ such that the corresponding inequality is not tight for $\vv$.

    Suppose that $i \in S$. Then, by the argument from the proof of \Cref{thm:facial_distance_1}, we immediately can argue that distance between the face and the complement polytope is at least $\frac{1}{\sqrt{n}}$, and thus at least $\frac{1}{\|\bC\|_\infty \sqrt{n}}$.

    Suppose that $i \in T$, and consider any point $\vx \in \cF$; necessarily $\bC' \vx = \vd'$. Necessarily, we also have that $\norm[\big]{\bC' \mathbf{v}}_{1} < \norm[\big]{\vd'}_1$, by the nonnegativity of $\bC$, $\vd$, and the polytope, and the fact that $\vv$ doesn't lie on the face we are considering. Furthermore, by integrality of the polytope of $\bC$, we must have that $\norm[\big]{\bC' \mathbf{v}}_{1} \leq \norm[\big]{\vd'}_1 - 1$.

    It follows that any convex combination of vertices on the complement polytope, $\vy$, would also satisfy this inequality: $\norm[\big]{\bC' \vy}_{1} \leq \norm[\big]{\vd'}_1 - 1$. Noting again that the polytope lies in the nonnegative orthant, we can subtract the two inequalities and apply the Cauchy-Schwarz inequality to obtain:  \[1 \leq \norm[\big]{\bC' (\vx-\vy)}_1 \leq \norm[\big]{\bC'}_\infty \norm[\big]{\vx-\vy}_1 \leq \sqrt{n} \norm[\big]{\bC'}_\infty \norm[\big]{\vx-\vy}_2.\]

    Thus, this means in either case that the distance between the chosen face and the complement polytope is at least $\frac{1}{\norm[\big]{\bC}_\infty \sqrt{n}}$.

    Since these bounds hold for any chosen face, this means the facial distance is bounded below by $\frac{1}{\norm[\big]{\bC}_\infty \sqrt{n}}$ as well.
\end{proof}

\section{Approximate \texttt{ROMD} Proofs}
\label{sec:approx_sc_omd_proofs}
First, we show the following crucial inequality, which we will repeatedly use to bound regret when approximate prox calls are used.

\begin{lemma}
    Let $\vx^* = \argmin_{\vx \in \cX} \inp[\big]{ \vg}{ \vx }   + \frac{1}{\eta} \div{\vx}{\vc}$ and $\hat{\vx}$ be such that $\inp[\big]{  \vg}{ \hat{\vx} } + \frac{1}{\eta} \div{\hat{\vx}}{\vc} \leq \inp[\big]{  \vg}{ \vx^* } + \frac{1}{\eta} \div{\vx^*}{\vc} + \eps$; $\hat{\vx}$ is an $\eps$ argmin to the prox computation.

    Then we have for any $\vd \in \cX$:
    \begin{equation*}
        \eta \inp[\big]{  \vg}{ \hat{\vx} - \vd } \leq \div{\vd}{ \vc} - \div{\vd}{ \hat{\vx}} - \div{\hat{\vx}}{ \vc} + \eps.
    \end{equation*}

    Furthermore, $\frac{1}{2 \eta} \norm[\big]{\vx^* - \hat{\vx}}^2 \leq \eps$.\label{lemma:approx_prox}
\end{lemma}
\begin{proof}
    By definition of $\hat{\vx}$, we have
    \begin{align*}
         & \eta \inp[\big]{  \vg}{ \hat{\vx} } + \vphi(\hat{\vx}) - \vphi(\vc) - \inp[\big]{ \nabla \vphi(\vc)}{  \hat{\vx} - \vc } \leq \eta \inp[\big]{  \vg}{ \vd } + \vphi(\vd) - \vphi(\vc) - \inp[\big]{  \nabla \vphi(\vc)}{ \vd - \vc } + \epsilon.
    \end{align*}
    Subtracting these inequalities, we have
    \begin{align*} \eta \inp[\big]{ \vg}{ \hat{\vx} - \vd  } & \leq \vphi(\vd) - \vphi(\hat{\vx}) - \inp[\big]{ \nabla \vphi(\vc)}{ \vd -\hat{\vx} } + \eps \\
                                                         & \leq  \inp[\big]{ \nabla\vphi(\vc) - \nabla \vphi(\hat{\vx})}{ \hat{\vx} - \vd} + \eps       \\
                                                         & = \div{\vd}{ \vc} - \div{\vd}{ \hat{\vx}} - \div{\hat{\vx}}{\vc} + \eps.
    \end{align*}
    The second inequality follows from the convexity of $\vphi$, and the equality follows from the three-point lemma.

    In the case that $\vphi$ is $1$-strongly convex with respect to the Euclidean norm, we have that
    \[\frac{1}{2 \eta} \norm[\big]{\vx^* - \hat{\vx}}^2 \leq \inp[\big]{  \vg}{ \hat{\vx} } + \frac{1}{\eta} \div{\hat{\vx}}{ \vc} - \inp[\big]{  \vg}{ \vx^* } - \frac{1}{\eta} \div{\vx^*}{ \vc} \leq \eps.\] where the first inequality follows from strong convexity of a function implying quadratic growth of the function~\citep{karimi2016linear}, and the second inequality by assumption on $\hat{\vx}$.
\end{proof}

Next, we show the following lemma characterizing an approximate first-order condition for \Cref{algo:approx_sc_omd}, which will be useful in analyzing the convergence rates of our algorithms.

\begin{lemma}
    \Cref{algo:approx_sc_omd} satisfies the following inequality:
    \begin{align*}
        \inp[\big]{ \eta (2 \bell\^{t-1}  - \bell\^{t-2}) + \nabla \vphi(\vx\^{{t}}) - \nabla \vphi(\vx\^{t-1})}{ \vx\^{{t}} - \vx } \leq \sqrt{2\eta\eps\^t} (2LD + 3 \eta).
    \end{align*}

    In fact, when \Cref{algo:approx_sc_omd} is instantiated with a \fw{} variant which uses the Wolfe gap as a termination criterion, as is the case for \fwromd{}, we have the following approximate first-order optimality condition:
    \begin{align*}
        \inp[\big]{ \eta (2 \bell\^{t-1}  - \bell\^{t-2}) + \nabla \vphi(\vx\^{{t}}) - \nabla \vphi(\vx\^{t-1})}{ \vx\^{{t}} - \vx } \leq  \eps\^t.
    \end{align*}
    \label{lemma:approx_variational_ineq}
\end{lemma}

\begin{proof}
    We define $\vx\^{t}_*$ to be the result of using an exact \romd{} update instead of using the \Cref{algo:approx_sc_omd} update at the $t^{th}$ iteration for Player $i$; this corresponds to assuming that we have an exact prox oracle instead of an approximate prox oracle.

    Using the first-order optimality condition, we have that for any $\vx \in \cX$
    \begin{align*}
        \inp[\big]{ \eta (2 \bell\^{t-1}  - \bell\^{t-2}) + \nabla \vphi(\vx\^{t}_*) - \nabla \vphi(\vx\^{t-1})}{ \vx\^{t}_* - \vx } \leq 0.
        \numberthis \label[ineq]{ineq:true_first_order_opt}
    \end{align*}.

    We now have the following:
    \begin{align*}
         & \inp[\big]{ \eta (2 \bell\^{t-1}  - \bell\^{t-2}) + \nabla \vphi(\vx\^{{t}}) - \nabla \vphi(\vx\^{t-1})}{ \vx\^{{t}} - \vx }                                             \\
         & \hspace{3cm} = \inp[\big]{ \eta (2 \bell\^{t-1}  - \bell\^{t-2}) + \nabla \vphi(\vx\^{t}_*) - \nabla \vphi(\vx\^{t-1})}{ \vx\^{t}_* - \vx }                          \\
         & \hspace{6cm} + \inp[\big]{ \nabla \vphi(\vx\^{t}) - \nabla \vphi(\vx\^{t}_*)}{ \vx\^{t}_* - \vx) }                                                                         \\
         & \hspace{6cm} + \inp[\big]{ \eta ( 2 \bell\^{t-1}  - \bell\^{t-2}) + \nabla \vphi(\vx\^{{t}}) - \nabla \vphi(\vx\^{t-1})}{ \vx\^{t} - \vx\^{t}_* }                      \\
         & \hspace{3cm} \leq 0 + \norm[\big]{ \nabla \vphi(\vx\^{t}) - \nabla \vphi(\vx\^{t}_*)  } \norm[\big]{ \vx\^{t}_* - \vx }
        \numberthis \label[ineq]{ineq:approx_first_order_opt_first_ineq}                                                                                                                              \\
         & \hspace{6cm} + \norm[\big]{ \eta(2 \bell\^{t-1}  - \bell\^{t-2}) + \nabla \vphi(\vx\^{{t}}) - \nabla \vphi(\vx\^{t-1}) } \norm[\big]{\vx\^{t} - \vx\^{t}_* }    \nonumber          \\
         & \hspace{3cm} \leq L \norm[\big]{\vx\^{t} - \vx\^{t}_*} \norm[\big]{ \vx\^{t}_* - \vx }
        \numberthis \label[ineq]{ineq:approx_first_order_opt_second_ineq}                                                                                                                             \\
         & \hspace{6cm} + \parens*{\eta \norm[\big]{ 2 \bell\^{t-1}  - \bell\^{t-2} } + \norm[\big]{ \nabla \vphi(\vx\^{{t}}) - \nabla \vphi(\vx\^{t-1})}}\norm[\big]{\vx\^{t} - \vx\^{t}_*}                \\
         & \hspace{3cm} \leq \sqrt{2\eta\eps\^t} L D  + \sqrt{2\eta\eps\^t} \parens*{\eta(2 \norm[\big]{\bell\^{t-1}} +\norm[\big]{ \bell\^{t-2}})  + L  \norm[\big]{\vx\^{t} - \vx\^{t-1}}}
        \numberthis \label[ineq]{ineq:approx_first_order_opt_third_ineq}                                                                                                                              \\
         & \hspace{3cm} \leq \sqrt{2\eta\eps\^t} (2LD + 3\eta).
        \numberthis \label[ineq]{ineq:approx_first_order_opt_fourth_ineq}
    \end{align*}
    We use \Cref{ineq:true_first_order_opt} and the Cauchy-Schwarz inequality in \Cref{ineq:approx_first_order_opt_first_ineq}, the triangle inequality and smoothness of $\vphi$ in \Cref{ineq:approx_first_order_opt_second_ineq}, \Cref{lemma:approx_prox} in \Cref{ineq:approx_first_order_opt_third_ineq}, and bounded losses in \Cref{ineq:approx_first_order_opt_fourth_ineq}.

    In the case that the Wolfe gap is used as a termination criterion, the stated approximate first-order optimality condition immediately follows because the left-hand side of the indicated inequality is precisely the Wolfe gap.
\end{proof}

\lemmaapproxscomdregret*
\begin{proof}

    Let $\vx' \in \cX$. If we apply \Cref{lemma:approx_prox}, then we obtain:

    \begin{align*}
         & \eta \inp[\big]{ \bell\^{t} + \vm\^{t+1}  - \vm\^{t}}{ \vx\^{t+1} - \vx' } \leq \divi{\vx'}{ \vx\^{t}} - \divi{\vx'}{ \vx\^{t+1}} - \divi{\vx\^{t+1}}{ \vx\^{t}} + \eps\^t. \numberthis \label[ineq]{ineq:approx_prox_romd_update}
    \end{align*}

    Summing the left side over $t = 1, \dots, T$, and noting that we can let $\vm\^{T+1} = 0$ without affecting losses at timesteps before $T+1$, we have:
    \begin{align*}
        &\sum_{t = 1}^{T} \inp[\big]{ \bell\^{t} + \vm\^{t+1} - \vm\^{t}}{ \vx\^{t+1} - \vx' } \\
        &\hspace{1cm} = \sum_{t = 1}^{T} \left[ \inp[\big]{ \bell\^{t}}{ \vx\^{t+1} - \vx' } + \inp[\big]{ \vm\^{t+1} - \vm\^{t}}{ \vx\^{t+1} - \vx' } \right]                                           \\
                                                                                                                 &\hspace{1cm} = \sum_{t = 1}^{T} \left[ \inp[\big]{ \bell\^{t}}{ \vx\^{t+1} - \vx' } + \inp[\big]{ \vm\^{t+1} - \vm\^{t}}{ \vx\^{t+1} } \right] + \inp[\big]{ \vm_{i}\^1 - \vm\^{T+1}}{ \vx' } \\
                                                                                                                 &\hspace{1cm} = \sum_{t = 1}^{T} \left[ \inp[\big]{ \bell\^{t}}{ \vx\^{t} - \vx' } + \inp[\big]{ \bell\^{t}}{ \vx\^{t+1} - \vx\^{t} } + \inp[\big]{ \vm\^{t+1} - \vm\^t}{ \vx\^{t+1} }\right] + \inp[\big]{\vm\^{1}}{\vx'} \\
                                                                                                                 &\hspace{1cm} = \sum_{t = 1}^{T} \left[ \inp[\big]{ \bell\^{t}}{ \vx\^{t} - \vx' } + \inp[\big]{ \bell\^{t}  - \vm\^{t} }{ \vx\^{t+1} - \vx\^{t} } \right] + \inp[\big]{\vm\^{1}}{\vx'} - \inp[\big]{ \vm\^1}{ \vx\^2 } + \inp[\big]{ \vm\^{T+1}}{ \vx\^{T+1} }                                                                                                            \\
                                                                                                                 &\hspace{1cm} = \sum_{t = 1}^{T} \left[ \inp[\big]{ \bell\^{t}}{ \vx\^{t} - \vx' } + \inp[\big]{ \bell\^{t}  - \vm\^{t} }{ \vx\^{t+1} - \vx\^{t} } \right] + \inp[\big]{\vm\^{1}}{\vx' - \vx\^2} \numberthis \label{eq:regret_decomposition}.
    \end{align*}

    This allows us to decompose the regret into two terms and apply \Cref{ineq:approx_prox_romd_update} to one of these terms:

    \allowdisplaybreaks
    \begin{align*}
        \Reg\^T & = \max_{\vx' \in \cX} \sum_{t = 1}^{T} \inp[\big]{ \bell\^t}{ \vx\^t - \vx' }                                                                                                                                                                                    \\
                & =  \sum_{t = 1}^{T} \brackets[\bigg]{\inp[\big]{ \bell\^{t}  - \vm\^{t} }{ \vx\^{t} - \vx\^{t+1} } + \inp[\big]{ \bell\^{t} + \vm\^{t+1} - \vm\^{t}}{ \vx\^{t+1} - \vx^* }} + \inp[\big]{\vm\^1}{\vx\^2 - \vx^*} \numberthis \label{eq:initial_rvu_bound_zeroth} \\
                & \leq  \sum_{t = 1}^{T} \bigg[ \inp[\big]{ \bell\^t  - \vm\^t }{ \vx\^{t} - \vx\^{t+1}} \bigg]  + \sum_{t=1}^{T} \frac{1}{\eta} \parens*{\divi{\vx^*}{\vx\^t} - \divi{\vx^*}{\vx\^{t+1}} - \divi{\vx\^{t+1}}{\vx\^t}}  \\
                & \hspace{2cm} + \inp[\big]{\vm\^1}{\vx\^2 - \vx^*} + \sum_{t=1}^{T} \frac{\eps\^t}{\eta} \numberthis     \label[ineq]{ineq:initial_rvu_bound_first}                                            \\
                & \leq  \sum_{t = 1}^{T} \bigg[ \inp[\big]{ \bell\^t  - \vm\^t }{ \vx\^{t} - \vx\^{t+1} } -\frac{1}{\eta} \divi{\vx\^{t+1}}{ \vx\^t} \bigg] + \frac{1}{\eta} \divi{\vx^*}{ \vx\^{0}}  + \inp[\big]{\vm\^1}{\vx\^2 - \vx^*} \\&\hspace{2cm}+ \sum_{t=1}^{T} \frac{\eps\^t}{\eta} \numberthis \label[ineq]{ineq:initial_rvu_bound_second}    \\
                & \leq  \sum_{t = 1}^{T} \brackets[\bigg]{\inp[\big]{ \bell\^t  - \vm\^t }{ \vx\^{t} - \vx\^{t+1} } -\frac{1}{2\eta} \norm[\big]{\vx\^{t+1} - \vx\^t}^2} + \frac{1}{\eta} \divi{\vx^*}{ \vx\^{0}} + \inp[\big]{\vm\^1}{\vx\^2 - \vx^*} \\&\hspace{2cm}+ \sum_{t=1}^{T} \frac{\eps\^t}{\eta} \numberthis \label[ineq]{ineq:initial_rvu_bound_third} \\
                & \leq \sum_{t = 1}^{T} \norm[\big]{\bell\^t  - \vm\^t }  \norm[\big]{\vx\^{t+1} - \vx\^t } - \frac{1}{2\eta} \sum_{t=1}^{T} \norm[\big]{\vx\^{t+1} - \vx\^t }^2 + \frac{1}{\eta} \divi{\vx^*}{ \vx\^{0}} + \inp[\big]{\vm\^1}{\vx\^2 - \vx^*} \\
                & \hspace{2cm} + \sum_{t=1}^{T} \frac{\eps\^t}{\eta} \numberthis \label[ineq]{ineq:initial_rvu_bound_fourth}.
    \end{align*}
\end{proof}

We apply \Cref{eq:regret_decomposition} to obtain \Cref{eq:initial_rvu_bound_zeroth}, \Cref{ineq:approx_prox_romd_update} to obtain \Cref{ineq:initial_rvu_bound_first}, drop a negative term to obtain \Cref{ineq:initial_rvu_bound_second}, apply strong convexity of $\vphi$ to obtain \Cref{ineq:initial_rvu_bound_third}, and finally apply the Cauchy-Schwarz inequality to obtain \Cref{ineq:initial_rvu_bound_fourth}.

\lemmarefinedrvu*
\begin{proof}
    Instantiating \Cref{lemma:approx_sc_omd_regret} with $\vm\^t = \bell\^{t-1}$, noting that $\bell\^0 = \mathbf{0}$ and using the definition of $\Omega$:
    \begin{align*}
        \Reg\^T  \leq \sum_{t = 1}^{T} \norm[\big]{\bell\^{t}  - \bell\^{t-1}} \norm[\big]{\vx\^{t+1} - \vx\^{t}} - \frac{1}{2\eta} \sum_{t=1}^{T} \norm[\big]{\vx\^{t+1} - \vx\^t}^2 + \frac{\Omega}{\eta}  + \sum_{t=1}^{T} \frac{\eps\^t}{\eta}.
    \end{align*} 

    Applying Young's inequality:
    \begin{align*}
        \norm[\big]{\bell\^{t}  - \bell\^{t-1}} \norm[\big]{\vx\^{t+1} - \vx\^{t}} \leq \frac{1}{2} \parens*{{2\eta} \norm[\big]{\bell\^{t}  - \bell\^{t-1}}^2 +  \frac{1}{2\eta} \norm[\big]{\vx\^{t+1} - \vx\^{t}}^2}.
    \end{align*}

    Furthermore, instantiating $\eps\^t = \frac{1}{t^2}$, we also have that:
    \begin{align*}
        \sum_{t=1}^T \frac{\eps\^t}{\eta} & = \frac{1}{\eta} \sum_{t=1}^T \frac{1}{t^2} \leq \frac{2}{\eta}.
    \end{align*}

    Combining the above, we have:
    \begin{align*}
        \Reg\^T & \leq \sum_{t = 1}^{T} \frac{1}{2} (2\eta \norm[\big]{\bell\^{t}  - \bell\^{t-1}}^2 +  \frac{1}{2\eta} \norm[\big]{\vx\^{t+1} - \vx\^{t}}^2)  - \frac{1}{2\eta} \sum_{t=1}^{T} \norm[\big]{\vx\^{t+1} - \vx\^t}^2  +  \frac{\Omega}{\eta} + \sum_{t=1}^{T} \frac{\eps\^t}{\eta} \\
                & =  \eta \sum_{t = 1}^{T}  \norm[\big]{\bell\^{t}  - \bell\^{t-1}}^2 - \frac{1}{4\eta} \sum_{t=1}^{T} \norm[\big]{\vx\^{t+1} - \vx\^t}^2  +  \frac{\Omega}{\eta}  + \sum_{t=1}^{T} \frac{\eps\^t}{\eta}                                                                               \\
                & \leq \frac{\Omega+2}{\eta} + {\eta} \sum_{t = 1}^{T}  \norm[\big]{\bell\^{t}  - \bell\^{t-1}}^2 - \frac{1}{4\eta} \sum_{t=1}^{T} \norm[\big]{\vx\^{t+1} - \vx\^t}^2.
    \end{align*}
\end{proof}

\subsubsection{Self-Play in Games}
In order to compute equilibria in games, we will assume that Player $i$ receives as its loss $\bell_i\^t = -\nabla_{\vx_i} u_i(\vx_1\^t, \dots , \vx_n\^t)$.

\begin{assumption}
    We assume that our games satisfy a smoothness condition:
    \begin{align*}
        \norm[\big]{\bell_i\^{t} - \bell_i\^{t-1} } \leq \sum_{j\neq i} \norm[\big]{\vx_j\^t - \vx_j\^{t-1} }.
    \end{align*}

    This can always be satisfied by rescaling the utility function for any game where the utility function is multilinear in the players' strategies (as is the case in NFGs and EFGs).
    \label{assumption:smoothness}
\end{assumption}

Next, we restate Theorem 4 from~\citet{syrgkanis2015fast}, which characterizes the stepsize required to achieve constant regret.
\begin{theorem}{\citep{syrgkanis2015fast}}
    If each player employs an algorithm satisfying the RVU property with parameters $\alpha$, $\beta$, and $\gamma$, such that $\beta \leq \gamma/(N-1)^2$, then $\sum_{i \in [N]} \Reg_i\^T \leq \alpha N$.
    \label{thm:constant_social_regret}
\end{theorem}
\begin{proof}
    By \Cref{assumption:smoothness} and Jensen's inequality we have:
    \begin{align*}
        \norm[\big]{\bell_i\^{t} - \bell_i\^{t-1} }^2 &\leq \parens[\bigg]{\sum_{j\neq i} \norm[\big]{\vx_j\^t - \vx_j\^{t-1}} }^2 \\
        &\leq (N-1) \sum_{j\neq i} \norm[\big]{\vx_j\^t - \vx_j\^{t-1} }^2. 
    \end{align*}

    Summing up the terms for all the players, we have that
    \begin{align*}
        \sum_{i \in [N]} \norm[\big]{\bell_i\^{t} - \bell_i\^{t-1} }^2 \leq (N-1)^2 \sum_{i \in [N]} \norm[\big]{\vx_j\^t - \vx_j\^{t-1} }^2.
    \end{align*}
    The theorem immediately follows by noting that the assumption on $\beta$ and $\gamma$ ensures that the latter two terms in the RVU bound can be dropped, and the inequality will still hold.
\end{proof}.

\begin{lemma}
    When running $T$ iterations of \fwromd{} using $\vm\^t = \bell\^{t-1}$\normalfont{:}
    \begin{enumerate}
        \item $\eps\^{t} = \frac{1}{t^2}$-optimal prox computations at each time step requires $O(T L \frac{D^2}{\delta^2} \log \left[ LDT \right] )$ LMO calls.
        \item $\eps\^{t} = \eps$-optimal prox computations at each time step requires $O(T L \frac{D^2}{\delta^2} \log \left[\frac{LD}{\eps} \right])$ LMO calls.
    \end{enumerate}
    \label{lemma:total_lmo_calls}
\end{lemma}

\begin{proof}

    In the first case, note that \afw{} can achieve a $\eps\^t$ optimal solution with $O\parens[\big]{\frac{LD^2}{\delta^2} \log \big[ \frac{LD}{\eps\^t} \big]}$ LMO calls, which means that \fwromd{} requires $O\parens[\big]{TL \frac{D^2}{\delta^2} \log \left[ LDT \right] }$ LMO calls in order to achieve constant cumulative regret, since we can lower bound $\eps\^t$ by $\frac{1}{T^2}$.

    In the second case, note that \afw{} can achieve an $\eps$ optimal solution with $O\parens[\big]{\frac{LD^2}{\delta^2} \log \big[ \frac{LD}{\eps\^t} \big]}$ LMO calls, which means that \fwromd{} requires $O\parens[\big]{TL \frac{D^2}{\delta^2} \log \left[ \frac{LD}{\eps} \right] }$ LMO calls in order to achieve constant cumulative regret.

    Additionally, note that at least $1$ LMO call is required by \afw{} at each iteration to check whether \afw{} has reached an $\eps\^t$-optimal solution. It follows that $N$ is also $O(T \log N)$, since $T$ is $O(N)$. It follows then that $\frac{1}{T}$ is in $O(\frac{\log N}{N})$, and thus with $N$ LMO calls we can achieve $O(\frac{\log N}{N})$ average regret.

\end{proof}

\thmneregretbound*

\begin{proof}
    For \fwromd{}, we know that the RVU property is satisfied with $\alpha_i = \frac{\Omega_i + 2}{\eta_i}$, $\beta_i = \eta_i$, $\gamma_i = \frac{1}{4\eta_i}$ for Player $i$. By \Cref{thm:constant_social_regret}, if we take $\eta_i \leq \frac{1}{2 (N-1)}$, then we have that $\sum_{i \in [N]} \Reg_i\^T \leq \alpha (N-1)$, where $\alpha = \max_{i \in [N]} \alpha_i$.

    It follows that we if we take $\xi$ as the Nash gap corresponding to the average strategies of the two players $\bar{\vx}_1 = \frac{1}{T} \sum_{t=1}^{T} \vx_1\^T$ and $\bar{\vx}_2 = \frac{1}{T} \sum_{t=1}^{T} \vx_2\^T$, then we have:

    \begin{align*}
        \xi = \max_{\vx_2 \in \cX_2} \inp[\big]{ \bA \bar{\vx}_1}{ \vx_2 } - \min_{\vx_1 \in \cX_1} \inp[\big]{ \bA\vx_1}{ \bar{\vx}_2 } = \frac{1}{T} \biggl(\Reg_2\^T + \Reg_1\^T\biggr) \leq \frac{\alpha}{T}
    \end{align*}
    since for any given $\vx_2 \in \cX_2$ we have $\inp[\big]{ \bA \bar{\vx}_1}{ \vx_2 } = \sum_{t=1}^T \inp[\big]{ \bA \vx\^t_1}{ \vx_2 } $
    and for any given $\vx_1 \in \cX_1$ we have $\inp[\big]{ \bA \vx_1}{ \bar{\vx}_2 } = \sum_{t=1}^T \inp[\big]{ \bA \vx_1}{ \vx\^t_2 }$.

    This demonstrates that we can achieve an $\eps'$-NE in a two-player zero-sum game in $O(1/\eps')$ iterations. By \Cref{lemma:total_lmo_calls}, if we use $\eps_i\^t = \frac{1}{t^2}$, since $T = O(1/\epsilon')$, we require $O(\max_{i \in [N]} \frac{1}{\eps'} \frac{L_iD_i^2}{\delta_i^2} \log \left[ \frac{L_iD_i}{\eps'} \right])$ LMO calls to achieve a $\epsilon'$-NE.
\end{proof}
\thmcceregretbound*
\begin{proof}
    We claim that letting $\eta_i = \frac{1}{T^{1/4}}$ allows for this result to hold.

    In order to prove this statement, we first prove a lemma about the \emph{stability} of our iterates.
    \begin{lemma}
        \Cref{algo:approx_sc_omd} with $\vm\^t = \bell\^{t-1}$ has the following property:

        \begin{align*}
            \norm[\big]{ \vx\^{t+1} - \vx\^t } \leq 3\eta + \sqrt{2\eta\eps\^t}
        \end{align*}
        \label{lemma:stability}
    \end{lemma}
    \begin{proof}
        As in the proof of \Cref{lemma:approx_variational_ineq}, we define $\vx\^{t}_*$ as the result of using an exact \romd{} update instead of using the \Cref{algo:approx_sc_omd} update at the $t^{th}$ iteration for Player $i$; this corresponds to assuming that we have an exact prox oracle instead of an approximate prox oracle.

        Using first-order optimality, we have that for any $\vx \in \cX$
        \begin{align*}
            \inp[\big]{ \eta (2 \bell\^{t-1}  - \bell\^{t-2}) + \nabla \vphi(\vx\^{t}_*) - \nabla \vphi(\vx)}{ \vx\^{t}_* - \vx } \leq 0 \numberthis \label[ineq]{ineq:true_first_order_opt_second_time}.
        \end{align*}
        
        Using this inequality, we have:

        \begin{align*}
            \norm[\big]{\vx\^{t}_* - \vx}^2 &\leq \inp[\big]{\nabla \vphi(\vx\^{t}_*) - \nabla \vphi(\vx)}{\vx\^{t}_* - \vx} \numberthis \label[ineq]{ineq:stability_zeroth} \\
            &\leq -\inp[\big]{\eta (2 \bell\^{t-1}  - \bell\^{t-2})}{\vx\^{t}_* - \vx} \numberthis \label[ineq]{ineq:stability_first}\\
            &\leq \eta \norm[\big]{2 \bell\^{t-1}  - \bell\^{t-2}}\norm[\big]{\vx\^{t}_* - \vx} \numberthis \label[ineq]{ineq:stability_second}\\
            &\leq \eta \parens[\Big]{2 \norm[\big]{\bell\^{t-1}} + \norm[\big]{\bell\^{t-2}}}\norm[\big]{\vx\^{t}_* - \vx} \numberthis \label[ineq]{ineq:stability_third}\\
            &\leq 3 \eta \norm[\big]{\vx\^{t}_* - \vx} \numberthis \label[ineq]{ineq:stability_fourth}
        \end{align*}

        In \Cref{ineq:stability_zeroth}, we apply the strong convexity of $\vphi$, in \Cref{ineq:stability_first} we apply \Cref{ineq:true_first_order_opt_second_time}, in \Cref{ineq:stability_second} we apply the Cauchy-Schwarz inequality, in \Cref{ineq:stability_third} we apply the triangle inequality, and finally in \Cref{ineq:stability_fourth} we apply the assumption on the norms of the losses. Dividing by $\norm[\big]{\vx\^{t}_* - \vx\^{t-1} }$ (assuming it is non-zero; otherwise, the below inequality trivially holds), we have that:
        \begin{align*}
            \norm[\big]{\vx\^{t}_* - \vx\^{t-1} } \leq 3\eta.
        \end{align*}

        Now using the triangle inequality and \Cref{lemma:approx_prox}, we have that
        \begin{align*}
            \norm[\big]{\vx\^{{t}} - \vx\^{t-1} } \leq 3\eta + \sqrt{2\eta\eps\^t}.
        \end{align*}
    \end{proof}

    By \Cref{assumption:smoothness}, Jensen's inequality, and the above, we have:
    \begin{align*}
        \norm[\big]{\bell_i\^{t} - \bell_i\^{t-1} }^2 \leq (N-1) \sum_{j\neq i} \norm[\big]{\vx_j\^t - \vx_j\^{t-1} }^2 \leq (N-1)^2 \max_{j \in [N]}\parens[\Bigg]{18\eta_j^2 + 4\eta_j\eps_j\^t}.
        \numberthis \label[ineq]{ineq:loss_stability_bound}
    \end{align*}

    Now we can use the refined RVU bound from \Cref{lemma:approx_sc_omd_refined_rvu}:
    
    \begin{align*}
        \Reg_i\^T & \leq \frac{\Omega_i + 2}{\eta_i} + \eta_i (N-1)^2 \max_{j \in [N]}\sum_{t=1}^{T} \parens[\Bigg]{18\eta_j^2 + 4\eta_j\eps_j\^t} - \frac{1}{4\eta_i} \sum_{t=1}^{T}\norm[\big]{\vx_i\^{t+1} - \vx_i\^{t}}^2 \numberthis  \label[ineq]{ineq:cce_regret_bound_one}  \\
                  & \leq \frac{\Omega_i + 2}{\eta_i} + \eta_i (N-1)^2 \max_{j \in [N]}\sum_{t=1}^{T} \parens[\Bigg]{18\eta_j^2 + 4\eta_j\eps_j\^t}     \\
                  &= \frac{\Omega_i + 2}{\eta_i} + \eta_i (N-1)^2 \max_{j \in [N]} \parens[\Bigg]{18T\eta_j^2 + 8\eta_j}  \numberthis \label{eq:cce_regret_bound_two}    \\
    \end{align*}
    
    To obtain \Cref{ineq:cce_regret_bound_one}, we plug \Cref{ineq:loss_stability_bound} into \Cref{lemma:approx_sc_omd_refined_rvu}, and for \Cref{eq:cce_regret_bound_two}, we use the fact that $\eps_i\^t = \frac{1}{t^2}$ so $\sum_{i=1}^{T} \eps_i\^t \leq 2$.

    Now if we let $\eta_i = \frac{1}{T^{1/4}}$, we get that $\Reg_i\^T$ is in $O(T^{1/4})$, showing that the average joint strategy of the players converges to a CCE, at a rate $O(T^{-3/4})$. Equivalently, to reach a $\eps'$-CCE, we require $O(\eps'^{-4/3})$ iterations of \Cref{algo:approx_sc_omd}. By \Cref{lemma:total_lmo_calls}, when using \fwromd{}, if we use $\eps_i\^t = \frac{1}{t^2}$, since $T = O(1/\eps'^{4/3})$, we require $O(\max_{i \in [N]} \frac{1}{\eps'^{4/3}} \frac{L_iD_i^2}{\delta_i^2} \log \left[ \frac{L_iD_i}{\eps'} \right])$ LMO calls to achieve a $\eps'$-CCE.

\end{proof}
\section{Last-Iterate Results}
\thmasymptoticlastiterate*
\begin{proof}
    We follow the proof of Theorem A.12 from~\citet{anagnostides2022on}.

    We rewrite the regret bound given by \Cref{lemma:approx_sc_omd_refined_rvu} for Player $i$:
    \begin{align*}
        \Reg_i\^T & \leq \frac{\Omega_i + 2}{\eta_i} + \eta_i \sum_{t = 1}^{T}  \norm[\big]{\bell\^{t}_i  - \bell\^{t-1}_i }^2 - \frac{1}{8\eta_i} \sum_{t=1}^{T} \norm[\big]{\vx\^{t+1}_i - \vx_i\^t}^2 -  \frac{1}{8\eta_i} \sum_{t=1}^{T} \norm[\big]{\vx_i\^{t+1} - \vx_i\^t}^2.
    \end{align*}

    Applying \Cref{assumption:smoothness} and Jensen's inequality, we have:
    \begin{align*}
        \Reg_i\^T & \leq \frac{\Omega_i + 2}{\eta_i} + (N-1) \eta_i \sum_{j\neq i} \sum_{t = 1}^{T}  \norm[\big]{\vx\^{t}_j - \vx\^{t-1}_j }^2 - \frac{1}{8\eta_i} \sum_{t=1}^{T} \norm[\big]{\vx\^{t+1}_i - \vx_i\^t}^2 \\ 
        &\hspace{1cm} -  \frac{1}{8\eta_i} \sum_{t=1}^{T} \norm[\big]{\vx_i\^{t+1} - \vx_i\^t}^2.
    \end{align*}

    Using the fact that $\eta_{\mathrm{max}} = \max_{i \in [N]} \eta_i$, we can rewrite the above as:
      \begin{align*}
        \Reg_i\^T & \leq \frac{\Omega_i + 2}{\eta_i} + (N-1) \eta_{\mathrm{max}} \sum_{j\neq i} \sum_{t = 1}^{T}  \norm[\big]{\vx\^{t}_j - \vx\^{t-1}_j }^2 - \frac{1}{8\eta_{\mathrm{max}}} \sum_{t=1}^{T} \norm[\big]{\vx\^{t+1}_i - \vx_i\^t}^2 \\ 
        &\hspace{1cm} -  \frac{1}{8\eta_{\mathrm{max}}} \sum_{t=1}^{T} \norm[\big]{\vx_i\^{t+1} - \vx_i\^t}^2.
    \end{align*}

    Summing these terms over all the players yields: 
    \begin{align*}
        \sum_{i=1}^N \Reg_i\^T & \leq \sum_{i=1}^N \frac{\Omega_i + 2}{\eta_i} + \parens[\Big]{(N-1)^2 \eta_{\mathrm{max}} - \frac{1}{8\eta_{\mathrm{max}}}} \sum_{i = 1}^N \sum_{t = 1}^{T}  \norm[\big]{\vx\^{t}_i - \vx\^{t-1}_i }^2 \\ 
        &\hspace{1cm} -  \frac{1}{8\eta_{\mathrm{max}}} \sum_{i=1}^N \sum_{t=1}^{T} \norm[\big]{\vx_i\^{t+1} - \vx_i\^t}^2.
    \end{align*}

    Using the assumptions $\eta_{\mathrm{max}} \leq \frac{1}{2\sqrt{2}(N-1)}$ and $\sum_{i=1}^N \Reg_i\^T \geq 0$, we can write
    \begin{align*}
        0 \leq \sum_{i=1}^N \Reg_i\^T \leq \sum_{i=1}^N \brackets[\bigg]{\frac{\Omega_i + 2}{\eta_i} -  \frac{1}{8\eta_{\mathrm{max}}} \sum_{t=1}^{T} \norm[\big]{\vx_i\^{t+1} - \vx_i\^t}^2}.
    \end{align*}
    Hence,
    \begin{align*}
        \sum_{t=1}^T \sum_{i=1}^{N} \norm[\big]{ \vx_i\^{t+1} - \vx_i\^{t} }^2  \leq 8 \eta_{\mathrm{max}} \sum_{i=1}^{N} \frac{\Omega_i + 2}{\eta_i}.
    \end{align*}

    Assuming for all $t \in [T]$ that $\sum_{i=1}^N \norm[\big]{ \vx_i\^{t+1} - \vx_i\^{t} }^2 \geq \eps^2$, we must have that $T \leq \frac{8 \eta_{\mathrm{max}}}{\eps^2} \sum_{i=1}^{N} \frac{\Omega_i + 2}{\eta_i}$.
    Thus, as long as $T > \left \lceil \frac{8 \eta_{\mathrm{max}}}{\eps^2} \sum_{i=1}^{N} \frac{\Omega_i + 2}{\eta_i} \right \rceil$, there must exist $t \in [T]$ such that $\frac{1}{2} \parens*{\sum_{i=1}^N  \norm[\big]{ \vx_i\^{t+1} - \vx_i\^{t} }^2 + \sum_{i=1}^N  \norm[\big]{ \vx_i\^{t} - \vx_i\^{t-1} }^2 } \leq \eps^2$.

    Next, we show that $\frac{1}{2} \parens[\Big]{\sum_{i=1}^N  \norm[\big]{ \vx_i\^{t+1} - \vx_i\^{t} }^2 + \sum_{i=1}^N  \norm[\big]{ \vx_i\^{t} - \vx_i\^{t-1} }^2} \leq \eps^2$ implies that we are at an approximate Nash equilibrium.

    Using \Cref{lemma:approx_variational_ineq}, we have for any $\vx_i \in \cX_i$:

    \begin{align*}
        \inp[\big]{ \eta_i (2 \bell\^{t}_i  - \bell\^{t-1}_i) + \nabla \vphi_i(\vx_i\^{{t+1}}) - \nabla \vphi_i(\vx_i\^{t})}{ \vx_i\^{{t+1}} - \vx_i } \leq \sqrt{2\eta_i\eps_i\^{t+1}} (2L_iD_i + 3 \eta_i).
    \end{align*}

    Rearranging, we have:
    \begin{align*}
        \eta_i \inp[\big]{ \bell^{t}_i}{ \vx_i\^{t+1} - \vx_i } & \leq \sqrt{2\eta_i\eps\^{t+1}_i} (2L_iD_i + 3 \eta_i) \\
        & \hspace{1cm} + \inp[\big]{ \eta_i (-\bell\^{t}_i  + \bell\^{t-1}_i) - \nabla \vphi_i(\vx_i\^{{t+1}}) + \nabla \vphi_i(\vx_i\^{t})}{\vx_i\^{t+1} - \vx_i}                                            \\
                                                                & \leq \sqrt{2\eta_i\eps\^{t+1}_i} (2L_iD_i + 3 \eta_i) \\
                                                                & \hspace{1cm} + \parens*{\norm[\big]{\bell\^{t-1}_i - \bell\^{t}_i } + \norm[\big]{\nabla \vphi_i(\vx_i\^{{t}}) - \nabla \vphi_i(\vx_i\^{t+1})} } \parens*{\norm[\big]{ \vx_i\^{t+1} - \vx_i }}
        \numberthis \label[ineq]{ineq:best_iterate_second_ineq}                                                                                                                                                                                                                                                \\
                                                                & \leq \sqrt{2\eta_i\eps\^{t+1}_i} (2L_iD_i + 3 \eta_i) + \parens*{\sum_{j\neq i} \norm[\big]{\vx_j\^{t-1} - \vx_j\^{t}} + L_i \norm[\big]{ \vx_i\^{t} - \vx_i\^{t+1} } }\Omega_i
        \numberthis \label[ineq]{ineq:best_iterate_third_ineq}                                                                                                                                                                                                                                                 \\
                                                                & \leq \sqrt{2\eta_i\eps\^{t+1}_i} (2L_iD_i + 3 \eta_i) + 2\Omega_i \eps(L_i + N -1).
        \numberthis \label[ineq]{ineq:best_iterate_fourth_ineq}
    \end{align*}

    We applied the Cauchy-Schwarz inequality in \Cref{ineq:best_iterate_second_ineq}, \Cref{assumption:smoothness} and smoothness of $\vphi_i$ in \Cref{ineq:best_iterate_third_ineq}, and the bound on the second-order path lengths in \Cref{ineq:best_iterate_fourth_ineq}.

    Furthermore, note that $\inp[\big]{ \bell_i\^t}{ \vx\^{t}_i - \vx\^{t+1}_i } \leq \norm[\big]{ \bell_i\^t} \norm[\big]{ \vx\^{t+1}_i - \vx\^{t}_i} \leq \eps$ due to the assumptions on the boundedness of the losses and the second-order path lengths.
    Thus, we have that:
    \begin{align*}
        \inp[\big]{ \bell^{t}_i}{ \vx_i\^{t} - \vx_i } & = \inp[\big]{ \bell^{t}_i}{ \vx_i\^{t+1} - \vx_i } + \inp[\big]{ \bell^{t}_i}{
        \vx\^{t}_i - \vx_i\^{t+1} }                                                                                                                                        \\
                                                       & \leq \sqrt{2\eta_i\eps_i\^{t+1}} \parens[\Big]{\frac{2L_iD_i}{\eta_i} + 3} +\frac{2\Omega_i}{\eta_i} \eps(L_i + N -1)  + \eps \\
                                                       & = \sqrt{2\eta_i\eps^2} \parens[\Big]{\frac{2L_iD_i}{\eta_i} + 3} +\frac{2\Omega_i}{\eta_i} \eps(L_i + N -1)  + \eps \\
                                                       & \leq \eps\parens*{\sqrt{2\eta_i} \parens[\Big]{\frac{2L_iD_i}{\eta_i} + 3} +\alpha_i}.
    \end{align*}

    The result follows by noting the definition of approximate Nash equilibrium. In the case that the players are employing \fwromd{}, by \Cref{lemma:approx_variational_ineq}, we have instead that for any $\vx_i \in \cX_i$:
      \begin{align*}
        \inp[\big]{ \eta_i (2 \bell\^{t}_i  - \bell\^{t-1}_i) + \nabla \vphi_i(\vx_i\^{{t+1}}) - \nabla \vphi_i(\vx_i\^{t})}{ \vx_i\^{{t+1}} - \vx_i } \leq \eps_i\^{t+1}.
    \end{align*} Using the same analysis as above and noting that $\eps^2 < \eps$ we have that

        \begin{align*}
        \inp[\big]{ \bell^{t}_i}{ \vx_i\^{t} - \vx_i } & = \inp[\big]{ \bell^{t}_i}{ \vx_i\^{t+1} - \vx_i } + \inp[\big]{ \bell^{t}_i}{
        \vx\^{t}_i - \vx_i\^{t+1} }                                                                                                                                        \\
                                                       & \leq \frac{\eps^2}{\eta_i} +\frac{2\Omega_i}{\eta_i} \eps(L_i + N -1)  + \eps \\
                                                       & \leq \eps \parens*{\frac{1}{\eta_i} + \frac{2 \Omega_i}{\eta_i}(L_i + N - 1) + 1} \\
                                                       & = \alpha_i \eps. \\
    \end{align*}

    Thus, it is sufficient to let $\eps \leq \min_{i \in [N]} \frac{\eps'}{\alpha_i}$ to ensure that the iterate corresponds to a $\eps'$-NE.
    The number of required LMO calls follows immediately from \Cref{lemma:total_lmo_calls}.
\end{proof}
\thmlinearrate*
We adapt arguments from \citet{wei2021last} and \citet{malitsky2015projected}.

\begin{proof}

    In this proof, for convenience, we define the following $\bell\^t \defeq (\bell\^t_\vx, \bell\^t_\vy)$, $\vm\^t \defeq (\vm\^t_\vx, \vm\^t_\vy)$, $\vphi(\vz) \defeq \vphi_\vx(\vx) + \vphi_\vy(\vy)$. We take $\eps\^t_\vx \defeq \eps\^t_\vy \defeq \eps$, and $\eps\^t_\vz \defeq \eps\^t_\vx + \eps\^t_\vy \defeq 2\eps$. Additionally, we let $\vw\^{t} \defeq 2\vz\^{t} - \vz\^{t-1}$.

    The calls to their respective \apo{}s for $\vx$ and $\vy$ in a single iteration of \Cref{algo:approx_sc_omd} can be written as

    \begin{align*}
        \vx\^{t+1} & = \mathrm{\apo{}}_\cX \parens[\big]{- \eta \inp[\big]{ \bell\^{t}_\vx + \vm\^{t}_\vx - \vm\^{t-1}_\vx}{ \cdot }, \vphi, \vx\^t, \eps}, \\
        \vy\^{t+1} & = \mathrm{\apo{}}_\cY\parens[\big]{- \eta \inp[\big]{ \bell\^{t}_\vy + \vm\^{t}_\vy - \vm\^{t-1}_\vy}{ \cdot }, \vphi, \vy\^t, \eps}.
        \numberthis \label{eq:xy_sc_omd_intermediate_iterate}
    \end{align*}

    This can be written as a single prox call for $\vz$ as follows.

    \begin{align*}
        \vz\^{t+1} & = \mathrm{\apo{}}_\cZ\parens[\big]{- \eta \inp[\big]{ \bell\^{t} + \vm\^{t} - \vm\^{t-1}}{ \cdot }, \vphi, \vz\^t, 2\eps}.
        \numberthis \label{eq:sc_omd_single_prox_call}\end{align*}

    We define $\vz\^{t+1}_*$ as true solution to the prox call in \Cref{eq:sc_omd_single_prox_call}.

    First, we prove a version of Lemma 1 from \citet{wei2021last} and Lemma 3.1 from \citet{malitsky2015projected}; this inequality will allow us to characterize the bound the current distance to optimality in terms of the distance to optimality at the previous iterate. We define $g(\eps) = \max_{i \in \{1, 2\}} \sqrt{2\eta_i\eps} (2L_iD_i + 3 \eta_i)$ when the players are assumed to be employing \Cref{algo:approx_sc_omd}, and $g(\eps) = \eps$ when the players are assumed to be employing \fwromd{}; the reason we make this definition is because \Cref{lemma:approx_variational_ineq} yields a simpler bound when the Wolfe gap is used as a stopping criterion for the approximate proximal computation.

    \begin{lemma} Under the same assumptions as \Cref{thm:linear_rate_approx_sc_omd}, 
        \begin{equation*}
            \dist(\vz\^{t+1}, \cZ^*)^2 + \frac12 \norm[\big]{\vw\^{t} - \vz\^{t+1}}^2 \leq \dist(\vz\^{t}, \cZ^*)^2 +  \frac12 \norm[\big]{\vw\^{t-1} - \vz\^{t}}^2 - \frac{1}{4}\norm[\big]{\vw\^{t} - \vz\^{t}}^2 - \frac{1}{4}\norm[\big]{\vw\^{t} - \vz\^{t+1}}^2 + 4\eps + 8g(\eps).
        \end{equation*}
        
    \end{lemma}
    \begin{corollary}
        Let $\Theta\^{t} \defeq \dist(\vz\^{t}, \cZ^*)^2 + \frac12 \norm[\big]{\vw\^{t-1} - \vz\^{t}}^2$.
        and $\zeta\^t \defeq \norm[\big]{\vw\^{t} - \vz\^{t}}^2 + \norm[\big]{\vw\^{t} - \vz\^{t+1}}^2$.

        Under the same assumptions as \Cref{thm:linear_rate_approx_sc_omd}, 
        \begin{equation*}
            \Theta\^{t+1} \leq \Theta\^t - \frac{1}{4} \zeta\^t + 4\eps + 8g(\eps).
        \end{equation*}
       
        \label{corr:recurrence}
    \end{corollary}

    \begin{proof}
        We use \Cref{lemma:approx_prox} to first note the following for any $\vz \in \cZ$:

        \begin{align*}
            \eta \inp[\big]{ \bell\^{t} + \vm\^{t} - \vm\^{t+1}}{ \vz\^{t+1} - \vz } & \leq \div{\vz\^{t}}{\vz} - \div{\vz\^{t+1}}{\vz} - \div{\vz\^{t+1}}{\vz\^{t}} + 2\eps.
        \end{align*}

        Note that by definition, $\bell\^{t} = \vec{F}(\vz\^{t})$, and additionally by assumption $\vm\^t = \bell\^{t-1} = \vec{F}(\vz\^{t-1})$, so we can rewrite \Cref{eq:sc_omd_single_prox_call}:
        \begin{align*}
            \eta \inp[\big]{ \vec{F}(\vw\^{t})}{ \vz\^{t+1} - \vz } & \leq \div{\vz\^{t}}{\vz} - \div{\vz\^{t+1}}{\vz} - \div{\vz\^{t+1}}{\vz\^{t}} + 2\eps.
            \numberthis \label[ineq]{eq:stacked_three_point_lemma}
        \end{align*}

        Now note that since $\inp[\big]{ \bA\vx}{ \vy }$ is convex with respect to $\vx$ and concave with respect to $\vy$, we have that
        $\eta \inp[\big]{ \vec{F}(\vw\^{t}) - \vec{F}(\vz)}{ \vw\^{t} - \vz } \geq 0$ so this term can be added to the right side of \Cref{eq:stacked_three_point_lemma} to yield the following:

        \begin{equation}
            \begin{aligned}
                \div{\vz\^{t+1}}{\vz} & \leq \div{\vz\^{t}}{\vz} - \div{\vz\^{t+1}}{\vz\^{t}} - \eta \inp[\big]{ \vec{F}(\vw\^{t})}{ \vz\^{t+1} - \vz } + \eta \inp[\big]{ \vec{F}(\vw\^{t}) - \vec{F}(\vz)}{ \vw\^{t} - \vz } + 2\eps \\
                                      & =\div{\vz\^{t}}{\vz} - \div{\vz\^{t+1}}{\vz\^{t}} + \eta \inp[\big]{ \vec{F}(\vw\^{t})}{ \vw\^{t} -\vz\^{t+1} } -\eta \inp[\big]{ \vec{F}(\vz)}{ \vw\^{t} - \vz } + 2\eps                      \\
                                      & = \div{\vz\^{t}}{\vz} - \div{\vz\^{t+1}}{\vz\^{t}} + \eta \inp[\big]{ \vec{F}(\vw\^{t}) - \vec{F}(\vw\^{t-1})}{ \vw\^{t} -\vz\^{t+1} }                                                         \\
                                      & \hspace{1cm} + \eta \inp[\big]{ \vec{F}(\vw\^{t-1})}{ \vw\^{t} -\vz\^{t+1} } -\eta \inp[\big]{ \vec{F}(\vz)}{ \vw\^{t} - \vz } + 2\eps.
                \label[ineq]{eq:sc_omd_regret_bound}
            \end{aligned}
        \end{equation}

        First, we attempt to bound the first inner product that appears on the right-hand side of \Cref{eq:sc_omd_regret_bound}.
        \begin{align*}
            \eta \inp[\big]{ \vec{F}(\vw\^{t}) - \vec{F}(\vw\^{t-1})}{ \vw\^{t} -\vz\^{t+1} } & \leq \eta \norm[\big]{ \vec{F}(\vw\^{t}) - \vec{F}(\vw\^{t-1}) } \norm[\big]{\vw\^{t} -\vz\^{t+1} }
            \numberthis \label[ineq]{ineq:sc_omd_bound_inner_product_first_ineq}                                                                                                                                                                      \\
                                                                                              & \leq  \eta  \norm[\big]{ \vw\^{t} - \vw\^{t-1} } \norm[\big]{\vw\^{t} -\vz\^{t+1} }
            \numberthis \label[ineq]{ineq:sc_omd_bound_inner_product_second_ineq}                                                                                                                                                                     \\
                                                                                              & \leq \frac{1}{2} \eta   \parens*{ \norm[\big]{ \vw\^{t} - \vw\^{t-1} }^2 +  \norm[\big]{ \vw\^{t} - \vz\^{t+1} }^2}
            \numberthis \label[ineq]{ineq:sc_omd_bound_inner_product_third_ineq}                                                                                                                                                                      \\
                                                                                              & \leq \frac{1}{2} \eta  \parens*{2\norm[\big]{ \vw\^{t} - \vz\^{t} }^2 + 2\norm[\big]{ \vz\^{t} - \vw\^{t-1} }^2 + \norm[\big]{ \vw\^{t} - \vz\^{t+1} }^2}.
            \numberthis \label[ineq]{ineq:sc_omd_bound_inner_product_1}
        \end{align*}

        Here we have used the Cauchy-Schwarz inequality in \eqref{ineq:sc_omd_bound_inner_product_first_ineq}, smoothness of $\vec{F}$ with modulus 1 in \eqref{ineq:sc_omd_bound_inner_product_second_ineq}, Young's inequality in \eqref{ineq:sc_omd_bound_inner_product_third_ineq}, and the triangle inequality and again Young's inequality in \Cref{ineq:sc_omd_bound_inner_product_1}.

        Next, we try to bound the second inner product that appears on the right-hand side of \Cref{eq:sc_omd_regret_bound}. 

        We apply \Cref{lemma:approx_variational_ineq} to note that we have:
        \begin{align*}
            \inp[\big]{ \vz\^{t} - \vz\^{t-1} + \eta \vec{F}(\vw\^{t-1})}{ \vz\^{t} -\vz\^{t+1} } & \leq 2g(\eps), \\
            \inp[\big]{ \vz\^{t} - \vz\^{t-1} + \eta \vec{F}(\vw\^{t-1})}{ \vz\^{t} -\vz\^{t-1} } & \leq 2g(\eps).
        \end{align*}

        Adding these two inequalities together we have:

        \begin{align*}
            \inp[\big]{ \vz\^{t} - \vz\^{t-1} + \eta \vec{F}(\vw\^{t-1})}{ \vw\^{t} - \vz\^{t+1} } \leq 4g(\eps).
        \end{align*}

        It follows that

        \begin{equation}
            \begin{aligned}
                \eta \inp[\big]{ \vec{F}(\vw\^{t-1})}{ \vw\^{t} -\vz\^{t+1} } & \leq \inp[\big]{ \vz\^{t} - \vz\^{t-1}}{ \vz\^{t+1} - \vw\^{t} } + 4g(\eps)                                                                  \\
                                                                              & = \inp[\big]{ \vw\^{t} - \vz\^{t}}{ \vz\^{t+1} - \vw\^{t} } + 4g(\eps)                                                                       \\
                                                                              & = \frac{1}{2} \parens*{\norm[\big]{\vz\^{t} - \vz\^{t+1}}^2 - \norm[\big]{\vw\^{t} - \vz\^{t}}^2 - \norm[\big]{\vw\^{t} - \vz\^{t+1}}^2} + 4g(\eps).
            \end{aligned}
            \numberthis \label[ineq]{ineq:sc_omd_bound_inner_product_2}
        \end{equation}

        Combining \Cref{ineq:sc_omd_bound_inner_product_1}
        and \Cref{ineq:sc_omd_bound_inner_product_2} with \Cref{eq:sc_omd_regret_bound}, and noting that $\vphi(\vz)$ = $\frac{1}{2} \norm[\big]{ \vz }^2$. we have

        \begin{align*}
            \div{\vz\^{t+1}}{ \vz} & \leq \div{\vz\^{t}}{\vz} - \div{\vz\^{t+1}}{\vz\^{t}} -\eta \inp[\big]{ \vec{F}(\vz)}{ \vw\^{t} - \vz } + 2\eps                                                                                                                                              \\
                                   &\hspace{1cm} + \frac{1}{2} \parens*{\norm[\big]{\vz\^{t} - \vz\^{t+1}}^2 - \norm[\big]{\vw\^{t} - \vz\^{t}}^2 - \norm[\big]{\vw\^{t} - \vz\^{t+1}}^2} + 4g(\eps)                                                                                                                     \\
                                   &\hspace{1cm} + \frac{1}{2} \eta  \parens*{2\norm[\big]{ \vw\^{t} - \vz\^{t} }^2 + 2\norm[\big]{ \vz\^{t} - \vw\^{t-1} }^2 + \norm[\big]{ \vw\^{t} - \vz\^{t+1} }^2}                                                                                 \\
                                   & \leq \div{\vz\^{t}}{\vz} - \parens*{\frac{1}{2} - \eta } \norm[\big]{\vw\^{t} - \vz\^{t}}^2 - \parens*{\frac{1}{2} - \frac{1}{2}\eta } \norm[\big]{\vw\^{t} - \vz\^{t+1}}^2 + \eta  \norm[\big]{\vz\^{t} - \vw\^{t-1}}^2                                                  \\
                                   & \hspace{1cm}-\eta \inp[\big]{ \vec{F}(\vz)}{ \vw\^{t} - \vz } + 2\eps + g(4\eps)                                                                                                                                                                                            \\
                                   & \leq \frac{1}{2}\norm[\big]{\vz\^{t} - \vz}^2 - \frac{1}{4} \norm[\big]{\vw\^{t} - \vz\^{t}}^2 - \frac{3}{8} \norm[\big]{\vw\^{t} - \vz\^{t+1}}^2 + \frac{1}{4} \norm[\big]{\vz\^{t} - \vw\^{t-1}}^2 \\
                                   &\hspace{1cm}-\eta \inp[\big]{ \vec{F}(\vz)}{ \vw\^{t} - \vz } + 2\eps + 4g(\eps).
        \end{align*}

        Now if we set $\vz = \Pi_{\cZ^*}(\vz\^t)$ above note that
        $-\eta \inp[\big]{ \vec{F}(\vz)}{ \vw\^{t} - \vz } \leq 0$ by convexity-concavity of $\inp[\big]{ \bA \vx }{ \vy }$ with respect to $\vx$ and $\vy$, and optimality of $\vz$ ($\vz \in \cZ^*$). Additionally, we have  that $\dist(\vz\^{t+1}, \cZ^*)^2 \leq \dist(\vz\^{t+1}, \Pi_{\cZ^*}(\vz\^t))^2$. Using these observations as well as multiplying both sides by $2$ and adding $\frac{1}{2} \norm[\big]{\vw\^{t} - \vz\^{t+1}}^2$ to both sides:

        \begin{align*}
                \dist(\vz\^{t+1}, \cZ^*)^2 + \frac12 \norm[\big]{\vw\^{t} - \vz\^{t+1}}^2 & \leq \norm[\big]{\vz\^{t} - \Pi_{\cZ^*}(\vz) }^2  - \frac{1}{2}\norm[\big]{\vw\^{t} - \vz\^{t}}^2 - \frac{1}{4}\norm[\big]{\vw\^{t} - \vz\^{t+1}}^2 \\
                                                                                          & \hspace{1cm}+ \frac{1}{2}\norm[\big]{\vz\^{t} - \vw\^{t-1}}^2 + 4\eps + 8g(\eps)                                                                                   \\
                                                                                          & = \dist(\vz\^{t}, \cZ^*)^2 +  \frac12 \norm[\big]{\vw\^{t-1} - \vz\^{t}}^2 - \frac{1}{2}\norm[\big]{\vw\^{t} - \vz\^{t}}^2                          \\
                                                                                          & \hspace{1cm} - \frac{1}{4}\norm[\big]{\vw\^{t} - \vz\^{t+1}}^2 + 4\eps + 8g(\eps)                                                                                  \\
                                                                                          & \leq \dist(\vz\^{t}, \cZ^*)^2 +  \frac12 \norm[\big]{\vw\^{t-1} - \vz\^{t}}^2 - \frac{1}{4}\norm[\big]{\vw\^{t} - \vz\^{t}}^2                       \\
                                                                                          & \hspace{1cm} - \frac{1}{4}\norm[\big]{\vw\^{t} - \vz\^{t+1}}^2 + 4\eps + 8g(\eps)                                                                                  \\
        \end{align*}

        Next, we define $\Theta\^{t} \defeq \dist(\vz\^{t}, \cZ^*)^2 + \frac12 \norm[\big]{\vw\^{t-1} - \vz\^{t}}^2$.
        and $\zeta\^t \defeq \norm[\big]{\vw\^{t} - \vz\^{t}}^2 + \norm[\big]{\vw\^{t} - \vz\^{t+1}}^2$. This allows us to write the above as
        \begin{equation*}
            \Theta\^{t+1} \leq \Theta\^t - \frac{1}{4} \zeta\^t + 4\eps + 8g(\eps)
            \numberthis \label[ineq]{eq:sc_omd_recurrence}
        \end{equation*} as desired.

    \end{proof}

    Next, we prove a version of Lemma 4 of \citet{wei2021last}:

    \begin{lemma} Under the same assumptions as \Cref{thm:linear_rate_approx_sc_omd}, for any $\vz' \in \cZ$ such that $\vz' \neq \vz\^{t+1}_*$
        \begin{equation*}
            \frac{8}{25} \eta^2 \frac{\brackets*{\inp*{\vec{F}\parens*{\vz\^{t+1}_*}}{\vz\^{t+1}_*-\vz'}}_+^2}{\norm*{\vz'-\vz\^{t+1}_*}^2} \leq \norm*{\vw\^{t}-{\vz}\^{t} }^2 + 2\norm*{\vw\^{t}-{\vz}\^{t+1}}^2 + 8 \eps.
        \end{equation*}
        \label{lemma:approx_sp_ms_bound} \end{lemma}

    \begin{proof}
        By first-order optimality of $\vz\^{t+1}_*$ we have:

        \begin{align*}
            \inp*{ \vz\^{t+1}_* - \vz\^{t} + \eta \vec{F}(\vw\^{t})}{ \vz' -\vz\^{t+1}_* } \geq 0.
            \numberthis \label[ineq]{eq:sc_omd_true_sol_first_order_opt}
        \end{align*}

        Rearranging
        \Cref{eq:sc_omd_true_sol_first_order_opt}, we have:

        \begin{align*}
            \inp*{ \vz\^{t+1}_* - \vz\^{t} }{ \vz' -\vz\^{t+1}_* } & \geq \inp[\big]{ \eta \vec{F}(\vw\^{t})}{ \vz\^{t+1}_* - \vz' }                                                                                                             \\
                                                                           & =  \eta \inp[\big]{ \vec{F}\parens*{\vz\^{t+1}_*}}{ \vz\^{t+1}_* - \vz' } - \eta \inp*{\vec{F}\parens*{\vz\^{t+1}_*} -  \vec{F}(\vw\^{t})}{ \vz\^{t+1}_* - \vz' }                \\
                                                                           & \geq \eta \inp[\big]{ \vec{F}\parens*{\vz\^{t+1}_*}}{ \vz\^{t+1}_* - \vz' } - \eta  \norm*{ \vec{F}(\vw\^{t}) -  \vec{F}\parens*{\vz\^{t+1}_*}} \norm*{ \vz\^{t+1}_* - \vz'}
            \numberthis \label[ineq]{ineq:wei_lemma_four_first_ineq}                                                                                                                                                                                      \\
                                                                           & \geq \eta  \inp[\big]{ \vec{F}\parens*{\vz\^{t+1}_*}}{ \vz\^{t+1}_* - \vz' } - \eta  \norm*{\vw\^{t} - \vz\^{t+1}_*} \norm*{ \vz\^{t+1}_* - \vz' }
            \numberthis \label[ineq]{ineq:wei_lemma_four_second_ineq}                                                                                                                                                                                     \\
                                                                           & \geq \eta  \inp[\big]{ \vec{F}\parens*{\vz\^{t+1}_*}}{ \vz\^{t+1}_* - \vz' } - \frac{1}{4}  \norm*{\vw\^{t} - \vz\^{t+1}_*} \norm*{ \vz\^{t+1}_* - \vz' }.
            \numberthis \label[ineq]{ineq:wei_lemma_four_third_ineq}
        \end{align*}

        Here we have applied Cauchy-Schwarz in \Cref{ineq:wei_lemma_four_first_ineq}, smoothness of $\vec{F}$ with modulus 1 in \Cref{ineq:wei_lemma_four_second_ineq}, and the condition on $\eta$ in \Cref{ineq:wei_lemma_four_third_ineq}.

        Next, we apply Cauchy-Schwarz to upper bound the left-hand side and rearrange to obtain the following:

        \begin{align*}
            \norm*{ \vz\^{t+1}_* - \vz' } \parens*{ \norm*{ \vz\^{t+1}_* - \vz\^{t}}  +  \frac{1}{4}  \norm*{\vw\^{t} - \vz\^{t+1}_*} }   \geq  \eta  \inp*{ \vec{F}\parens*{\vz\^{t+1}_*}}{ \vz\^{t+1}_* - \vz'}.
        \end{align*}

        Then squaring (and taking care in case the right-hand side is negative), we have:

        \begin{align*}
            \parens*{ \norm*{\vz\^{t+1}_*-{\vz}\^{t} }  + \frac{1}{4}\norm*{\vw\^{t}-\vz\^{t+1}_*} }^2  \geq \eta^2 \frac{\brackets*{\inp*{\vec{F}\parens*{\vz\^{t+1}_*}}{\vz\^{t+1}_*-\vz'}}_+^2}{\norm*{\vz'-\vz\^{t+1}_*}^2}.
        \end{align*}

        Now, note that
        \begin{align*}
            \parens*{ \norm*{\vz\^{t+1}_*-{\vz}\^{t} }  + \frac{1}{4}\norm*{\vw\^{t}-\vz\^{t+1}_*} }^2 & \leq  \parens*{\norm*{\vw\^{t}-{\vz}\^{t} }  + \frac{5}{4}\norm*{\vw\^{t}-\vz\^{t+1}_*}}^2          \numberthis \label[ineq]{ineq:spms_ineq_one} \\
                                                                                                           & \leq \parens*{\frac{5}{4}\norm*{\vw\^{t}-{\vz}\^{t} }  + \frac{5}{4}\norm*{\vw\^{t}-\vz\^{t+1}_*}}^2 \\
                                                                                                           & \leq \frac{25}{8} \parens*{\norm*{\vw\^{t}-{\vz}\^{t} }^2 + \norm*{\vw\^{t}-\vz\^{t+1}_*}^2} \numberthis \label[ineq]{ineq:spms_ineq_two}
        \end{align*}

        and so combining with the above, we have

        \begin{align*}
            \frac{8}{25} \eta^2 \frac{\brackets*{F\left(\vz\^{t+1}_*\right)^{\top}\left(\vz\^{t+1}_*-\vz^{\prime}\right)}_+^2}{\norm*{\vz^{\prime}-\vz\^{t+1}_*}^2} & \leq \norm*{\vw\^{t}-{\vz}\^{t} }^2 + \norm*{\vw\^{t}-\vz\^{t+1}_*}^2                                               \\
                                                                                                                                                                     & \leq \norm[\big]{\vw\^{t}-{\vz}\^{t} }^2 + 2\norm[\big]{\vw\^{t}-{\vz}\^{t+1}}^2 + 2\norm[\big]{\vz\^{t+1}_* - \vz\^{t+1} }^2 \numberthis \label[ineq]{ineq:spms_ineq_three}\\
                                                                                                                                                                     & \leq \norm[\big]{\vw\^{t}-{\vz}\^{t} }^2 + 2\norm[\big]{\vw\^{t}-{\vz}\^{t+1}}^2 + 8 \eps \numberthis \label[ineq]{ineq:spms_ineq_four}.
        \end{align*}
        
        In \Cref{ineq:spms_ineq_one}, we use the triangle inequality, in \Cref{ineq:spms_ineq_two}, we use Young's inequality, in \Cref{ineq:spms_ineq_three}, we use the triangle inequality and Young's inequality, and in \Cref{ineq:spms_ineq_four} we use \Cref{lemma:approx_prox}.
    \end{proof}

    Finally, we follow the Proof of Theorem 8 of~\citet{wei2021last} to prove our main result.

    \begin{align*}
        \zeta\^{t} & \geq \frac{1}{2} \norm*{\vz\^{t+1}-\vw\^{t}}^{2} + \frac{1}{4}\left(2\norm*{\vz\^{t+1}-\vw\^{t}}^{2}+\norm*{\vw\^{t}-\vz\^{t}}^{2} \right)                                                                                                                                                        \\
                   & \geq \frac{1}{2} \norm*{\vz\^{t+1}-\vw\^{t}}^{2} + \frac{1}{4} \parens*{\sup_{z' \in \cZ} \frac{8}{25} \eta^2 \frac{\brackets*{\vec{F}\left(\vz\^{t+1}_*\right)^{\top}\left(\vz\^{t+1}_*-\vz^{\prime}\right)}_+^2}{\norm*{\vz^{\prime}-\vz\^{t+1}_*}^2} - 8\eps} \numberthis \label[ineq]{ineq:final_proof_one}  \\
                   & = \frac{1}{2} \norm*{\vz\^{t+1}-\vw\^{t}}^{2} + \sup_{z' \in \cZ} \frac{2}{25} \eta^2 \frac{\brackets*{\vec{F}\left(\vz\^{t+1}_*\right)^{\top}\left(\vz\^{t+1}_*-\vz^{\prime}\right)}_+^2}{\norm*{\vz^{\prime}-\vz\^{t+1}_*}^2} - 2\eps                                                                          \\
                   & \geq \frac{1}{2} \norm*{\vz\^{t+1}-\vw\^{t}}^{2}+  \frac{2 \eta^{2} \nu^{2}}{25}\norm*{\vz\^{t+1}_*-\Pi_{\mathcal{Z}^{*}}\left(\vz\^{t+1}_*\right)}^{2} - 2\eps  \numberthis \label[ineq]{ineq:final_proof_two}                                                                                      \\
                   & \geq \frac{1}{2} \norm*{\vz\^{t+1}-\vw\^{t}}^{2}+  \frac{2 \eta^{2} \nu^{2}}{25} \parens*{\norm*{\vz\^{t+1}-\Pi_{\mathcal{Z}^{*}}\left(\vz_*\^{t+1}\right)} - \norm*{\vz\^{t+1}_* - \vz\^{t+1}}}^2 - 2\eps \numberthis \label[ineq]{ineq:final_proof_three}                                                                                           \\
                   & \geq \frac{1}{2} \norm*{\vz\^{t+1}-\vw\^{t}}^{2}+  \frac{2 \eta^{2} \nu^{2}}{25} \parens*{\norm*{\vz\^{t+1}-\Pi_{\mathcal{Z}^{*}}\left(\vz\^{t+1}\right)} - \norm*{\vz\^{t+1}_* - \vz\^{t+1}}}^2 - 2\eps \numberthis \label[ineq]{ineq:final_proof_four}                                                                                           \\
                   & \geq \frac{1}{2} \norm*{\vz\^{t+1}-\vw\^{t}}^{2}+  \frac{ \eta^{2} \nu^{2}}{25} \parens*{\norm*{\vz\^{t+1}-\Pi_{\mathcal{Z}^{*}}\left(\vz\^{t+1}\right)}^{2} - 2\norm*{\vz\^{t+1}_* - \vz\^{t+1}}^2} - 2\eps \numberthis \label[ineq]{ineq:final_proof_five}                                                                                           \\
                   & \geq \frac{1}{2} \norm*{\vz\^{t+1}-\vw\^{t}}^{2}+  \frac{ \eta^{2} \nu^{2}}{25} \parens*{\norm*{\vz\^{t+1}-\Pi_{\mathcal{Z}^{*}}\left(\vz\^{t+1}\right)}^{2} - 8\eps} - 2\eps                                                                                                                                 \\
                   & \geq C_2 \Theta\^{t+1} - C_1 \eps.
    \end{align*}
    where $C_2 = \min(\frac{1}{2}, \frac{\eta^2\nu^2}{25})$ and $C_1 = 2 \parens*{1+\frac{4\eta^2\nu^2}{25}}$.

    We use \Cref{lemma:approx_sp_ms_bound} in \Cref{ineq:final_proof_one}, the \ref{ineq:spms} condition in \Cref{ineq:final_proof_two}, the triangle inequality in  \Cref{ineq:final_proof_three}, the definition of the projection operator in \Cref{ineq:final_proof_four}, and Young's inequality in \Cref{ineq:final_proof_five}.

    By \Cref{corr:recurrence}:
    \begin{align*}
        \Theta\^{t+1} & \leq \Theta\^t - \frac{1}{4} \zeta\^t + 4\eps + 8g(\eps)                                \\
                      & \leq \Theta\^t  - \frac{1}{4}C_2 \Theta\^{t+1} + \parens*{\frac{1}{4} C_1 + 4} \eps + 8g(\eps).
    \end{align*}

    Rearranging, we have that

    \[\Theta\^{t+1} \parens*{1 + \frac{1}{4} C_2} \leq \Theta\^t + \parens*{\frac{1}{4} C_1 + 4} \eps + 8g(\eps). \]

    Then define $\Xi\^t = \Theta\^t - \frac{(16 + C_1)\eps + 32g(\eps)}{C_2}$. We can then write the above as

    \[\parens*{\Xi\^{t+1} +  \frac{(16 + C_1)\eps + 32g(\eps)}{C_2}}\parens*{1+\frac{1}{4}C_2} \leq \Xi\^t +  \frac{(16 + C_1)\eps + 32g(\eps)}{C_2} + \parens*{\frac{1}{4} C_1 + 4} \eps + 8g(\eps). \]

    Rearranging we have

    \begin{align*}
        \parens*{1+\frac{C_2}{4}} \Xi\^{t+1} \leq \Xi\^t.
    \end{align*}

    Thus we have
    \begin{align*}
        \Xi\^t \leq \parens*{1+\frac{C_2}{4}}^{-t+1} \Xi\^1 \leq 2(1+\frac{C_2}{4})^{-t} \Xi\^1.
    \end{align*}

    Note by construction $\Xi\^1 \leq \Theta\^1 = \dist(\vz\^1, \cZ^*)^2$ and $\Xi\^t \geq \dist(\vz\^1, \cZ^*)^2  -  \frac{(16 + C_1)\eps + 32g(\eps)}{C_2}$ so from the above we have that

    \begin{align*}
        \dist(\vz\^{t}, \cZ^*)^2 -  \frac{(16 + C_1)\eps + 32g(\eps)}{C_2} & \leq \Xi\^{t}                                        \\
                                                             & \leq 2 \parens*{1 + \frac{C_2}{4}}^{-t} \Xi\^1               \\
                                                             & \leq 2 \parens*{1 + \frac{C_2}{4}}^{-t} \Theta\^1            \\
                                                             & = 2 \parens*{1 + \frac{C_2}{4}}^{-t} \dist(\vz\^1, \cZ)^2
    \end{align*}

    so that

    \begin{align*}
        \dist(\vz\^{t}, \cZ^*) \leq 2 \parens*{1 + \frac{C_2}{4}}^{-t} \dist(\vz\^1, \cZ^*)^2 +  \frac{(16 + C_1)\eps + 32g(\eps)}{C_2}
    \end{align*}
    as desired (using the appropriate $g$ based on whether the players are assumed to be employing \Cref{algo:approx_sc_omd} or \fwromd{}).

    Next, we argue why the runtime is as stated when using \fwromd. Note, that \afw{} for Player $i$ can achieve a $\eps$ optimal solution with $O\parens*{L_i \frac{D_i^2}{\delta_i^2} \log \frac{L_iD_i}{\eps}}$ LMO calls. Given the convergence rate bound, note that we can achieve a $\frac{2C_2 + 48 + C_1}{C_2} \eps$ solution we need at most $\frac{\log \frac{1}{\eps}}{\log \frac{4+C_2}{4}}$ iterations of \Cref{algo:approx_sc_omd}.
    It follows that we need $O\parens*{\frac{\log \frac{1}{\eps}}{\log \frac{4+C_2}{4}}L_i\frac{D_i^2}{\delta_i^2} \log \frac{L_iD_i}{\eps}}$ calls for Player $i$. If we let $\eps' = \frac{2C_2 + 48 + C_1}{C_2} \eps$, then $\eps = \frac{C_2}{2C_2+ 48 + C_1} \eps' $ so we can compute a $\eps'$-NE in $O\parens*{\max_{i \in \{1, 2\}} \frac{\log \frac{2C_2+ 48 + C_1}{C_2\eps'}}{\log \frac{4+C_2}{4}} \frac{L_iD_i^2}{\delta_i^2} \log \frac{(2C_2+ 48 + C_1)L_iD_i}{C_2\eps'}}$  calls.

    Finally, the size of the support follows from the fact that \afw{} adds at most one pure strategy at every iteration, which means that at every iteration of our approximate method, \afw{} will return a strategy with $O\parens*{\max_{i \in \{1, 2\}} L_i \frac{D_i^2}{\delta_i^2} \log \frac{L_iD_i}{\eps}} = O\parens*{\max_{i \in \{1, 2\}} \frac{L_iD_i^2}{\delta^2_i} \log \frac{L_iD_i(2C_2+ 48 + C_1)}{C_2\eps'}}$ in the support, and in particular this is yes for the last strategy returned by \fwromd{}.

\end{proof}

\section{Game Descriptions}
\label{sec:game_descriptions}
\paragraph{Two- and Three-Player Kuhn Poker}
Two-player Kuhn poker was originally proposed by \citet{kuhn1950simplified}. We employ the three-player variation described in \citet{farina2018ex}. In a three-player
Kuhn poker game with rank $r$ there are $r$ possible cards. At
the beginning of the game, each player pays one chip to the
pot, and each player is dealt a single private card. The first
player can check or bet, i.e., putting an additional chip in
the pot. Then, the second player can check or bet after a first
player’s check or fold/call the first player’s bet. If no bet
was previously made, the third player can either check or
bet. Otherwise, the player has to fold or call. After a bet of
the second player (resp., third player), the first player (resp.,
the first and the second players) still has to decide whether
to fold or to call the bet. At the showdown, the player with
the highest card who has not folded wins all the chips in the
pot.

\paragraph{Two- and Three-Player Liar's Dice}
Liar's dice is another standard benchmark
introduced by \citet{lisy2015online}. At the beginning of the game, each of the
players privately rolls an unbiased $k$-face die. Then, the
players alternate in making (potentially no) claims
about their toss. The first player begins bidding, announcing
any face value up to $k$ and the minimum number of dice
that the player believes are showing that value among the
dice of all the players. Then, each player has two choices
during their turn: to make a higher bid or to challenge the
previous bid by declaring the previous bidder a ``liar''. A
bid is higher than the previous one if either the face value
is higher, or the number of dice is higher. If the current
player challenges the previous bid, all dice are revealed. If
the bid is valid, the last bidder wins and obtains a reward
of $+1$ while the challenger obtains a negative payoff of $-1$.
Otherwise, the challenger wins and gets reward $+1$, and the
last bidder obtains reward of $-1$. All the other players obtain
reward $0$. We use parameter $k=6$ in the two-player version (this is a standard value used in several papers) and
$k=3$ in the three-player version.

\paragraph{Two-Player Leduc Poker}
Leduc poker is another classic two-player benchmark game introduced by \citet{southey2012bayes}. We employ game instances of rank 3, in which the
deck consists of three suits with three cards each. The maximum number of raises per betting round can
be either 1 or 2. As the game starts, players pay one chip
to the pot. There are two betting rounds. In the first one, a
single private card is dealt to each player, while in the second
round, a single board card is revealed. The raise amount is
set to 2 and 4 in the first and second rounds, respectively.

\paragraph{Three-Player Goofspiel}
This bidding game was originally introduced
by \citet{ross1971goofspiel}. We use a 3-rank variant; that is, each player
has a hand of cards with values ${-1, 0, 1}$. A third stack
of cards with values ${-1, 0, 1}$ is shuffled and placed on
the table. At each turn, a prize card is revealed, and each
player privately chooses one of his or her cards to bid, with the
highest card winning the current prize. In case of a tie, the
prize is split evenly among the winners. After three turns, all
the prizes have been dealt out, and the payoff of each player
is computed as follows: each prize card's value is equal to
its face value, and the players' scores are computed as the
sum of the values of the prize cards they have won. For our experiments, we used the limited information variant~\citep{lanctot2009monte}. In this variant, instead of the players revealing the cards they have chosen to play, the bid cards are submitted to a referee (who is fair and trusted by all the players), who simulates the gameplay as before (the highest card wins the prize, and in the case of a tie the prize is split evenly among the winners).

\section{Pseudocode of Algorithms used in Experiments}
\label{sec:experimental_algos_pseudocode}
In this section, we describe \fwomd{} and the algorithms we compare against in our experiments. In all of our experiments, optimistic variants use the previous loss as the prediction: $\vm\^t = \bell\^{t-1}$.
We drop the subscript $i$ throughout most of this section because we take the point of view of a generic agent applying the algorithm, except for the description of averaging and restarting in \Cref{sec:restarting_pseudocode}.

\subsection{\fwomd}
In \Cref{algo:approx_omd}, we present a non-optimistic version of \Cref{algo:approx_sc_omd}. \fwomd{} is \Cref{algo:approx_omd} instantiated with \afw{} (\Cref{algo:afw}) as the \apo.

\begin{figure}[H]
    \removelatexerror\begin{algorithm}[H]
        \DontPrintSemicolon
        \caption{\texttt{OMD} with Approximate Proximal Computations}
        \label{algo:approx_omd}
        \KwData{%
        $\cX \subseteq \R^n$: convex and compact set \;
        $\vphi: \cX \to \R_{\geq 0}$: $L$-smooth, 1-strongly convex \;
        $\eta > 0$: step-size parameter\;
        $\eps\^{t}$: desired accuracy of prox call at each $t$\;
        $\operatorname{\apo{}_\cX}$: an \apo{} for $\cX$\;
        $\vx\^0 \in \cX$\;
        $\bell\^0 = \mathbf{0}$ }
        \Fn{NextStrategy()}{
            \Return{$\operatorname{APO}_\cX(-\eta \inp[\big]{ \bell\^{t-1}}{ \cdot },
                    \vphi, \vx\^{t-1}, \eps)$}
        }
    \end{algorithm}
\end{figure}

\subsection{\normalfont{(\texttt{O})\texttt{FTPL}}}
In \Cref{algo:ftpl,algo:oftpl}, we present \ftpl{} and \oftpl{}, as we implemented for our experiments. We used the Gumbel distribution to generate noise, with location 0, and scale $\eta$, since this corresponds to multiplicative weights using a stepsize of $\frac{1}{\eta}$~\citep{abernethy2016perturbation, suggala2020follow}.

\begin{figure}[H]
    \removelatexerror\begin{algorithm}[H]
        \DontPrintSemicolon
        \caption{\ftpl{}~\citep{kalai2005efficient}}
        \label{algo:ftpl}
        \KwData{%
            $\cX \subseteq \R^n$: convex and compact set \;
            $\mathrm{LMO}_\cX$: LMO for $\cX$ \;
            $\eta > 0$: noise parameter\;
            $m$: number of samples\;
            $\vx\^0 \in \cX$\;
            $\bell\^0 = \mathbf{0}$ }
        \Fn{NextStrategy()}{
            \Return{$\operatorname{LMO}_\cX(\sum_{i=0}^{t-1} \bell\^{i} - \operatorname{Gumbel}(0, \eta))$}
        }
    \end{algorithm}
\end{figure}

\begin{figure}[H]
    \removelatexerror\begin{algorithm}[H]
        \DontPrintSemicolon
        \caption{Optimistic \ftpl{} (\oftpl{})~\citep{suggala2020follow}}
        \label{algo:oftpl}
        \KwData{%
            $\cX \subseteq \R^n$: convex and compact set \;
            $\mathrm{LMO}_\cX$: LMO for $\cX$ \;
            $\eta > 0$: noise parameter\;
            $m$: number of samples\;
            $\vx\^0 \in \cX$\;
            $\bell\^0 = \vm\^0 = \mathbf{0}$ }
        \Fn{NextStrategy($\vm\^t$)}{
            \Return{$\operatorname{LMO}_\cX(\sum_{i=0}^{t-1} \bell\^{i} + \vm\^t - \operatorname{Gumbel}(0, \eta))$}
        }
    \end{algorithm}
\end{figure}

\subsection{\normalfont{(\texttt{O})\texttt{FP}}}
Next, in \Cref{algo:fp,algo:ofp}, we present \fp{} and \ofp{}, as we implemented for our experiments. \ofp{}, is an optimistic generalization of \fp{}, based on the same idea used to generalize \ftpl{} and \ftrl{} to \oftpl{} and \oftrl{}. \fp{} can be thought of as letting the regularization/perturbation term go to 0 in \ftrl{} and \ftpl{} (letting the stepsize go to infinity and the noise go to 0, respectively), and similarly with \ofp{} and \oftrl{}/\oftpl{}.

\begin{figure}[H]
    \removelatexerror\begin{algorithm}[H]
        \DontPrintSemicolon
        \caption{Fictitious Play (\fp{})~\citep{brown1951iterative}}
        \label{algo:fp}
        \KwData{%
            $\cX \subseteq \R^n$: convex and compact set \;
            $\mathrm{LMO}_\cX$: LMO for $\cX$ \;
            $\bell\^0 = \mathbf{0}$ }
        \Fn{NextStrategy()}{
            \Return{$\operatorname{LMO}_\cX(\sum_{i=0}^{t-1} \bell\^{i})$}
        }
    \end{algorithm}
\end{figure}

\begin{figure}[H]
    \removelatexerror\begin{algorithm}[H]
        \DontPrintSemicolon
        \caption{Optimistic Fictitious Play (\ofp{})}
        \label{algo:ofp}
        \KwData{%
            $\cX \subseteq \R^n$: convex and compact set \;
            $\mathrm{LMO}_\cX$: LMO for $\cX$ \;
            $\bell\^0 = \vm\^0 = \mathbf{0}$ }
        \Fn{NextStrategy($\vm\^t$)}{
            \Return{$\operatorname{LMO}_\cX(\sum_{i=0}^{t-1} \bell\^{i} + \vm\^t)$}
        }
    \end{algorithm}
\end{figure}

\subsection{\normalfont{\texttt{O}(\texttt{BR})}}
Finally, in \Cref{algo:br,algo:obr}, we present \br{} and \obr{}, as we implemented for our experiments. \obr{}, is an optimistic generalization of \br{}, based on the same idea used to generalize \ftpl{} and \ftrl{} to \oftpl{} and \oftrl{}. \br{} and \obr{} can be thought of as letting the regularization term go to 0 in \texttt{OMD} and \romd{} respectively (letting the stepsize go to infinity).

\begin{figure}[H]
    \removelatexerror\begin{algorithm}[H]
        \DontPrintSemicolon
        \caption{Best Response (\br{})}
        \label{algo:br}
        \KwData{%
            $\cX \subseteq \R^n$: convex and compact set \;
            $\mathrm{LMO}_\cX$: LMO for $\cX$ \;
            $\bell\^0 = \mathbf{0}$ }
        \Fn{NextStrategy()}{
            \Return{$\operatorname{LMO}_\cX(\bell\^{t-1})$}
        }
    \end{algorithm}
\end{figure}

\begin{figure}[H]
    \removelatexerror\begin{algorithm}[H]
        \DontPrintSemicolon
        \caption{Optimistic Best Response (\obr{})}
        \label{algo:obr}
        \KwData{%
            $\cX \subseteq \R^n$: convex and compact set \;
            $\mathrm{LMO}_\cX$: LMO for $\cX$ \;
            $\bell\^0 = \vm\^0 = \mathbf{0}$ }
        \Fn{NextStrategy($\vm\^t$)}{
            \Return{$\operatorname{LMO}_\cX(\bell\^{t-1} + \vm\^t - \vm\^{t-1})$}
        }
    \end{algorithm}
\end{figure}

\subsection{Averaging and Restarting Pseudocode}
\label{sec:restarting_pseudocode}
When using different averaging schemes, the duality gap is computed with respect to the average iterate (as defined by that averaging scheme).

When using uniform averaging, the weight placed on the new iterate is $f(t) = \frac{1}{t+1}$.

When using linear averaging, the weight placed on the new iterate is $f(t) = \frac{2}{t+2}$.

When using quadratic averaging, the weight placed on the new iterate is $f(t) = \frac{6t + 6}{(t+2)(2t+3)}$.

When using last iterate, the weight placed on the new iterate is $f(t) = 1$.

The ``average'' iterate is then set as follows:
$\bar{\vx}_i\^{t+1} = \bar{\vx}_i\^{t} + f(t) (\vx_i\^{t+1} - \bar{\vx}_i\^{t})$.

\begin{figure}[H]
    \removelatexerror
    \begin{algorithm}[H]
        \DontPrintSemicolon
        \caption{Adaptive Restarting of Algorithm}
        \label{algo:adaptive_restarting}
        \KwData{
            $T$: number of iterations to run algorithm\;
            $\bar{\vx}_1\^0 = \vx_1\^0 \in \cX_1,\, \bar{\vx}_2\^0 = \vx_2\^0 \in \cX_2$\;
            $\operatorname{Alg}_i$ for $i \in \{1, 2\}$: algorithm which generates iterates for each of the players\;
        }
        $r = 0$\;
        $\xi = \max_{\vx_2 \in \cX_2} \inp[\big]{ \bA \bar{\vx}_1\^0}{ \vx_2 } - \min_{\vx_1 \in \cX_1} \inp[\big]{ \bA\vx_1}{ \bar{\vx}_2\^0 }$ \;
        \For{$t = 0, \dots , T-1$}
        {
        $\vx\^{t+1}_1 = \operatorname{Alg}_{1}()$ \;
        $\vx\^{t+1}_2 = \operatorname{Alg}_{2}()$ \;
        $\bar{\vx}_1\^{t+1} =  \bar{\vx}_1\^{t} + f(r) (\vx\^{t+1}_1 - \bar{\vx}_1\^{t})$ \;
        $\bar{\vx}_2\^{t+1} =  \bar{\vx}_2\^{t} + f(r) (\vx\^{t+1}_2 - \bar{\vx}_2\^{t})$ \;
        \uIf{$\max_{\vx_2 \in \cX_2} \inp[\big]{ \bA \bar{\vx}_1\^{t+1}}{ \vx_2 } - \min_{\vx_1 \in \cX_1} \inp[\big]{ \bA\vx_1}{ \bar{\vx}_2\^{t+1} } \leq \frac{\xi}{2}$}{
        $\xi = \max_{\vx_2 \in \cX_2} \inp[\big]{ \bA \bar{\vx}_1\^{t+1}}{ \vx_2 } - \min_{\vx_1 \in \cX_1} \inp[\big]{ \bA\vx_1}{ \bar{\vx}_2\^{t+1} }$\;
        $r = 0$\;
        }
        }
    \end{algorithm}
\end{figure}

\begin{figure*}[t]
    \centering
    \includegraphics[width=\textwidth]{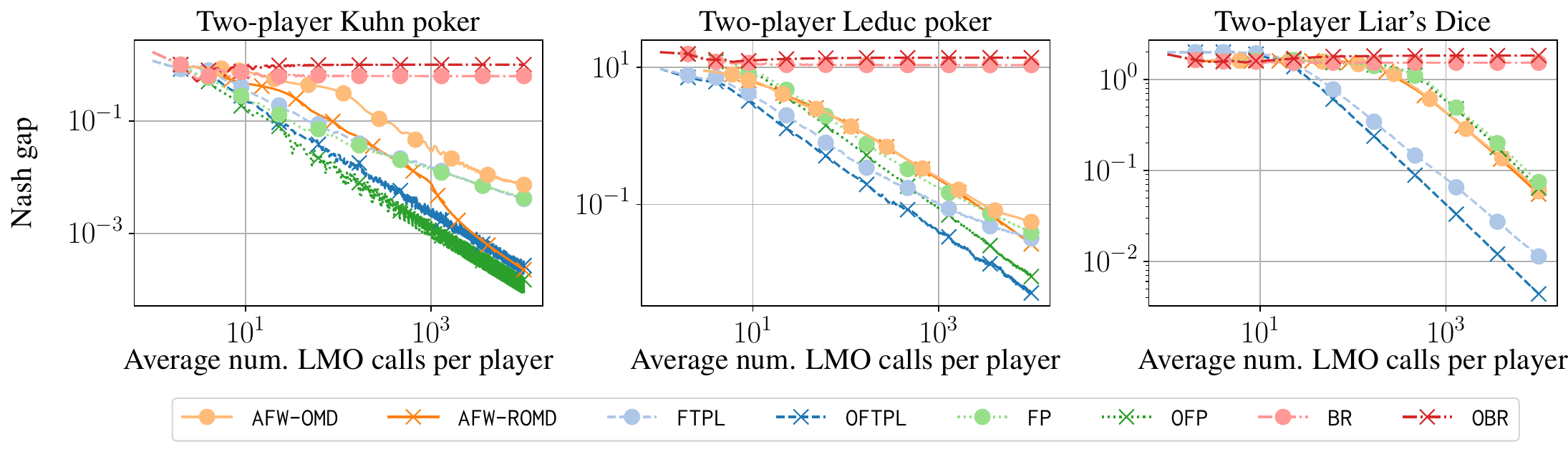}\\
    \includegraphics[width=\textwidth]{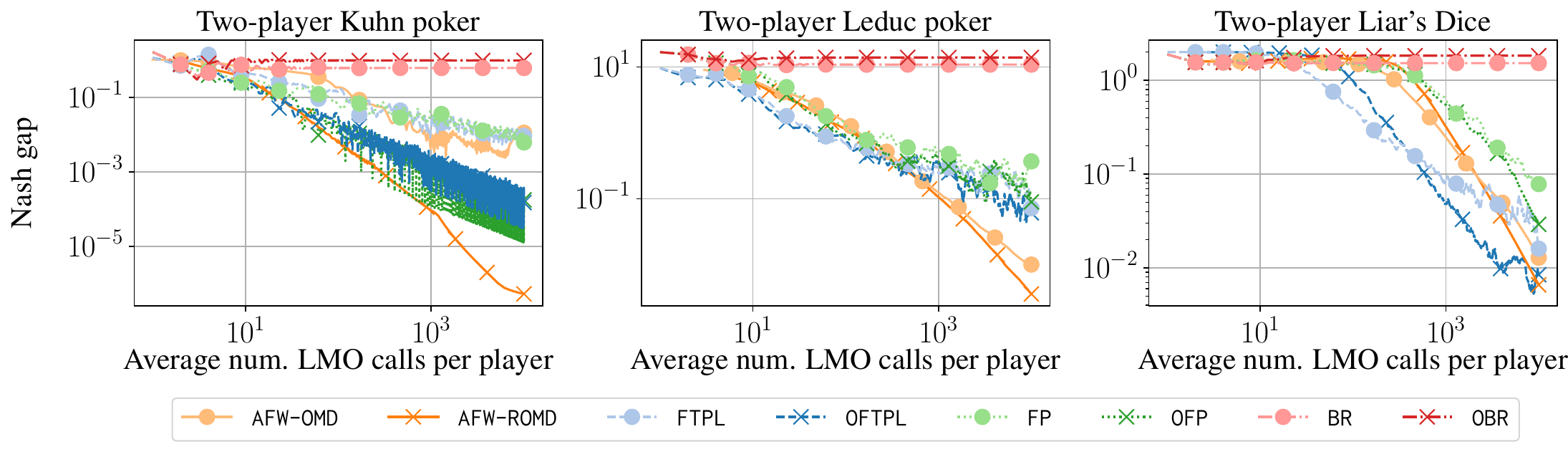}\\
    \includegraphics[width=\textwidth]{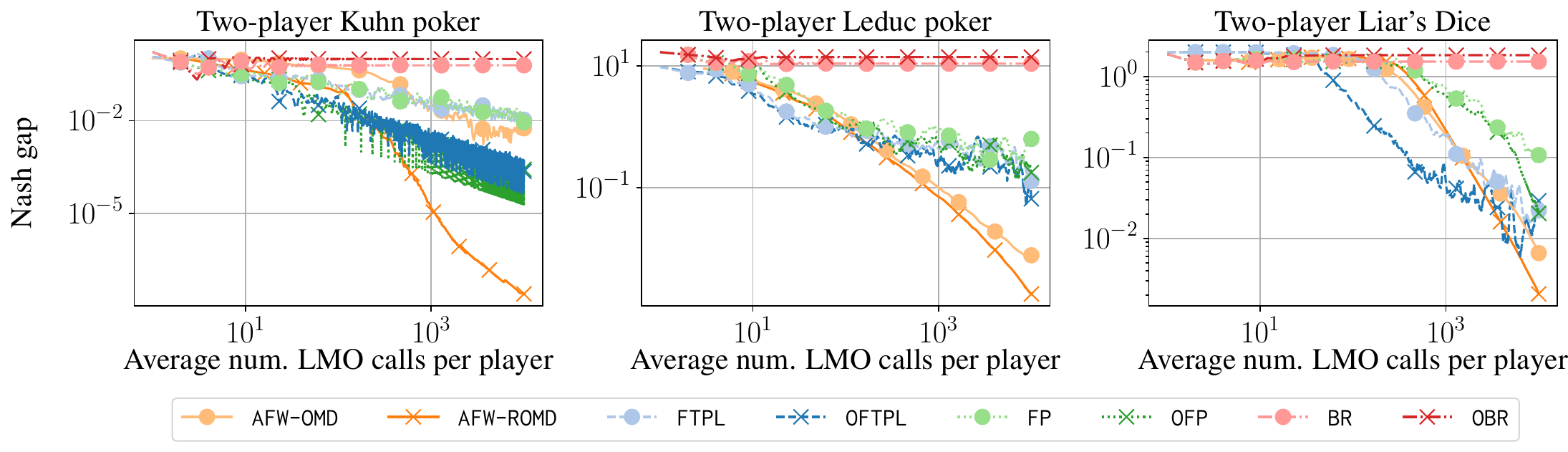}\\
    \includegraphics[width=\textwidth]{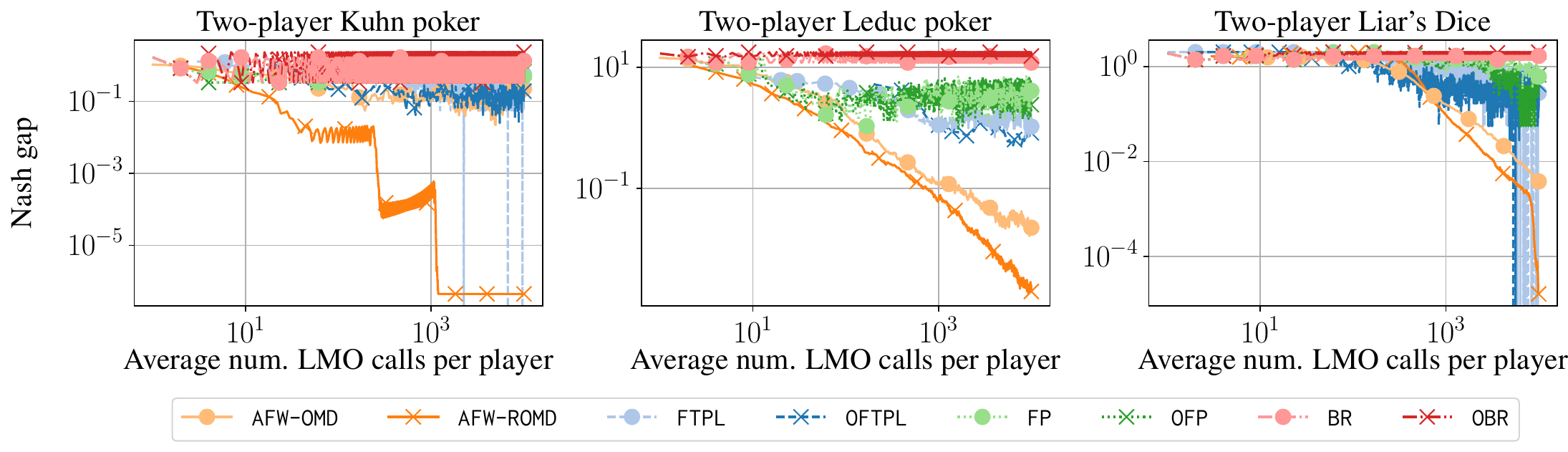}\\
    \caption{Convergence to NE as a function of average LMO calls per player for \fwomd{}, \fwromd{}, \ftpl{}, \oftpl{}, \fp{}, \ofp{}, \br{}, and \obr{}, for, from top to bottom, uniform, linear, quadratic, and last averaging, without using restarting.}
    \vspace{-5mm}
    \label{fig:NE_avg_comparison_no_restarts}
\end{figure*}

\begin{figure*}[t]
    \centering
    \includegraphics[width=\textwidth]{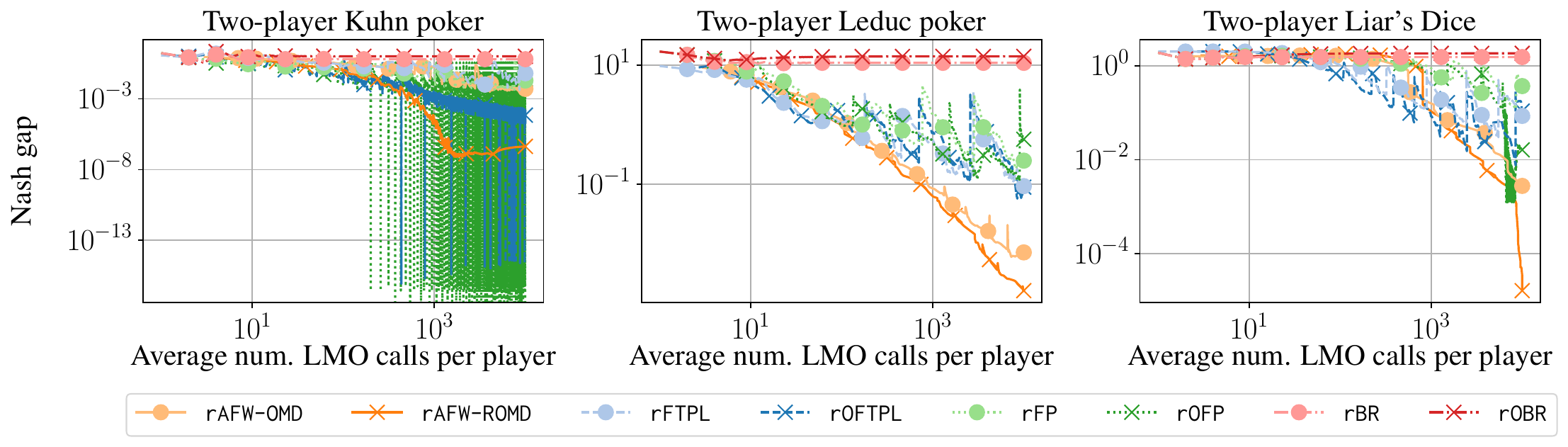}\\
    \includegraphics[width=\textwidth]{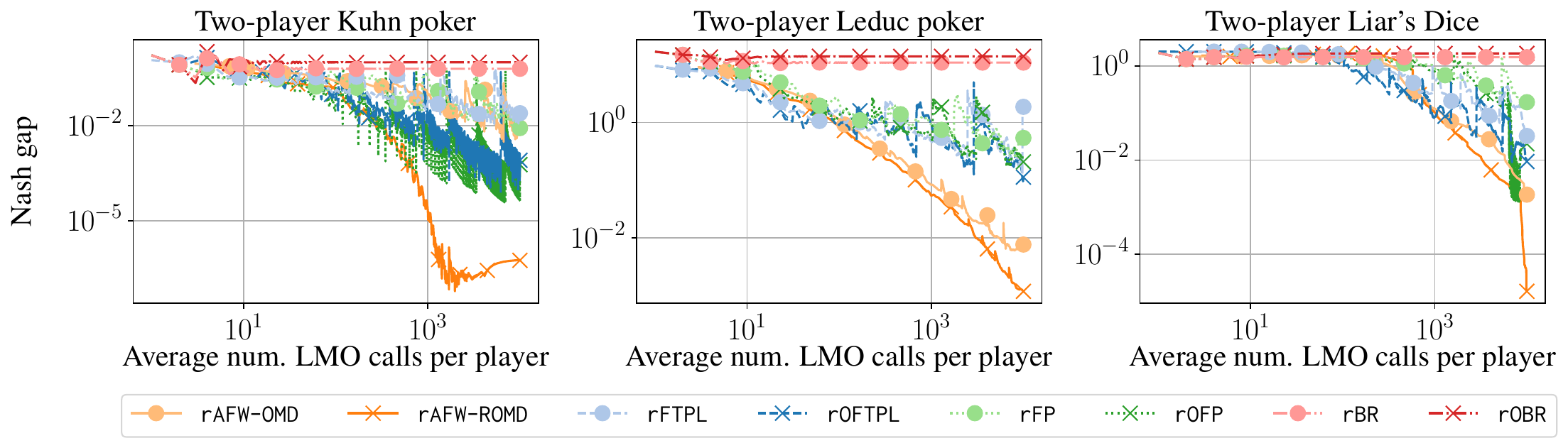}\\
    \includegraphics[width=\textwidth]{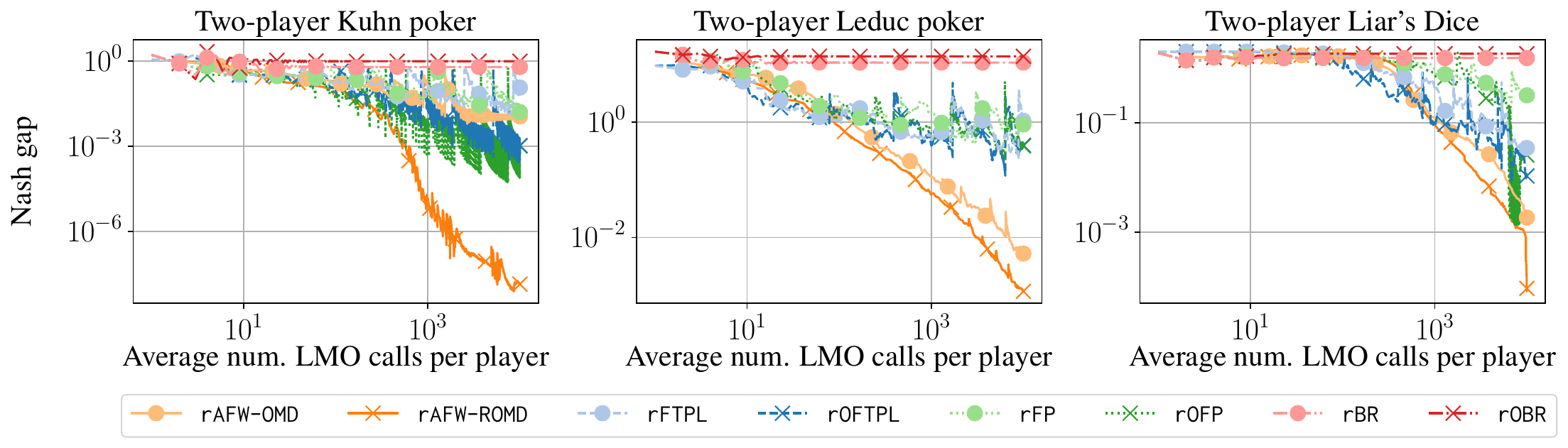}\\
    \caption{Convergence to NE as a function of average LMO calls per player for \fwomd{}, \fwromd{}, \ftpl{}, \oftpl{}, \fp{}, \ofp{}, \br{}, and \obr{}, for, from top to bottom, uniform, linear, and quadratic averaging, when using restarting.}
    \vspace{-5mm}
    \label{fig:NE_avg_comparison_restarts}
\end{figure*}
\section{Additional Experimental Details}
\label{sec:further_experimental_details}

\subsection{Comparisons of Algorithms across Averaging and Restarting Schemes}
In this section, we present plots of all the algorithms that we evaluate for different choices of averaging and restarting. This is only relevant for two-player games since, in the other settings, our performance measure is the maximum (uniform) average of an individual player's regret (so iterate averaging is not relevant).

We implement adaptive restarting by resetting the averaging process every time the duality gap halves. Adaptive restarting was recently shown effective in practice for EFGs~\citep{chakrabarti2023blockcoordinate}. Since adaptive restarting is applied as a heuristic in our experiments that has not previously applied to the algorithms we present as well as the ones we compare against, we label restarted variants of algorithms with a prepended ``r'' (e.g., the adaptive restarting heuristic applied to \oftpl{} is labeled as \roftpl{}) to distinguish them from the original presentation of the algorithm.

In \Cref{fig:NE_avg_comparison_no_restarts}, we compare the algorithm on different choices of averaging for all algorithms when restarting is not used. In \Cref{fig:NE_avg_comparison_restarts}, we compare the algorithm on different choices of averaging for all algorithms when restarting is used.

It can be seen that our algorithms are generally much more stable when averaging or restarting is applied relative to the other algorithms. This is to be expected because while our algorithm has last-iterate convergence guarantees, the other algorithms do not. Thus, even though \normalfont{(\texttt{O})\texttt{FTPL}} and \normalfont{(\texttt{O})\texttt{FP}} seem to perform well in the last-iterate case or with restarting on Kuhn and Liar's Dice, their behavior is extremely erratic. Observe that our algorithms generally outperform the other algorithms across different averaging and restarting schemes, on Leduc and Liar's Dice. As mentioned before, \normalfont{(\texttt{O})\texttt{FTPL}} and \normalfont{(\texttt{O})\texttt{FP}} appear to be very unstable, and in Kuhn, while they are often able to find low duality gap solutions, they oscillate quite a bit.

\begin{figure*}[t]
    \centering
    \includegraphics[width=\textwidth]{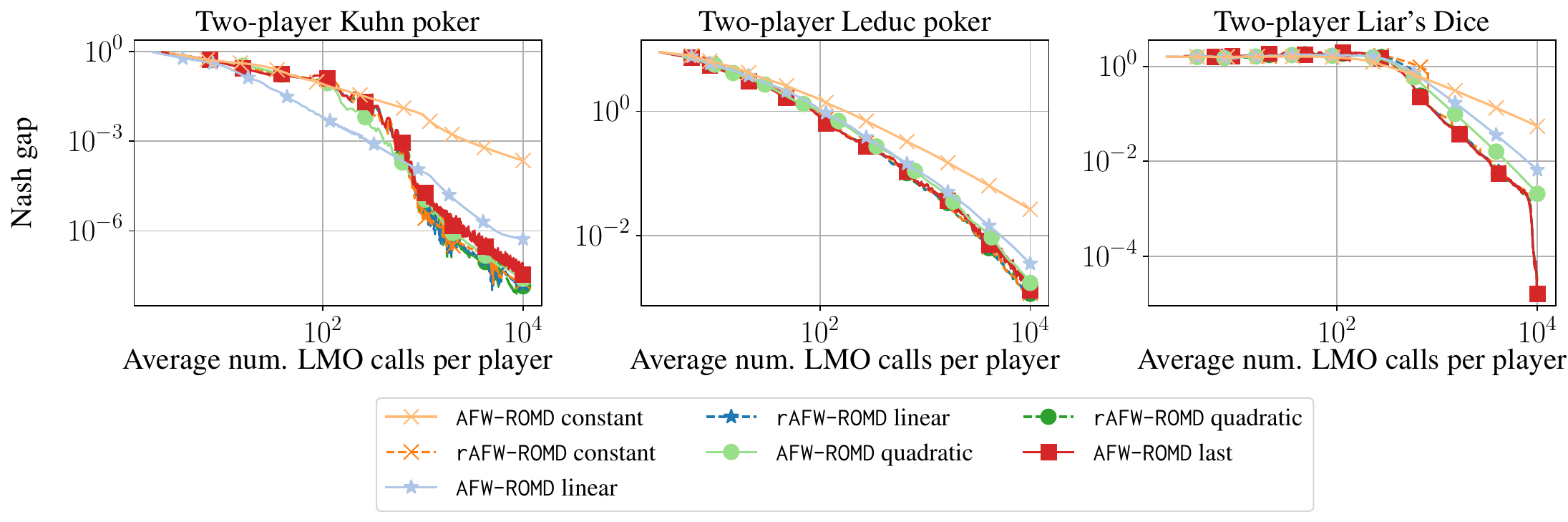}\\
    \caption{Convergence to NE as a function of average LMO calls per player for \fwomd{} for all combinations of averaging schemes and restarting.}
    \vspace{-5mm}
    \label{fig:NE_ablation_averaging_no_optimistic}
\end{figure*}

\begin{figure*}[t]
    \centering
    \includegraphics[width=\textwidth]{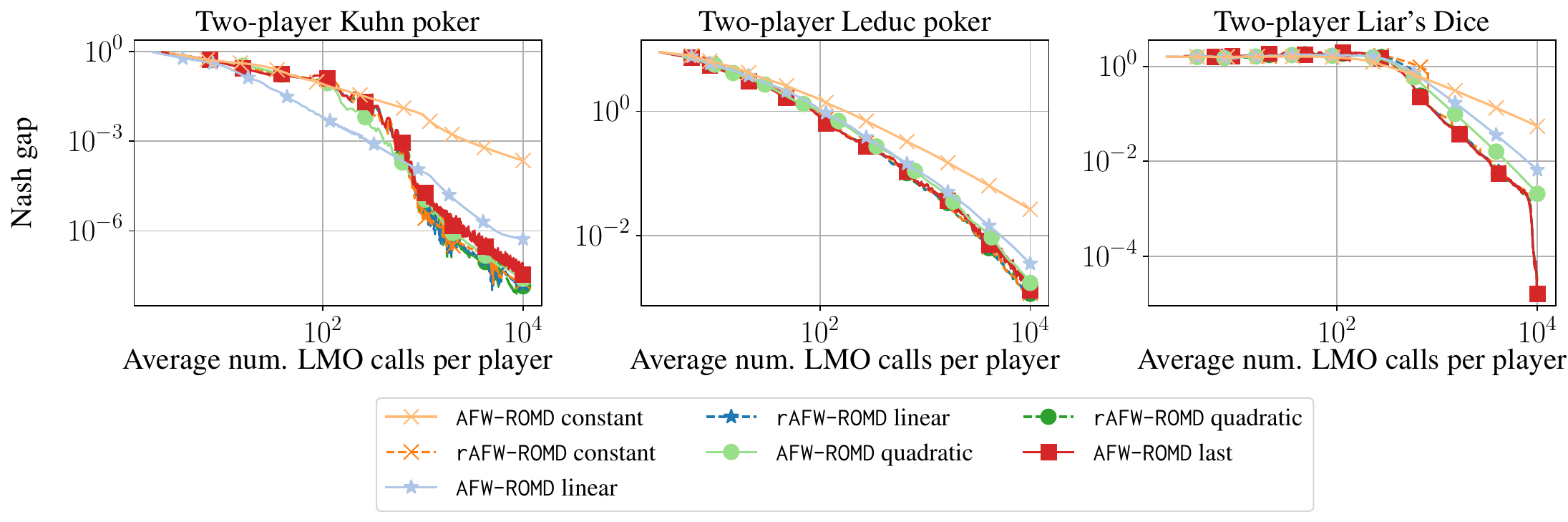}
    \caption{Convergence to NE as a function of average LMO calls per player for \fwromd{} for all combinations of averaging schemes and restarting.}
    \vspace{-5mm}
    \label{fig:NE_ablation_averaging_optimistic}
\end{figure*}
\subsection{Comparison of Averaging Schemes for \fwomd{} and \fwromd{}}
In \Cref{fig:NE_ablation_averaging_no_optimistic,fig:NE_ablation_averaging_optimistic}, we evaluate \fwomd{} and \fwromd{} on two-player games on all combinations of averaging and restarting. As above, we label restarted variants of algorithms with a prepended ``r'' to distinguish them from the original presentation of the algorithm.

We observe here that for \fwomd{}, the restarted schemes (whether they are using uniform averaging, linear averaging, or quadratic averaging) are all very similar and generally perform best.

For \fwromd{}, we observe that the restarted schemes and the last iterate perform quite similarly and generally better than the non-restarted schemes.

\begin{figure*}[t]
    \centering
    \includegraphics[width=\textwidth]{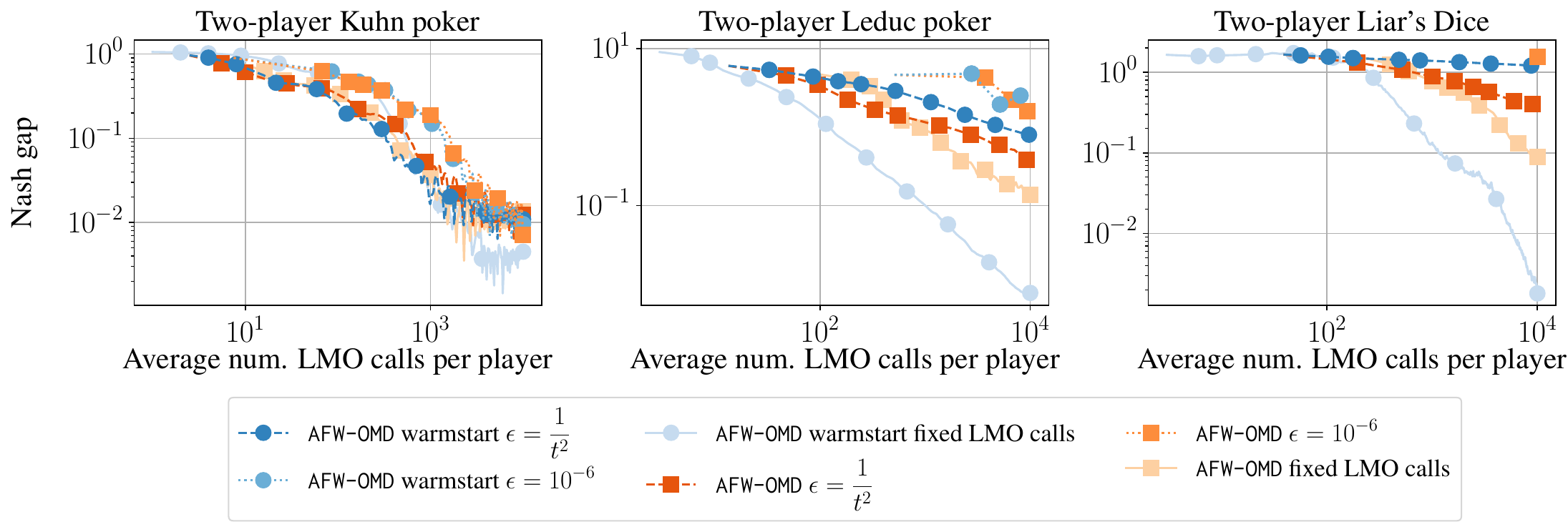}
    \caption{Convergence to NE as a function of average LMO calls per player for \fwomd{} for different choices of warmstarting and termination criteria.}
    \vspace{-5mm}
    \label{fig:NE_no_restarts_ablation_no_optimistic}
\end{figure*}

\begin{figure*}[t]
    \centering
    \includegraphics[width=\textwidth]{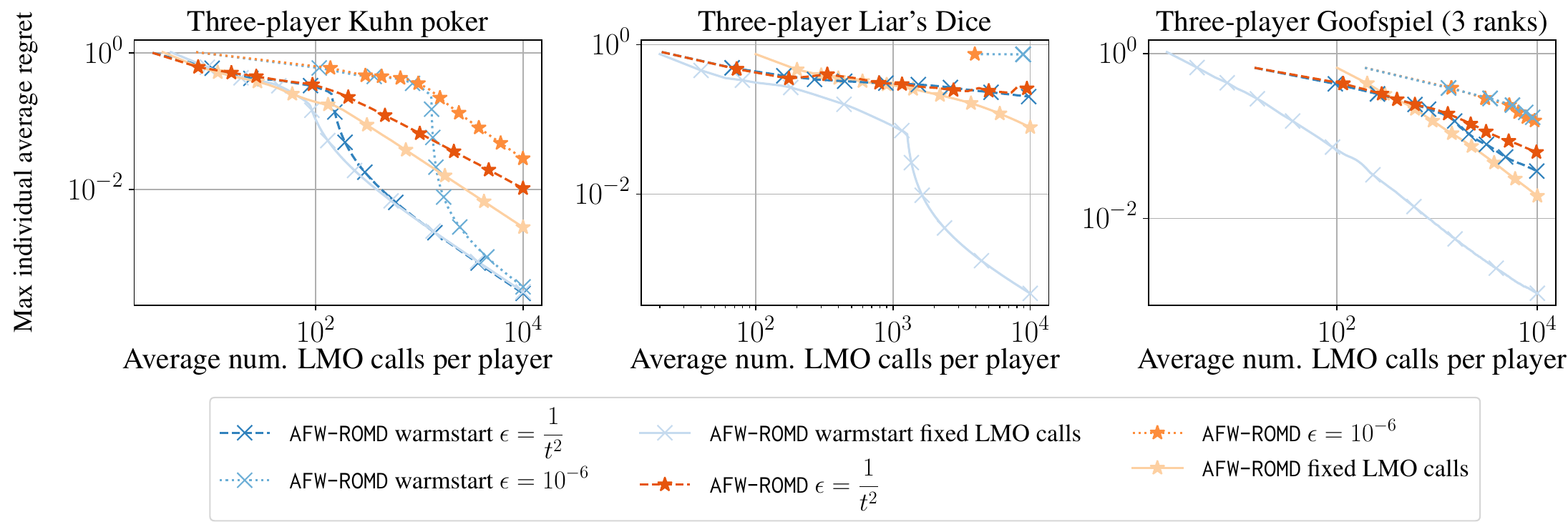}
    \caption{Convergence to CCE as a function of average LMO calls per player for \fwromd{} for different choices of warmstarting and termination criteria.}
    \vspace{-5mm}
    \label{fig:CCE_ablation_optimistic}
\end{figure*}

\subsection{Comparison of Warmstart and \apo{} Termination Criteria}
In \Cref{fig:NE_no_restarts_ablation_no_optimistic}, we evaluate \fwomd{} on two-player games, and in \Cref{fig:CCE_ablation_optimistic}, we evaluate \fwromd{} on multiplayer games, when using different combinations of warmstarting and termination criteria for \apo{} calls, since these combinations were not explored in \Cref{sec:experiments}. We choose the best averaging scheme for each warm starting and termination criteria combination.

It is quite clear from these figures that warmstarting and using a fixed number of LMO calls per iteration leads to the best performance of our algorithms.

\subsection{Parameter Choices}
In Tables \ref{table:NE_no_restarts} to \ref{table:CCE_ablation_optimistic}, for each figure presented on experimental results, we provide the specific parameter choices used for each algorithm.
The columns, in order, correspond to the game, the algorithm, the averaging scheme, whether restarting was used or not, the max number of LMO calls allowed during each iteration of the algorithm, the stepsize or noise (the latter in the case of \ftpl{} and \oftpl{}, and the former in all other cases), whether warmstarting was used (for our algorithms), and the APO termination criterion (for our algorithms).

\begin{longtable}{llllrrll}
    \toprule
    \bf Game & \bf Algorithm & \bf Averaging & \bf Restarts & $m$ & $\eta$ & \bf Warmstart & \bf APO term. crit. \\
    \midrule
    \multirow{8}{*}{\parbox[m][][c]{2.0cm}{Two-player                                                            \\Kuhn poker}} & \fwomd{} & quadratic & no & 1 & 0.08 & yes &  fixed LMO calls\\
             & \fwromd{}      & quadratic     & no           & 5   & 1.28   & yes           & fixed LMO calls     \\
             & \ftpl{}          & last          & no           & 3   & 20.48  & N/A           & N/A                 \\
             & \oftpl{}         & last          & no           & 3   & 20.48  & N/A           & N/A                 \\
             & \fp{}            & constant      & no           & 1   & 1.28   & N/A           & N/A                 \\
             & \ofp{}           & linear        & no           & 1   & 1.28   & N/A           & N/A                 \\
             & \br{}            & quadratic     & no           & 1   & 1.28   & N/A           & N/A                 \\
             & \obr{}           & quadratic     & no           & 1   & 1.28   & N/A           & N/A                 \\
    \midrule
    \multirow{8}{*}{\parbox[m][][c]{2.0cm}{Two-player                                                            \\Leduc poker}} & \fwomd{} & quadratic & no & 3 & 1.28 & yes &  fixed LMO calls\\
             & \fwromd{}      & last          & no           & 2   & 1.28   & yes           & fixed LMO calls     \\
             & \ftpl{}          & constant      & no           & 1   & 0.32   & N/A           & N/A                 \\
             & \oftpl{}         & constant      & no           & 1   & 0.01   & N/A           & N/A                 \\
             & \fp{}            & constant      & no           & 1   & 1.28   & N/A           & N/A                 \\
             & \ofp{}           & constant      & no           & 1   & 1.28   & N/A           & N/A                 \\
             & \br{}            & quadratic     & no           & 1   & 1.28   & N/A           & N/A                 \\
             & \obr{}           & linear        & no           & 1   & 1.28   & N/A           & N/A                 \\
    \midrule
    \multirow{8}{*}{\parbox[m][][c]{2.0cm}{Two-player                                                            \\Liar's Dice}} & \fwomd{} & last & no & 3 & 10.24 & yes &  fixed LMO calls\\
             & \fwromd{}      & last          & no           & 3   & 10.24  & yes           & fixed LMO calls     \\
             & \ftpl{}          & last          & no           & 1   & 0.32   & N/A           & N/A                 \\
             & \oftpl{}         & last          & no           & 1   & 0.08   & N/A           & N/A                 \\
             & \fp{}            & constant      & no           & 1   & 1.28   & N/A           & N/A                 \\
             & \ofp{}           & linear        & no           & 1   & 1.28   & N/A           & N/A                 \\
             & \br{}            & last          & no           & 1   & 1.28   & N/A           & N/A                 \\
             & \obr{}           & last          & no           & 1   & 1.28   & N/A           & N/A                 \\
    \bottomrule
    \caption{The parameters for all algorithms shown in \Cref{fig:NE_CCE_alg_comparison} for two-player games. The final two columns are irrelevant for algorithms that are not ours and thus are marked as N/A. Additionally, if the number of LMO calls is not fixed for our algorithms, then N/A is used in the maximum number of LMO calls per iteration ($m$) column.}
    \label[table]{table:NE_no_restarts}
\end{longtable}

\begin{longtable}{llllrrll}
    \toprule
    \bf Game & \bf Algorithm & \bf Averaging & \bf Restarts & $m$ & $\eta$ & \bf Warmstart & \bf APO term. crit. \\
    \midrule
    \multirow{8}{*}{\parbox[m][][c]{2.0cm}{Three-player                                                          \\Kuhn poker}} & \fwomd{} & constant & no & 4 & 40.96 & yes &  fixed LMO calls\\
             & \fwromd{}      & constant      & no           & 5   & 1.28   & yes           & fixed LMO calls     \\
             & \ftpl{}          & constant      & no           & 2   & 0.02   & N/A           & N/A                 \\
             & \oftpl{}         & constant      & no           & 2   & 0.01   & N/A           & N/A                 \\
             & \fp{}            & constant      & no           & 1   & 1.28   & N/A           & N/A                 \\
             & \ofp{}           & constant      & no           & 1   & 1.28   & N/A           & N/A                 \\
             & \br{}            & constant      & no           & 1   & 1.28   & N/A           & N/A                 \\
             & \obr{}           & constant      & no           & 1   & 1.28   & N/A           & N/A                 \\
    \midrule
    \multirow{8}{*}{\parbox[m][][c]{2.0cm}{Three-player                                                          \\Liar's Dice}} & \fwomd{} & constant & no & 20 & 40.96 & yes &  fixed LMO calls\\
             & \fwromd{}      & constant      & no           & 4   & 20.48  & yes           & fixed LMO calls     \\
             & \ftpl{}          & constant      & no           & 1   & 0.01   & N/A           & N/A                 \\
             & \oftpl{}         & constant      & no           & 1   & 0.01   & N/A           & N/A                 \\
             & \fp{}            & constant      & no           & 1   & 1.28   & N/A           & N/A                 \\
             & \ofp{}           & constant      & no           & 1   & 1.28   & N/A           & N/A                 \\
             & \br{}            & constant      & no           & 1   & 1.28   & N/A           & N/A                 \\
             & \obr{}           & constant      & no           & 1   & 1.28   & N/A           & N/A                 \\
    \midrule
    \multirow{8}{*}{\parbox[m][][c]{2.0cm}{Three-player                                                          \\Goofspiel (3 ranks)}} & \fwomd{} & constant & no & 1 & 81.92 & yes &  fixed LMO calls\\
             & \fwromd{}      & constant      & no           & 2   & 40.96  & yes           & fixed LMO calls     \\
             & \ftpl{}          & constant      & no           & 1   & 0.04   & N/A           & N/A                 \\
             & \oftpl{}         & constant      & no           & 2   & 0.04   & N/A           & N/A                 \\
             & \fp{}            & constant      & no           & 1   & 1.28   & N/A           & N/A                 \\
             & \ofp{}           & constant      & no           & 1   & 1.28   & N/A           & N/A                 \\
             & \br{}            & constant      & no           & 1   & 1.28   & N/A           & N/A                 \\
             & \obr{}           & constant      & no           & 1   & 1.28   & N/A           & N/A                 \\
    \bottomrule
    \caption{The parameters for all algorithms shown in \Cref{fig:NE_CCE_alg_comparison} for multiplayer games. The final two columns are irrelevant for algorithms that are not ours and thus are marked as N/A. Additionally, if the number of LMO calls is not fixed for our algorithms, then N/A is used in the maximum number of LMO calls per iteration ($m$) column.}
    \label[table]{table:CCE}
\end{longtable}

\begin{longtable}{llllrrll}
    \toprule
    \bf Game                                & \bf Algorithm & \bf Averaging & \bf Restarts & $m$ & $\eta$ & \bf Warmstart & \bf APO term. crit.        \\
    \midrule
    \multirow{6}{*}{\parbox[m][][c]{2.0cm}{Two-player \\Kuhn poker}}  & \fwromd{}      & quadratic     & no           & N/A & 1.28   & yes           & $\epsilon = \frac{1}{t^2}$ \\
                                            & \fwromd{}      & quadratic     & no           & N/A & 1.28   & yes           & $\epsilon = 10^{-6}$       \\
                                            & \fwromd{}      & quadratic     & no           & 5   & 1.28   & yes           & fixed LMO calls            \\
                                            & \fwromd{}      & last          & no           & N/A & 1.28   & no            & $\epsilon = \frac{1}{t^2}$ \\
                                            & \fwromd{}      & last          & no           & N/A & 1.28   & no            & $\epsilon = 10^{-6}$       \\
                                            & \fwromd{}      & last          & no           & 20  & 1.28   & no            & fixed LMO calls            \\
    \midrule
    \multirow{6}{*}{\parbox[m][][c]{2.0cm}{Two-player \\Leduc poker}} & \fwromd{}      & last          & no           & N/A & 0.64   & yes           & $\epsilon = \frac{1}{t^2}$ \\
                                            & \fwromd{}      & last          & no           & N/A & 2.56   & yes           & $\epsilon = 10^{-6}$       \\
                                            & \fwromd{}      & last          & no           & 3   & 1.28   & yes           & fixed LMO calls            \\
                                            & \fwromd{}      & last          & no           & N/A & 0.64   & no            & $\epsilon = \frac{1}{t^2}$ \\
                                            & \fwromd{}      & last          & no           & N/A & 2.56   & no            & $\epsilon = 10^{-6}$       \\
                                            & \fwromd{}      & quadratic     & no           & 100 & 0.64   & no            & fixed LMO calls            \\
    \midrule
    \multirow{6}{*}{\parbox[m][][c]{2.0cm}{Two-player \\Liar's Dice}} & \fwromd{}      & last          & no           & N/A & 5.12   & yes           & $\epsilon = \frac{1}{t^2}$ \\
                                            & \fwromd{}      & quadratic     & no           & N/A & 5.12   & yes           & $\epsilon = 10^{-6}$       \\
                                            & \fwromd{}      & last          & no           & 3   & 10.24  & yes           & fixed LMO calls            \\
                                            & \fwromd{}      & last          & no           & N/A & 5.12   & no            & $\epsilon = \frac{1}{t^2}$ \\
                                            & \fwromd{}      & quadratic     & no           & N/A & 0.04   & no            & $\epsilon = 10^{-6}$       \\
                                            & \fwromd{}      & last          & no           & 200 & 5.12   & no            & fixed LMO calls            \\
    \bottomrule
    \caption{The parameters for all algorithms shown in \Cref{fig:NE_CCE_ablation} for two-player games. The final two columns are irrelevant for algorithms that are not ours and thus are marked as N/A. Additionally, if the number of LMO calls is not fixed for our algorithms, then N/A is used in the maximum number of LMO calls per iteration ($m$) column.}
    \label[table]{table:NE_no_restarts_ablation}
\end{longtable}

\begin{longtable}{llllrrll}
    \toprule
    \bf Game                                  & \bf Algorithm & \bf Averaging & \bf Restarts & $m$ & $\eta$ & \bf Warmstart & \bf APO term. crit.        \\
    \midrule
    \multirow{6}{*}{\parbox[m][][c]{2.0cm}{Three-player                                                                                                  \\Kuhn poker}} & \fwomd{} & constant & no & N/A & 40.96 & yes &  $\epsilon = \frac{1}{t^2}$\\
                                              & \fwomd{}       & constant      & no           & N/A & 40.96  & yes           & $\epsilon = 10^{-6}$       \\
                                              & \fwomd{}       & constant      & no           & 4   & 40.96  & yes           & fixed LMO calls            \\
                                              & \fwomd{}       & constant      & no           & N/A & 81.92  & no            & $\epsilon = \frac{1}{t^2}$ \\
                                              & \fwomd{}       & constant      & no           & N/A & 40.96  & no            & $\epsilon = 10^{-6}$       \\
                                              & \fwomd{}       & constant      & no           & 1   & 40.96  & no            & fixed LMO calls            \\
    \midrule
    \multirow{6}{*}{\parbox[m][][c]{2.0cm}{Three-player \\Liar's Dice}} & \fwomd{}       & constant      & no           & N/A & 40.96  & yes           & $\epsilon = \frac{1}{t^2}$ \\
                                              & \fwomd{}       & constant      & no           & N/A  & 0.04  & yes           & $\epsilon = 10^{-6}$            \\
                                              & \fwomd{}       & constant      & no           & 20  & 40.96  & yes           & fixed LMO calls            \\
                                              & \fwomd{}       & constant      & no           & N/A & 40.96  & no            & $\epsilon = \frac{1}{t^2}$ \\
                                              & \fwomd{}       & constant      & no           & N/A  & 0.02  & no           & $\epsilon = 10^{-6}$            \\
                                              & \fwomd{}       & constant      & no           & 100 & 40.96  & no            & fixed LMO calls            \\
    \midrule
    \multirow{6}{*}{\parbox[m][][c]{2.0cm}{Three-player                                                                                                  \\Goofspiel (3 ranks)}} & \fwomd{} & constant & no & N/A & 40.96 & yes &  $\epsilon = \frac{1}{t^2}$\\
                                              & \fwomd{}       & constant      & no           & N/A & 81.92  & yes           & $\epsilon = 10^{-6}$       \\
                                              & \fwomd{}       & constant      & no           & 1   & 81.92  & yes           & fixed LMO calls            \\
                                              & \fwomd{}       & constant      & no           & N/A & 81.92  & no            & $\epsilon = \frac{1}{t^2}$ \\
                                              & \fwomd{}       & constant      & no           & N/A & 81.92  & no            & $\epsilon = 10^{-6}$       \\
                                              & \fwomd{}       & constant      & no           & 20  & 40.96  & no            & fixed LMO calls            \\
    \bottomrule
    \caption{The parameters for all algorithms shown in \Cref{fig:NE_CCE_ablation} for multiplayer games. The final two columns are irrelevant for algorithms that are not ours and thus are marked as N/A. Additionally, if the number of LMO calls is not fixed for our algorithms, then N/A is used in the maximum number of LMO calls per iteration ($m$) column.}
    \label[table]{table:CCE_ablation}
\end{longtable}

\begin{longtable}{llllrrll}
    \toprule
    \bf Game & \bf Algorithm & \bf Averaging & \bf Restarts & $m$ & $\eta$ & \bf Warmstart & \bf APO term. crit. \\
    \midrule
    \multirow{8}{*}{\parbox[m][][c]{2.0cm}{Two-player                                                            \\Kuhn poker}} & \fwomd{} & constant & no & 3 & 0.32 & yes &  fixed LMO calls\\
             & \fwromd{}      & constant      & no           & 4   & 1.28   & yes           & fixed LMO calls     \\
             & \ftpl{}          & constant      & no           & 1   & 0.04   & N/A           & N/A                 \\
             & \oftpl{}         & constant      & no           & 1   & 0.08   & N/A           & N/A                 \\
             & \fp{}            & constant      & no           & 1   & 1.28   & N/A           & N/A                 \\
             & \ofp{}           & constant      & no           & 1   & 1.28   & N/A           & N/A                 \\
             & \br{}            & constant      & no           & 1   & 1.28   & N/A           & N/A                 \\
             & \obr{}           & constant      & no           & 1   & 1.28   & N/A           & N/A                 \\
    \midrule
    \multirow{8}{*}{\parbox[m][][c]{2.0cm}{Two-player                                                            \\Leduc poker}} & \fwomd{} & constant & no & 3 & 2.56 & yes &  fixed LMO calls\\
             & \fwromd{}      & constant      & no           & 3   & 1.28   & yes           & fixed LMO calls     \\
             & \ftpl{}          & constant      & no           & 1   & 0.32   & N/A           & N/A                 \\
             & \oftpl{}         & constant      & no           & 1   & 0.01   & N/A           & N/A                 \\
             & \fp{}            & constant      & no           & 1   & 1.28   & N/A           & N/A                 \\
             & \ofp{}           & constant      & no           & 1   & 1.28   & N/A           & N/A                 \\
             & \br{}            & constant      & no           & 1   & 1.28   & N/A           & N/A                 \\
             & \obr{}           & constant      & no           & 1   & 1.28   & N/A           & N/A                 \\
    \midrule
    \multirow{8}{*}{\parbox[m][][c]{2.0cm}{Two-player                                                            \\Liar's Dice}} & \fwomd{} & constant & no & 3 & 10.24 & yes &  fixed LMO calls\\
             & \fwromd{}      & constant      & no           & 2   & 5.12   & yes           & fixed LMO calls     \\
             & \ftpl{}          & constant      & no           & 1   & 0.02   & N/A           & N/A                 \\
             & \oftpl{}         & constant      & no           & 1   & 0.01   & N/A           & N/A                 \\
             & \fp{}            & constant      & no           & 1   & 1.28   & N/A           & N/A                 \\
             & \ofp{}           & constant      & no           & 1   & 1.28   & N/A           & N/A                 \\
             & \br{}            & constant      & no           & 1   & 1.28   & N/A           & N/A                 \\
             & \obr{}           & constant      & no           & 1   & 1.28   & N/A           & N/A                 \\
    \midrule
    \multirow{8}{*}{\parbox[m][][c]{2.0cm}{Two-player                                                            \\Kuhn poker}} & \fwomd{} & linear & no & 1 & 0.32 & yes &  fixed LMO calls\\
             & \fwromd{}      & linear        & no           & 2   & 2.56   & yes           & fixed LMO calls     \\
             & \ftpl{}          & linear        & no           & 1   & 0.64   & N/A           & N/A                 \\
             & \oftpl{}         & linear        & no           & 1   & 0.04   & N/A           & N/A                 \\
             & \fp{}            & linear        & no           & 1   & 1.28   & N/A           & N/A                 \\
             & \ofp{}           & linear        & no           & 1   & 1.28   & N/A           & N/A                 \\
             & \br{}            & linear        & no           & 1   & 1.28   & N/A           & N/A                 \\
             & \obr{}           & linear        & no           & 1   & 1.28   & N/A           & N/A                 \\
    \midrule
    \multirow{8}{*}{\parbox[m][][c]{2.0cm}{Two-player                                                            \\Leduc poker}} & \fwomd{} & linear & no & 3 & 1.28 & yes &  fixed LMO calls\\
             & \fwromd{}      & linear        & no           & 5   & 1.28   & yes           & fixed LMO calls     \\
             & \ftpl{}          & linear        & no           & 1   & 0.32   & N/A           & N/A                 \\
             & \oftpl{}         & linear        & no           & 1   & 0.08   & N/A           & N/A                 \\
             & \fp{}            & linear        & no           & 1   & 1.28   & N/A           & N/A                 \\
             & \ofp{}           & linear        & no           & 1   & 1.28   & N/A           & N/A                 \\
             & \br{}            & linear        & no           & 1   & 1.28   & N/A           & N/A                 \\
             & \obr{}           & linear        & no           & 1   & 1.28   & N/A           & N/A                 \\
    \midrule
    \multirow{8}{*}{\parbox[m][][c]{2.0cm}{Two-player                                                            \\Liar's Dice}} & \fwomd{} & linear & no & 3 & 10.24 & yes &  fixed LMO calls\\
             & \fwromd{}      & linear        & no           & 2   & 10.24  & yes           & fixed LMO calls     \\
             & \ftpl{}          & linear        & no           & 1   & 0.02   & N/A           & N/A                 \\
             & \oftpl{}         & linear        & no           & 2   & 0.08   & N/A           & N/A                 \\
             & \fp{}            & linear        & no           & 1   & 1.28   & N/A           & N/A                 \\
             & \ofp{}           & linear        & no           & 1   & 1.28   & N/A           & N/A                 \\
             & \br{}            & linear        & no           & 1   & 1.28   & N/A           & N/A                 \\
             & \obr{}           & linear        & no           & 1   & 1.28   & N/A           & N/A                 \\
    \midrule
    \multirow{8}{*}{\parbox[m][][c]{2.0cm}{Two-player                                                            \\Kuhn poker}} & \fwomd{} & quadratic & no & 1 & 0.08 & yes &  fixed LMO calls\\
             & \fwromd{}      & quadratic     & no           & 5   & 1.28   & yes           & fixed LMO calls     \\
             & \ftpl{}          & quadratic     & no           & 1   & 0.32   & N/A           & N/A                 \\
             & \oftpl{}         & quadratic     & no           & 1   & 0.04   & N/A           & N/A                 \\
             & \fp{}            & quadratic     & no           & 1   & 1.28   & N/A           & N/A                 \\
             & \ofp{}           & quadratic     & no           & 1   & 1.28   & N/A           & N/A                 \\
             & \br{}            & quadratic     & no           & 1   & 1.28   & N/A           & N/A                 \\
             & \obr{}           & quadratic     & no           & 1   & 1.28   & N/A           & N/A                 \\
    \midrule
    \multirow{8}{*}{\parbox[m][][c]{2.0cm}{Two-player                                                            \\Leduc poker}} & \fwomd{} & quadratic & no & 3 & 1.28 & yes &  fixed LMO calls\\
             & \fwromd{}      & quadratic     & no           & 3   & 1.28   & yes           & fixed LMO calls     \\
             & \ftpl{}          & quadratic     & no           & 1   & 0.32   & N/A           & N/A                 \\
             & \oftpl{}         & quadratic     & no           & 1   & 0.16   & N/A           & N/A                 \\
             & \fp{}            & quadratic     & no           & 1   & 1.28   & N/A           & N/A                 \\
             & \ofp{}           & quadratic     & no           & 1   & 1.28   & N/A           & N/A                 \\
             & \br{}            & quadratic     & no           & 1   & 1.28   & N/A           & N/A                 \\
             & \obr{}           & quadratic     & no           & 1   & 1.28   & N/A           & N/A                 \\
    \midrule
    \multirow{8}{*}{\parbox[m][][c]{2.0cm}{Two-player                                                            \\Liar's Dice}} & \fwomd{} & quadratic & no & 2 & 10.24 & yes &  fixed LMO calls\\
             & \fwromd{}      & quadratic     & no           & 2   & 10.24  & yes           & fixed LMO calls     \\
             & \ftpl{}          & quadratic     & no           & 1   & 0.32   & N/A           & N/A                 \\
             & \oftpl{}         & quadratic     & no           & 1   & 0.08   & N/A           & N/A                 \\
             & \fp{}            & quadratic     & no           & 1   & 1.28   & N/A           & N/A                 \\
             & \ofp{}           & quadratic     & no           & 1   & 1.28   & N/A           & N/A                 \\
             & \br{}            & quadratic     & no           & 1   & 1.28   & N/A           & N/A                 \\
             & \obr{}           & quadratic     & no           & 1   & 1.28   & N/A           & N/A                 \\
    \midrule
    \multirow{8}{*}{\parbox[m][][c]{2.0cm}{Two-player                                                            \\Kuhn poker}} & \fwomd{} & last & no & 1 & 0.32 & yes &  fixed LMO calls\\
             & \fwromd{}      & last          & no           & 2   & 2.56   & yes           & fixed LMO calls     \\
             & \ftpl{}          & last          & no           & 3   & 20.48  & N/A           & N/A                 \\
             & \oftpl{}         & last          & no           & 20  & 0.32   & N/A           & N/A                 \\
             & \fp{}            & last          & no           & 1   & 1.28   & N/A           & N/A                 \\
             & \ofp{}           & last          & no           & 1   & 1.28   & N/A           & N/A                 \\
             & \br{}            & last          & no           & 1   & 1.28   & N/A           & N/A                 \\
             & \obr{}           & last          & no           & 1   & 1.28   & N/A           & N/A                 \\
    \midrule
    \multirow{8}{*}{\parbox[m][][c]{2.0cm}{Two-player                                                            \\Leduc poker}} & \fwomd{} & last & no & 1 & 1.28 & yes &  fixed LMO calls\\
             & \fwromd{}      & last          & no           & 2   & 1.28   & yes           & fixed LMO calls     \\
             & \ftpl{}          & last          & no           & 10  & 2.56   & N/A           & N/A                 \\
             & \oftpl{}         & last          & no           & 200 & 0.16   & N/A           & N/A                 \\
             & \fp{}            & last          & no           & 1   & 1.28   & N/A           & N/A                 \\
             & \ofp{}           & last          & no           & 1   & 1.28   & N/A           & N/A                 \\
             & \br{}            & last          & no           & 1   & 1.28   & N/A           & N/A                 \\
             & \obr{}           & last          & no           & 1   & 1.28   & N/A           & N/A                 \\
    \midrule
    \multirow{8}{*}{\parbox[m][][c]{2.0cm}{Two-player                                                            \\Liar's Dice}} & \fwomd{} & last & no & 4 & 5.12 & yes &  fixed LMO calls\\
             & \fwromd{}      & last          & no           & 3   & 10.24  & yes           & fixed LMO calls     \\
             & \ftpl{}          & last          & no           & 1   & 0.32   & N/A           & N/A                 \\
             & \oftpl{}         & last          & no           & 2   & 0.02   & N/A           & N/A                 \\
             & \fp{}            & last          & no           & 1   & 1.28   & N/A           & N/A                 \\
             & \ofp{}           & last          & no           & 1   & 1.28   & N/A           & N/A                 \\
             & \br{}            & last          & no           & 1   & 1.28   & N/A           & N/A                 \\
             & \obr{}           & last          & no           & 1   & 1.28   & N/A           & N/A                 \\
    \bottomrule
    \caption{The parameters for all algorithms shown in \Cref{fig:NE_avg_comparison_no_restarts}. The final two columns are irrelevant for algorithms that are not ours and thus are marked as N/A. Additionally, if the number of LMO calls is not fixed for our algorithms, then N/A is used in the maximum number of LMO calls per iteration ($m$) column. }
    \label[table]{table:NE_avg_comparison_no_restarts}
\end{longtable}

\begin{longtable}{llllrrll}
    \toprule
    \bf Game & \bf Algorithm & \bf Averaging & \bf Restarts & $m$ & $\eta$ & \bf Warmstart & \bf APO term. crit. \\
    \midrule
    \multirow{8}{*}{\parbox[m][][c]{2.0cm}{Two-player                                                            \\Kuhn poker}} & \rfwomd{} & constant & yes & 4 & 0.32 & yes &  fixed LMO calls\\
             & \rfwromd{}     & constant      & yes          & 5   & 1.28   & yes           & fixed LMO calls     \\
             & \rftpl{}         & constant      & yes          & 1   & 5.12   & N/A           & N/A                 \\
             & \roftpl{}        & constant      & yes          & 1   & 0.01   & N/A           & N/A                 \\
             & \rfp{}           & constant      & yes          & 1   & 1.28   & N/A           & N/A                 \\
             & \rofp{}          & constant      & yes          & 1   & 1.28   & N/A           & N/A                 \\
             & \rbr{}           & constant      & yes          & 1   & 1.28   & N/A           & N/A                 \\
             & \robr{}          & constant      & yes          & 1   & 1.28   & N/A           & N/A                 \\
    \midrule
    \multirow{8}{*}{\parbox[m][][c]{2.0cm}{Two-player                                                            \\Leduc poker}} & \rfwomd{} & constant & yes & 3 & 1.28 & yes &  fixed LMO calls\\
             & \rfwromd{}     & constant      & yes          & 4   & 1.28   & yes           & fixed LMO calls     \\
             & \rftpl{}         & constant      & yes          & 1   & 0.64   & N/A           & N/A                 \\
             & \roftpl{}        & constant      & yes          & 2   & 0.16   & N/A           & N/A                 \\
             & \rfp{}           & constant      & yes          & 1   & 1.28   & N/A           & N/A                 \\
             & \rofp{}          & constant      & yes          & 1   & 1.28   & N/A           & N/A                 \\
             & \rbr{}           & constant      & yes          & 1   & 1.28   & N/A           & N/A                 \\
             & \robr{}          & constant      & yes          & 1   & 1.28   & N/A           & N/A                 \\
    \midrule
    \multirow{8}{*}{\parbox[m][][c]{2.0cm}{Two-player                                                            \\Liar's Dice}} & \rfwomd{} & constant & yes & 2 & 10.24 & yes &  fixed LMO calls\\
             & \rfwromd{}     & constant      & yes          & 3   & 10.24  & yes           & fixed LMO calls     \\
             & \rftpl{}         & constant      & yes          & 1   & 0.08   & N/A           & N/A                 \\
             & \roftpl{}        & constant      & yes          & 2   & 0.02   & N/A           & N/A                 \\
             & \rfp{}           & constant      & yes          & 1   & 1.28   & N/A           & N/A                 \\
             & \rofp{}          & constant      & yes          & 1   & 1.28   & N/A           & N/A                 \\
             & \rbr{}           & constant      & yes          & 1   & 1.28   & N/A           & N/A                 \\
             & \robr{}          & constant      & yes          & 1   & 1.28   & N/A           & N/A                 \\
    \midrule
    \multirow{8}{*}{\parbox[m][][c]{2.0cm}{Two-player                                                            \\Kuhn poker}} & \rfwomd{} & linear & yes & 4 & 0.32 & yes &  fixed LMO calls\\
             & \rfwromd{}     & linear        & yes          & 2   & 2.56   & yes           & fixed LMO calls     \\
             & \rftpl{}         & linear        & yes          & 1   & 0.08   & N/A           & N/A                 \\
             & \roftpl{}        & linear        & yes          & 2   & 0.04   & N/A           & N/A                 \\
             & \rfp{}           & linear        & yes          & 1   & 1.28   & N/A           & N/A                 \\
             & \rofp{}          & linear        & yes          & 1   & 1.28   & N/A           & N/A                 \\
             & \rbr{}           & linear        & yes          & 1   & 1.28   & N/A           & N/A                 \\
             & \robr{}          & linear        & yes          & 1   & 1.28   & N/A           & N/A                 \\
    \midrule
    \multirow{8}{*}{\parbox[m][][c]{2.0cm}{Two-player                                                            \\Leduc poker}} & \rfwomd{} & linear & yes & 3 & 1.28 & yes &  fixed LMO calls\\
             & \rfwromd{}     & linear        & yes          & 3   & 1.28   & yes           & fixed LMO calls     \\
             & \rftpl{}         & linear        & yes          & 1   & 0.32   & N/A           & N/A                 \\
             & \roftpl{}        & linear        & yes          & 1   & 0.01   & N/A           & N/A                 \\
             & \rfp{}           & linear        & yes          & 1   & 1.28   & N/A           & N/A                 \\
             & \rofp{}          & linear        & yes          & 1   & 1.28   & N/A           & N/A                 \\
             & \rbr{}           & linear        & yes          & 1   & 1.28   & N/A           & N/A                 \\
             & \robr{}          & linear        & yes          & 1   & 1.28   & N/A           & N/A                 \\
    \midrule
    \multirow{8}{*}{\parbox[m][][c]{2.0cm}{Two-player                                                            \\Liar's Dice}} & \rfwomd{} & linear & yes & 2 & 10.24 & yes &  fixed LMO calls\\
             & \rfwromd{}     & linear        & yes          & 3   & 10.24  & yes           & fixed LMO calls     \\
             & \rftpl{}         & linear        & yes          & 2   & 0.16   & N/A           & N/A                 \\
             & \roftpl{}        & linear        & yes          & 1   & 0.16   & N/A           & N/A                 \\
             & \rfp{}           & linear        & yes          & 1   & 1.28   & N/A           & N/A                 \\
             & \rofp{}          & linear        & yes          & 1   & 1.28   & N/A           & N/A                 \\
             & \rbr{}           & linear        & yes          & 1   & 1.28   & N/A           & N/A                 \\
             & \robr{}          & linear        & yes          & 1   & 1.28   & N/A           & N/A                 \\
    \midrule
    \multirow{8}{*}{\parbox[m][][c]{2.0cm}{Two-player                                                            \\Kuhn poker}} & \rfwomd{} & quadratic & yes & 3 & 0.64 & yes &  fixed LMO calls\\
             & \rfwromd{}     & quadratic     & yes          & 5   & 1.28   & yes           & fixed LMO calls     \\
             & \rftpl{}         & quadratic     & yes          & 1   & 0.08   & N/A           & N/A                 \\
             & \roftpl{}        & quadratic     & yes          & 3   & 0.04   & N/A           & N/A                 \\
             & \rfp{}           & quadratic     & yes          & 1   & 1.28   & N/A           & N/A                 \\
             & \rofp{}          & quadratic     & yes          & 1   & 1.28   & N/A           & N/A                 \\
             & \rbr{}           & quadratic     & yes          & 1   & 1.28   & N/A           & N/A                 \\
             & \robr{}          & quadratic     & yes          & 1   & 1.28   & N/A           & N/A                 \\
    \midrule
    \multirow{8}{*}{\parbox[m][][c]{2.0cm}{Two-player                                                            \\Leduc poker}} & \rfwomd{} & quadratic & yes & 2 & 1.28 & yes &  fixed LMO calls\\
             & \rfwromd{}     & quadratic     & yes          & 3   & 1.28   & yes           & fixed LMO calls     \\
             & \rftpl{}         & quadratic     & yes          & 1   & 0.32   & N/A           & N/A                 \\
             & \roftpl{}        & quadratic     & yes          & 1   & 0.32   & N/A           & N/A                 \\
             & \rfp{}           & quadratic     & yes          & 1   & 1.28   & N/A           & N/A                 \\
             & \rofp{}          & quadratic     & yes          & 1   & 1.28   & N/A           & N/A                 \\
             & \rbr{}           & quadratic     & yes          & 1   & 1.28   & N/A           & N/A                 \\
             & \robr{}          & quadratic     & yes          & 1   & 1.28   & N/A           & N/A                 \\
    \midrule
    \multirow{8}{*}{\parbox[m][][c]{2.0cm}{Two-player                                                            \\Liar's Dice}} & \rfwomd{} & quadratic & yes & 2 & 10.24 & yes &  fixed LMO calls\\
             & \rfwromd{}     & quadratic     & yes          & 2   & 10.24  & yes           & fixed LMO calls     \\
             & \rftpl{}         & quadratic     & yes          & 1   & 0.32   & N/A           & N/A                 \\
             & \roftpl{}        & quadratic     & yes          & 1   & 0.16   & N/A           & N/A                 \\
             & \rfp{}           & quadratic     & yes          & 1   & 1.28   & N/A           & N/A                 \\
             & \rofp{}          & quadratic     & yes          & 1   & 1.28   & N/A           & N/A                 \\
             & \rbr{}           & quadratic     & yes          & 1   & 1.28   & N/A           & N/A                 \\
             & \robr{}          & quadratic     & yes          & 1   & 1.28   & N/A           & N/A                 \\
    \bottomrule
    \caption{The parameters for all algorithms shown in \Cref{fig:NE_avg_comparison_restarts}. The final two columns are irrelevant for algorithms that are not ours and thus are marked as N/A. Additionally, if the number of LMO calls is not fixed for our algorithms, then N/A is used in the maximum number of LMO calls per iteration ($m$) column. }
    \label[table]{table:NE_avg_comparison_restarts}
\end{longtable}

\begin{longtable}{llllrrll}
\toprule
\bf Game & \bf Algorithm & \bf Averaging & \bf Restarts & $m$ & $\eta$ & \bf Warmstart & \bf APO term. crit.\\
 \midrule
\multirow{7}{*}{\parbox[m][][c]{2.0cm}{Two-player\\Kuhn poker}} & \fwomd{} & constant & no & 1 & 0.32 & yes &  fixed LMO calls\\
 & \rfwomd{} & constant & yes & 1 & 0.32 & yes &  fixed LMO calls\\
 & \fwomd{} & linear & no & 1 & 0.08 & yes &  fixed LMO calls\\
 & \rfwomd{} & linear & yes & 1 & 0.32 & yes &  fixed LMO calls\\
 & \fwomd{} & quadratic & no & 5 & 0.32 & yes &  fixed LMO calls\\
 & \rfwomd{} & quadratic & yes & 1 & 0.32 & yes &  fixed LMO calls\\
 & \fwomd{} & last & no & 1 & 0.32 & yes &  fixed LMO calls\\
 \midrule
\multirow{7}{*}{\parbox[m][][c]{2.0cm}{Two-player\\Leduc poker}} & \fwomd{} & constant & no & 3 & 1.28 & yes &  fixed LMO calls\\
 & \rfwomd{} & constant & yes & 3 & 0.64 & yes &  fixed LMO calls\\
 & \fwomd{} & linear & no & 3 & 1.28 & yes &  fixed LMO calls\\
 & \rfwomd{} & linear & yes & 3 & 0.64 & yes &  fixed LMO calls\\
 & \fwomd{} & quadratic & no & 3 & 1.28 & yes &  fixed LMO calls\\
 & \rfwomd{} & quadratic & yes & 3 & 0.64 & yes &  fixed LMO calls\\
 & \fwomd{} & last & no & 4 & 0.64 & yes &  fixed LMO calls\\
 \midrule
\multirow{7}{*}{\parbox[m][][c]{2.0cm}{Two-player\\Liar's Dice}} & \fwomd{} & constant & no & 3 & 10.24 & yes &  fixed LMO calls\\
 & \rfwomd{} & constant & yes & 3 & 10.24 & yes &  fixed LMO calls\\
 & \fwomd{} & linear & no & 2 & 10.24 & yes &  fixed LMO calls\\
 & \rfwomd{} & linear & yes & 2 & 10.24 & yes &  fixed LMO calls\\
 & \fwomd{} & quadratic & no & 2 & 10.24 & yes &  fixed LMO calls\\
 & \rfwomd{} & quadratic & yes & 2 & 10.24 & yes &  fixed LMO calls\\
 & \fwomd{} & last & no & 3 & 10.24 & yes &  fixed LMO calls\\
\bottomrule
\caption{The parameters for all algorithms shown in \Cref{fig:NE_ablation_averaging_no_optimistic}. The final two columns are irrelevant for algorithms that are not ours and thus are marked as N/A. Additionally, if the number of LMO calls is not fixed for our algorithms, then N/A is used in the maximum number of LMO calls per iteration ($m$) column.}
\label[table]{table:NE_ablation_averaging_no_optimistic}
\end{longtable}

\begin{longtable}{llllrrll}
\toprule
\bf Game & \bf Algorithm & \bf Averaging & \bf Restarts & $m$ & $\eta$ & \bf Warmstart & \bf APO term. crit.\\
 \midrule
\multirow{7}{*}{\parbox[m][][c]{2.0cm}{Two-player\\Kuhn poker}} & \fwromd{} & constant & no & 4 & 1.28 & yes &  fixed LMO calls\\
 & \rfwromd{} & constant & yes & N/A & 1.28 & yes &  $\epsilon = \frac{1}{t^2}$\\
 & \fwromd{} & linear & no & 2 & 2.56 & yes &  fixed LMO calls\\
 & \rfwromd{} & linear & yes & N/A & 1.28 & yes &  $\epsilon = \frac{1}{t^2}$\\
 & \fwromd{} & quadratic & no & N/A & 1.28 & yes &  $\epsilon = \frac{1}{t^2}$\\
 & \rfwromd{} & quadratic & yes & N/A & 1.28 & yes &  $\epsilon = \frac{1}{t^2}$\\
 & \fwromd{} & last & no & N/A & 1.28 & yes &  $\epsilon = \frac{1}{t^2}$\\
 \midrule
\multirow{7}{*}{\parbox[m][][c]{2.0cm}{Two-player\\Leduc poker}} & \fwromd{} & constant & no & 3 & 1.28 & yes &  fixed LMO calls\\
 & \rfwromd{} & constant & yes & 3 & 1.28 & yes &  fixed LMO calls\\
 & \fwromd{} & linear & no & 3 & 1.28 & yes &  fixed LMO calls\\
 & \rfwromd{} & linear & yes & 3 & 1.28 & yes &  fixed LMO calls\\
 & \fwromd{} & quadratic & no & 5 & 1.28 & yes &  fixed LMO calls\\
 & \rfwromd{} & quadratic & yes & 3 & 1.28 & yes &  fixed LMO calls\\
 & \fwromd{} & last & no & 3 & 1.28 & yes &  fixed LMO calls\\
 \midrule
\multirow{7}{*}{\parbox[m][][c]{2.0cm}{Two-player\\Liar's Dice}} & \fwromd{} & constant & no & 2 & 5.12 & yes &  fixed LMO calls\\
 & \rfwromd{} & constant & yes & 3 & 10.24 & yes &  fixed LMO calls\\
 & \fwromd{} & linear & no & 2 & 10.24 & yes &  fixed LMO calls\\
 & \rfwromd{} & linear & yes & 3 & 10.24 & yes &  fixed LMO calls\\
 & \fwromd{} & quadratic & no & 2 & 10.24 & yes &  fixed LMO calls\\
 & \rfwromd{} & quadratic & yes & 3 & 10.24 & yes &  fixed LMO calls\\
 & \fwromd{} & last & no & 3 & 10.24 & yes &  fixed LMO calls\\
\bottomrule
\caption{The parameters for all algorithms shown in \Cref{fig:NE_ablation_averaging_optimistic}. The final two columns are irrelevant for algorithms that are not ours and thus are marked as N/A. Additionally, if the number of LMO calls is not fixed for our algorithms, then N/A is used in the maximum number of LMO calls per iteration ($m$) column.}
\label[table]{table:NE_ablation_averaging_optimistic}
\end{longtable}

\begin{longtable}{llllrrll}
\toprule
\bf Game & \bf Algorithm & \bf Averaging & \bf Restarts & $m$ & $\eta$ & \bf Warmstart & \bf APO term. crit.\\
 \midrule
\multirow{6}{*}{\parbox[m][][c]{2.0cm}{Two-player\\Kuhn poker}} & \fwomd{} & linear & no & N/A & 0.64 & yes &  $\epsilon = \frac{1}{t^2}$\\
& \fwomd{} & linear & no & N/A & 0.64 & yes &  $\epsilon = 10^{-6}$\\
 & \fwomd{} & linear & no & 1 & 0.08 & yes &  fixed LMO calls\\
 & \fwomd{} & linear & no & N/A & 0.64 & no &  $\epsilon = \frac{1}{t^2}$\\
 & \fwomd{} & linear & no & N/A & 0.64 & no &  $\epsilon = 10^{-6}$\\
 & \fwomd{} & linear & no & 10 & 0.64 & no &  fixed LMO calls\\
 \midrule
\multirow{6}{*}{\parbox[m][][c]{2.0cm}{Two-player\\Leduc poker}} & \fwomd{} & quadratic & no & N/A & 2.56 & yes &  $\epsilon = \frac{1}{t^2}$\\
 & \fwomd{} & last & no & N/A & 10.24 & yes &  $\epsilon = 10^{-6}$\\
 & \fwomd{} & quadratic & no & 3 & 1.28 & yes &  fixed LMO calls\\
 & \fwomd{} & quadratic & no & N/A & 2.56 & no &  $\epsilon = \frac{1}{t^2}$\\
 & \fwomd{} & last & no & N/A & 5.12 & no &  $\epsilon = 10^{-6}$\\
 & \fwomd{} & quadratic & no & 100 & 2.56 & no &  fixed LMO calls\\
 \midrule
\multirow{6}{*}{\parbox[m][][c]{2.0cm}{Two-player\\Liar's Dice}} & \fwomd{} & linear & no & N/A & 20.48 & yes &  $\epsilon = \frac{1}{t^2}$\\
 & \fwomd{} & last & no & N/A & 0.64 & yes &  $\epsilon = 10^{-6}$\\
 & \fwomd{} & last & no & 3 & 10.24 & yes &  fixed LMO calls\\
 & \fwomd{} & linear & no & N/A & 20.48 & no &  $\epsilon = \frac{1}{t^2}$\\
 & \fwomd{} & linear & no & N/A & 0.02 & no &  $\epsilon = 10^{-6}$\\
 & \fwomd{} & quadratic & no & 200 & 10.24 & no &  fixed LMO calls\\
\bottomrule
\caption{The parameters for all algorithms shown in \Cref{fig:NE_no_restarts_ablation_no_optimistic}. The final two columns are irrelevant for algorithms that are not ours and thus are marked as N/A. Additionally, if the number of LMO calls is not fixed for our algorithms, then N/A is used in the maximum number of LMO calls per iteration ($m$) column.}
\label[table]{table:NE_no_restarts_ablation_no_optimistic}
\end{longtable}

\begin{longtable}{llllrrll}
\toprule
\bf Game & \bf Algorithm & \bf Averaging & \bf Restarts & $m$ & $\eta$ & \bf Warmstart & \bf APO term. crit.\\
 \midrule
\multirow{6}{*}{\parbox[m][][c]{2.0cm}{Three-player\\Kuhn poker}} & \fwromd{} & constant & no & N/A & 1.28 & yes &  $\epsilon = \frac{1}{t^2}$\\
 & \fwromd{} & constant & no & N/A & 1.28 & yes &  $\epsilon = 10^{-6}$\\
 & \fwromd{} & constant & no & 5 & 1.28 & yes &  fixed LMO calls\\
 & \fwromd{} & constant & no & N/A & 1.28 & no &  $\epsilon = \frac{1}{t^2}$\\
 & \fwromd{} & constant & no & N/A & 1.28 & no &  $\epsilon = 10^{-6}$\\
 & \fwromd{} & constant & no & 4 & 1.28 & no &  fixed LMO calls\\
 \midrule
\multirow{6}{*}{\parbox[m][][c]{2.0cm}{Three-player\\Liar's Dice}} & \fwromd{} & constant & no & N/A & 5.12 & yes &  $\epsilon = \frac{1}{t^2}$\\
 & \fwromd{} & constant & no & N/A & 0.02 & yes &  $\epsilon = 10^{-6}$\\
 & \fwromd{} & constant & no & 20 & 10.24 & yes &  fixed LMO calls\\
 & \fwromd{} & constant & no & N/A & 10.24 & no &  $\epsilon = \frac{1}{t^2}$\\
 & \fwromd{} & constant & no & N/A & 20.48 & no &  $\epsilon = 10^{-6}$\\
 & \fwromd{} & constant & no & 100 & 5.12 & no &  fixed LMO calls\\
 \midrule
\multirow{6}{*}{\parbox[m][][c]{2.0cm}{Three-player\\Goofspiel (3 ranks)}} & \fwromd{} & constant & no & N/A & 20.48 & yes &  $\epsilon = \frac{1}{t^2}$\\
 & \fwromd{} & constant & no & N/A & 40.96 & yes &  $\epsilon = 10^{-6}$\\
 & \fwromd{} & constant & no & 2 & 40.96 & yes &  fixed LMO calls\\
 & \fwromd{} & constant & no & N/A & 20.48 & no &  $\epsilon = \frac{1}{t^2}$\\
 & \fwromd{} & constant & no & N/A & 40.96 & no &  $\epsilon = 10^{-6}$\\
 & \fwromd{} & constant & no & 100 & 20.48 & no &  fixed LMO calls\\
\bottomrule
\caption{The parameters for all algorithms shown in \Cref{fig:CCE_ablation_optimistic}. The final two columns are irrelevant for algorithms that are not ours and thus are marked as N/A. Additionally, if the number of LMO calls is not fixed for our algorithms, then N/A is used in the maximum number of LMO calls per iteration ($m$) column.}
\label[table]{table:CCE_ablation_optimistic}
\end{longtable}

\end{document}